\documentclass[11pt]{article}
\usepackage{graphicx}
\usepackage{tabularx, booktabs}
\newcolumntype{Y}{>{\centering\arraybackslash}X}
\newcolumntype{B}{>{\centering\arraybackslash\hsize=1.5\hsize}X}
\newcolumntype{Q}{>{\centering\arraybackslash\hsize=2\hsize}X}
\usepackage{titlesec}
\usepackage{wrapfig}
\usepackage{epstopdf}
\usepackage{setspace}
\usepackage[top=1.5in, bottom=1.5in, left=1in, right=1in]{geometry}
\usepackage{amsmath,amssymb}
\usepackage[title,titletoc,toc]{appendix}
\usepackage{color}
\newcommand{\1}[1]{\mathds{1}_{ {#1} }} 
 
\usepackage{algorithm}
\usepackage{algorithmic}
\usepackage{caption, subcaption}
\usepackage[round]{natbib}
\usepackage{lipsum}
\usepackage{indentfirst}
\usepackage{hyperref}          

\usepackage{leading} 
\hypersetup{
	colorlinks=true,
	citecolor=blue
}
\graphicspath{ {Figures/} }

\usepackage{dsfont}

\usepackage{setspace}

\title{\vspace{-1.0cm} \bf Fast Forecasting of Unstable Data Streams for On-Demand Service Platforms}

\author{Yu Jeffrey Hu$^{a}$, Jeroen Rombouts$^{b}$\footnote{Corresponding author. E-mail: rombouts@essec.edu, jeffrey.hu@scheller.gatech.edu, i.wilms@maastrichtuniversity.nl.}, and Ines Wilms$^c$ 
	\\ \textit{\small $^{a}$ Daniels School of Business, Purdue University, West Lafayette, Indiana}
	\\ \textit{\small $^{b}$ Essec Business School, France}
	\\ \textit{\small $^{c}$ Department of Quantitative Economics, Maastricht University, The Netherlands}
}

\begin{document}
	
	\begin{titlepage}
		\clearpage\thispagestyle{empty}
		\maketitle
		
		\begin{abstract}
			\begin{singlespace} {\small
					On-demand service platforms face a challenging problem of forecasting a large collection of high-frequency regional demand data streams that exhibit instabilities. 
					This paper develops a novel forecast framework that is fast and scalable, and automatically assesses changing environments without human intervention.
					We empirically test our framework on a large-scale demand data set from a leading on-demand delivery platform in Europe, and find strong performance gains from using our framework against several industry benchmarks, across all geographical regions, loss functions, and both pre- and post-Covid periods. 
					We translate forecast gains to economic impacts for this on-demand service platform by computing financial gains and reductions in computing costs.
					\\
				}
			\end{singlespace}

			\noindent {JEL Classification}: G12 
			\medskip
			
			\noindent {Keywords}: E-commerce; Platform econometrics; Streaming data; Forecast breakdown.
			
		\end{abstract}
		
		\thispagestyle{empty}
	\end{titlepage}

	\doublespacing
	
	\clearpage

\section{Introduction} \label{Introduction}
On-demand service platforms have experienced explosive growth in recent years.
Well-known examples include Uber, Lyft, and Didi for ride hailing,
Deliveroo, DoorDash, Uber Eats, and new players such as Getir and Gorillas for food deliveries, or Glamsquad and Zeel for in-home healthcare and beauty services. 
On-demand service platforms nowadays receive a lot of research attention in economics and management\footnote{Early fundamental work on two-sided platforms in economics and management can be found respectively in
\cite{Rochet_Tirole_2003} and \cite{Parker_VanAlstyne_2005}.}, e.g.,  \cite{Chen_JPE_2019}, \cite{Guda_Subramanian_mansci2019}, and \cite{BARRIOS202222}. In this paper, we focus on the demand forecasting problem faced by on-demand service platforms. This problem is critically important because a platform's success crucially hinges upon its ability of making fast and accurate demand forecasts such that its service providers are at the right time and location to serve consumer demand promptly.\footnote{In a more general way, consumers have preferences on their willingness to wait and to pay, allowing strategic timing and dynamic pricing such as proposed by \cite{Abhishek_2021}.} Faster and more accurate demand forecasts can lead to a better matching of demand and supply through smarter order assignment or surge pricing strategies (e.g., \citealp{Liu_mansci_2021}). 

The demand forecasting problem faced by on-demand service platforms is one of the wicked problems listed by \cite{Hevner_March_Park_Ram_2004} due to the salient features of demand data generated by such platforms. First, it consists of high-frequency streaming time series as data are typically sequentially updated through the frequent release of new data streams.
Second, on-demand service platforms divide the marketplace in which they operate into more granular regions, 
thereby giving rise to a large heterogeneous geographical collection of high-frequency demand series to be managed and forecast.
Finally, on-demand service platforms usually operate in unstable, rapidly changing environments and face irregular growth patterns which requires agility when forecasting demand. 
Slow reactions to such instabilities cause forecast performance to break down. 
Existing forecast methods only address these three issues (high-frequency streaming time series, large-scale settings, forecast breakdown caused by instabilities) mostly in isolation. 
Therefore, it is of high practical relevance and research value to design an integral solution tailored towards these notable features of demand data for on-demand service platforms, enabling better and faster decision-making.

This paper takes a design science approach as defined by \cite{Hevner_March_Park_Ram_2004}; it develops a novel forecast framework that is fast and scalable, and automatically assesses changing environments without human intervention. 
We collaborate with a leading on-demand delivery platform in Europe that faces the challenge of forecasting high-frequency and unstable demand for many geographical regions. First, we introduce a novel approach to account for unstable environments in which agile demand forecasts need to be obtained.
On-demand service platforms typically operate in a changing environment with shocks such as entry of competitors, external events, or internal campaigns. 
We demonstrate that our platform demand data show strong evidence of ``change-points" (also called ``structural breaks") which have to be taken into account when forecasting demand.
Earlier work by, for instance, \cite{kremer2011demand} has shown that forecasters typically over-react to forecast errors in stable environments but under-react in unstable environments.
We develop an innovative data-driven approach to detect breaks in forecast performance by adapting state-of-the-art, yet fast-to-compute, change-point techniques \citep{killick2012optimal}. 
We define a forecast breakdown as a break in the streaming forecast error loss. 
When such a break is detected, we, in the spirit of \cite{PESARAN2007134}, combine forecasts from our model based on the full-sample and post-break estimation windows. 
As such, break information is incorporated into the forecasts but we still use the full historical sample to stabilize the forecasts.

Second, 
we develop an innovative demand forecast framework built on regression models that incorporate seasonality, trends, and autoregressive dynamics that may change throughout the business day.  
The model is parametric and can be estimated in closed-form with ``renewable" estimators so that parameters can be efficiently updated sequentially when new data batches come in, e.g., \cite{Luo_2020_renewable}. 
Given the large heterogeneity across different geographical regions, we estimate a parametric model for each geographical region separately.
Computational speed is an issue because, for the purpose of operational planning, each geographical region requires real-time forecasting for the next few hours. 
Our forecast framework runs very fast and uses the same parametric model structure for each geographic region, thereby allowing deploying it at scale and centralized performance monitoring.  

\begin{figure}[t]
\centering
\includegraphics[width = 0.45\textwidth]{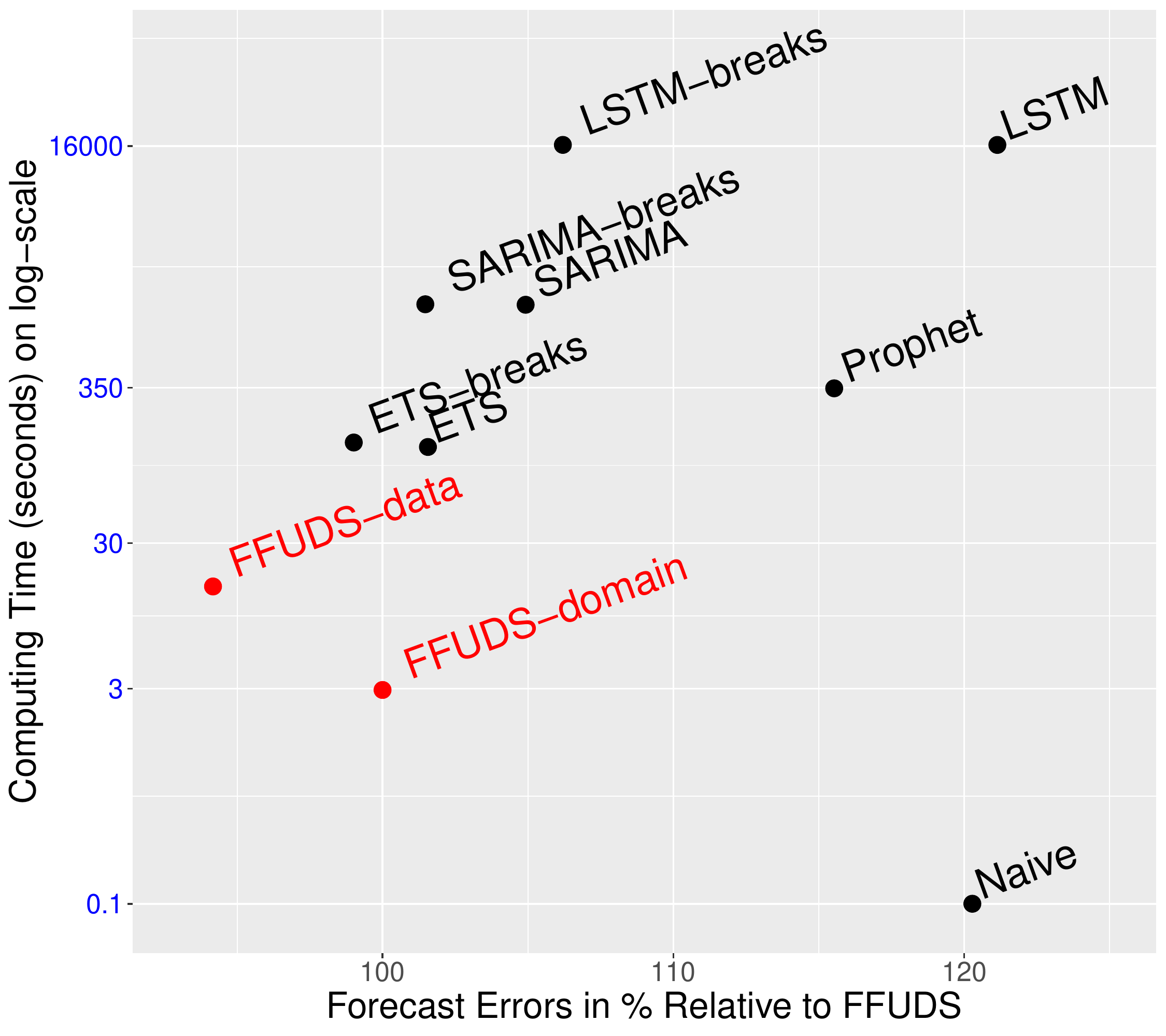}
\caption{Performance of the new method.}  
\label{fig:2D_plot_errors_vs_time}
\footnotesize \raggedright Notes:  Forecast errors in percentages versus computing time in seconds (in blue) for the forecasting methods: FFUDS-domain and FFUDS-data (ours in red), LSTM (standard and with breaks), Prophet, SARIMA (standard and with breaks), ETS (standard and with breaks) and the Naive method.\\[0.1cm]
\end{figure}

Finally, we name our forecast framework ``Fast Forecasting of Unstable Data Streams (FFUDS)" and empirically test it on a large-scale demand data set from a leading on-demand delivery platform. Our application hereby contributes to the information systems literature on smart sustainable mobility as mentioned by \cite{Ketter_ISR_2023}. The platform faces hourly demand between January 2019 and March 2021 for 294 geographical regions that constitute the UK market. The large amount of regions that we separately model allows for a confident assessment of the method's performance. 
As for most companies, the Covid-19 pandemic has a profound impact on the platform and gives rise to a rapidly changing, challenging environment in which forecast performance needs to be optimized.
We detect streaming forecast breakdowns and evaluate forecast performance using three loss functions, ranging from  popular statistical losses (squared errors and symmetric absolute percentage errors) towards economic losses that penalizes more when actual demand is larger than the forecast.
Besides, we offer two versions of FFUDS: one that allows for domain expertise to be incorporated in the demand model and one that is fully data-driven.
We compare their performance  against a popular industry benchmark that does not require parameter estimation, which we label as the ``Naive" model,  and a scalable state-of-the-art forecast method, namely the forecast package Prophet proposed by Facebook \citep{prophet_2018} which is currently employed as the demand forecast procedure at this platform. As final benchmarks, we also compare  against traditional time series forecasting models SARIMA and ETS and recurrent neural networks of the long-short-term-memory (LSTM) type.
It turns out that overall substantial forecast gains are obtained compared to the above benchmarks across all geographical regions, loss functions, and both pre- and post-Covid periods. For the post-Covid period in particular, i.e., from April 2020 to March 2021, the benchmarks 
perform on average (across all areas) in between 1\% (ETS) to 20\% (Naive) 
worse than our framework, see horizontal axis in Figure \ref{fig:2D_plot_errors_vs_time}, which will be further discussed in Section \ref{sec:forecast_results}.\footnote{{For data confidentiality reasons, we scale the forecast performance of FFUDS (with domain expertise) to 100\% and report the performance of the other methods relative to it.}}
Importantly, the key backbone of FFUDS, namely its ability to handle forecast breakdowns, can be paired with any forecast method, including the benchmarks LSTM, ETS and SARIMA, whose standard implementation does not offer breakdown detection. For all three methods, Figure \ref{fig:2D_plot_errors_vs_time} reveals that this addition may substantially improve forecast performance, thereby underwriting the generalizability of our proposed framework for a wide range of forecast methods. Finally,
we also translate these forecast gains to economic impact across all UK areas by computing monetary differences in cumulative monetary loss between our framework and the benchmarks and find major differences up to roughly 
\pounds3,000,000 per year.
In terms of computing cost, measured as the time it takes for one overnight sequential forecast update across all geographical regions, we find that 
the fastest implementation of FFUDS is over 
45 (compared to ETS) to 5,500 (compared to LSTM) times faster
(see vertical axis in Figure \ref{fig:2D_plot_errors_vs_time}), thereby freeing up substantial computing resources and/or computing time. To highlight its generality, a second empirical test of the FFUDS framework is carried out to a public bicycle sharing system serving the New York City area.

This paper makes several contributions. First, there is an emerging literature in information systems that takes a design science approach to solving forecasting problems in various business contexts. We extend this literature to a business context with high-frequency and unstable time series data for many geographical regions, by developing a novel, integral forecasting framework tailored towards the notable features of demand data for on-demand service platforms. The framework developed in this paper performs better than the popular approach in this literature that often uses some components of recurrent neural networks of the LSTM type (e.g.,  \citealp{Liu_Wang_Fan_Zhang_2020}, \citealp{Wang_Currim_Ram_2022}, \citealp{Wu_Zheng_Zhao_2021}). 

Second, on-demand service platforms are typically subject to a changing environment due to factors such as expansion, competition, or external events; as a result, their demand data shows strong evidence of  ``change-points"  which must be accounted for when forecasting demand. There is a recent literature in both economics and management on ``forecast breakdown" and how to improve the forecasting ability of models in the presence of instabilities (see \citealp{Rossi_JEL_2021} for a recent survey of this literature). This paper extends this literature by using a data-driven approach to detect breaks in forecast performance, leading to advantages such as offering a fast and scalable forecast procedure and alleviating the need to rely on human judgment to assess changing environments. 

The rest of the paper is organized as follows. 
Section \ref{literature} reviews the relevant literature streams to which we contribute.
Section \ref{PlatformData} describes the typical nature of platform streaming data and discusses how the proposed FFUDS framework is set up to address the prediction challenges these data face. 
Section \ref{Model} describes our proposed forecast framework FFUDS 
for streaming forecast breakdowns.
Section \ref{sec:platform-application} introduces the specific platform application we consider.  
Section \ref{sec:forecast_results} contains the empirical assessment of our approach  to all UK delivery areas. 
Section \ref{sec:implications} quantifies economic impact and discusses managerial insights of implementing an enhanced forecast procedure as FFUDS. 
To showcase its generality, Section \ref{app:citibike}  considers a second application of FFUDS on a public bicycle sharing system.
Section \ref{Conclusion} concludes and lays out future research directions.

\section{Literature Review} \label{literature}
This paper is related to several lines of research. 
First, there is a burgeoning literature in information systems that takes a design science approach to solving forecasting problems in various business scenarios, as described by \cite{Hevner_March_Park_Ram_2004}. In this literature, 
recurrent neural networks for sequential data and in particular Long-Short-Term-Memory (LSTM) models have gained popularity in recent years. For instance, \cite{Wang_Currim_Ram_2022} develops a deep learning model that combines LSTM components and embedding results to forecast passenger flows between pairs of city regions. LSTM models have been applied by \cite{Wu_Zheng_Zhao_2021} to predict customer misbehavior in social media, by \cite{Liu_Wang_Fan_Zhang_2020} to identify useful solutions in online knowledge communities. 
LSTM models, however, require substantial computing time to estimate the model, even more so if hyper-parameter tuning is considered. Moreover, they do not account for breakdowns in forecast performance. As such, they are not tailored towards the above specificities of the forecast problem faced by on-demand service platforms operating in changing environments. 

In the supply chain and operations management literature, demand forecasting is crucial to any company as it forms the basis for its planning and control activities. Accurate demand forecasts are therefore crucial elements of sales and operations planning whose purpose is to balance supply and demand (e.g., \citealp{petropoulos2022forecasting} and references therein). Traditional time series methods such as ARIMA and ETS (e.g., \citealp{forecastR_v2}) have a long-standing tradition to deliver such demand forecasts when historic demand data are available. Today, however, there is a growing need for forecasting at large-scale because of the large volume of related products and customers whose demand is recorded at a high velocity due to the many transactions that are continuously processed \citep{seyedan2020predictive}. 
A growing body of research therefore looks into the usage of machine learning methods  such as recurrent neural networks and deep learning  to forecast demand  because of their advanced learning and forecasting capabilities (e.g., \citealp{zhang2023proactive}), see
\cite{aamer2020data} for a recent review.

While there is a growing interest in using machine learning methods for demand forecasting tasks, recent studies also point towards the potential of incorporating human judgment since domain experts can add value to machine generated forecasts, see \cite{perera2019human} or \cite{brau2023demand}  for supply chain forecasting reviews, and  \cite{Yang_ISR_2023} for a deep learning application to text-based measurement of personality and its use to forecasting future firm performance. It is, however, often hard to understand how machine learning systems generate forecasts.   As a consequence, results on the effectiveness of  judgmental forecast adjustments are  often mixed. For instance, \cite{khosrowabadi2022evaluating} report that large positive forecast adjustments typically occur more  frequently but are often inaccurate, whereas large negative adjustments occur less frequently but are generally more accurate;  \cite{kremer2011demand} find that
forecasters typically over-react to forecast errors in stable environments but under-react in unstable environments.  Yet, deciding the right time to refresh knowledge to support decision making from massive data sets is a fundamental challenge in a wide range of critical applications, see \cite{SE_OR_2013}.

In economics, management, and statistics, there is a literature on how to detect forecast breakdowns and how to improve the forecasting ability of models in the presence of instabilities (e.g., \citealp{Rossi_JEL_2021} for a literature survey). \cite{Giacomini_Rossi_2009} have pioneered the ``forecast breakdown" research by proposing a breakdown test that requires the entire historic time series; however, such a method is not suitable in our streaming data setting in which only part of the time series is observable. More recently, one stream in this literature considers break/regime detection in the \textit{original} time series (i.e., the demand data time series in our case). For instance, \cite{killick2012optimal}, \cite{fryzlewicz2014},  \cite{Aminikhanghahi_2019} and \cite{Chen_Wang_Samworth_2022} develop nonparametric change-point tests, thereby including real-time change-point detection settings. Another stream in this literature studies break detection in the \textit{parameters} modeling the dynamics in time series (e.g., \citealp{Dufays_Rombouts_JoE_2020}). Finally, there are also papers such as \cite{Chib98} that study Markovian dynamics that regulate the transition between breaks. 

In computer science and machine learning, the incorporation of nonstationarity and change-points also receives  increased attention in various settings. For example, \cite{NEURIPS2018_a3f390d8} propose a robust Bayesian change-point detection algorithm for nonstationary streaming data. \cite{Luo_JASA_2022_batches} consider models for streaming clustered data with possible abnormal data batches.
\cite{Federici_neurips_2021} and \cite{Zhang_neurips_2021} provide theoretical frameworks for dealing with data distribution shifts.
Another important application is dynamic pricing, or more generally multi-armed bandit problems, where \cite{Cheung_mansci_2022} demonstrate that drifting parameters can be estimated with a sliding window approach instead of using all available historical data. Finally, \cite{Cao_2019} address piecewise-stationary bandits with applications to digital marketing. Recent overviews on change-point detection methods can be found in \cite{aminikhanghahi2017survey} and \cite{vandenburg2020evaluation}.

In the context of detecting breaks in steaming data, the 
Facebook Prophet method \citep{prophet_2018} forms an interesting benchmark to highlight. It is
a popular forecasting tool among practitioners that decomposes a time series into a trend, seasonal and holiday parts, handles out-of-sample forecast uncertainty, and in addition 
offers automatic change-point detection by including a sparse prior on the trend change parameter. 
These features require, however, Bayesian inference and thereby come at the cost of losing fast (closed-form) streaming estimators as desired in our case. Moreover, as opposed to parameter instability,  we are interested in detecting \textit{forecast breakdowns}, so instability in the forecast performance as measured by a forecast error loss function over time. 

Our literature review therefore reveals several important research gaps. 
In summary, there is no integral  framework tailored towards the forecasting challenges faced by on-demand service platforms. 
We therefore contribute to the information systems literature by offering such a framework that combines three main methodological contributions:
(i) a fast and scalable demand forecasting model, with
(ii) automatic forecast breakdown detection and 
(iii) a forecast combination approach where break information is incorporated to adjust forecasts to a changing environment and full historic data is utilized to stabilize the forecasts. Therefore, our forecasting framework is both accurate and fast-to-compute, and tailored towards the streaming data setting of on-demand service platforms.
In the next section, we highlight the typical features of  platform streaming data and discuss how the proposed FFUDS architecture is generally set up to tackle these.

\section{Framework Overview}
\label{PlatformData}

\subsection{Distinctive Features of Platform Streaming Data}\label{subsec:platform_features}
We refer to platform data when this data is used for real-time decision making at the digital core of an on-demand service platform. Platform data is different from cross-sectional or time series data because of the following  reasons.
First, platform data are streaming data, i.e.\ they are generated sequentially, typically with new batches entering at high frequency. The speed of integrating such incoming data for automated decision making is vital at any  digital platform.
Second, platform data is high-dimensional as they operate in hundreds to thousands of 
geographical areas (markets), sectors, product categories, etc. Hence, decision making requires forecasting at scale. 
Third, given their high frequency and granularity, platform data can be sparse meaning that low to zero volume for specific or extended time periods are observed. Such a feature has to be taken into account when modeling the data.
Fourth,  platform data streams are  subject to frequent changes since they highly depend on volatile market conditions due to competition, events and changing customer preferences. Algorithms using platform data therefore need to rapidly adapt to prevent performance drops. This final characteristic makes platform data  different from regular streaming data which are pervasive in practice and for which sequential updating and inference has been put forward by the seminal work of \cite{Luo_2020_renewable}.

The above-described features of platform data pose considerable challenges to generating accurate, automated forecasts in a timely way. Next, we provide a high-level overview of the FFUDS framework we propose, thereby explaining how the framework's design is set up to handle each of the data challenges specific to platform streaming data.

\subsection{Outline of FFUDS} \label{subsec:FFUDS-framework}
We propose an innovative data-driven method to deliver accurate forecasts for unstable data streams at large scale.
We name our framework FFUDS, Fast Forecasting of Unstable Data Streams, to reflect this overall aim.

\begin{figure}[t]
\centering
{\includegraphics[width=0.85\textwidth]{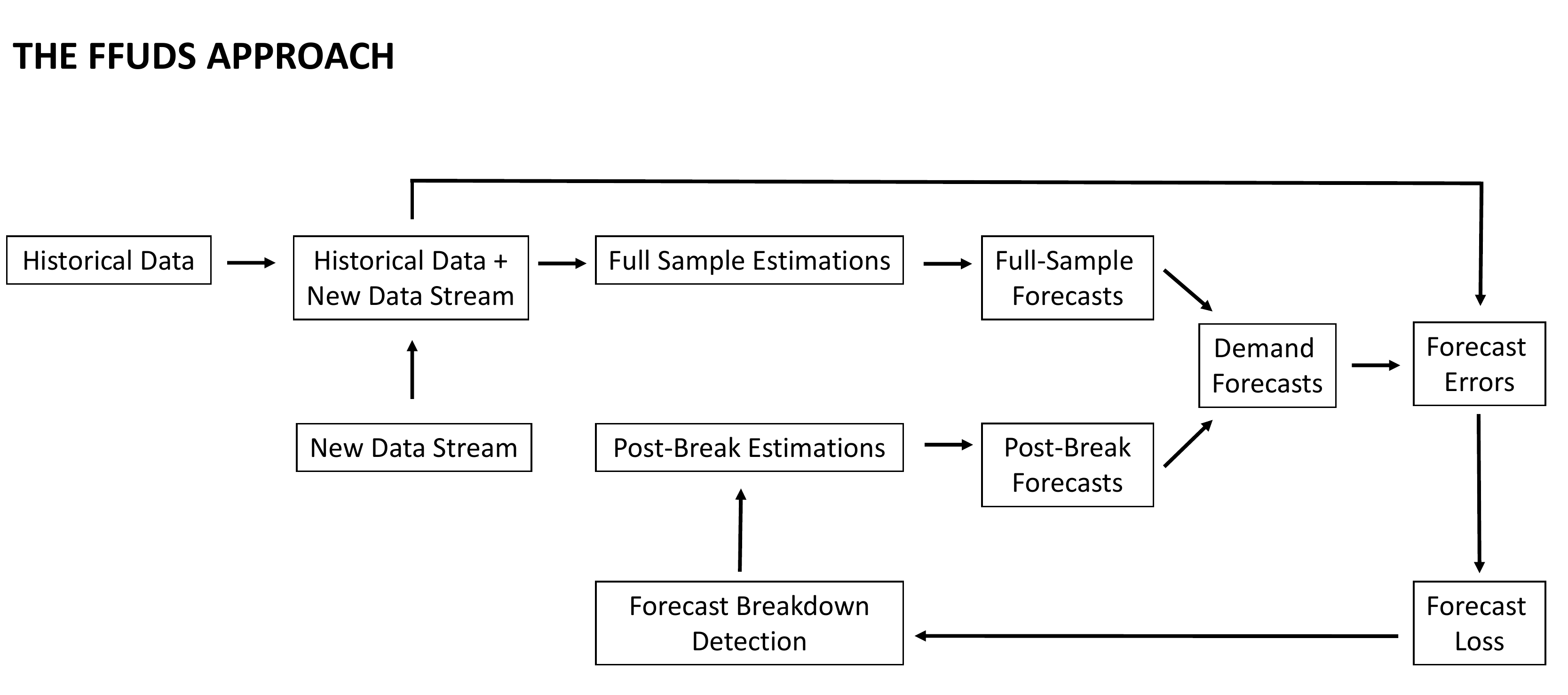}}	
\caption{The Fast Forecasting of Unstable Data Streams (FFUDS) Approach.}\label{fig:FFUDS_approach}
\end{figure}

Figure \ref{fig:FFUDS_approach} provides a schematic outline of the  FFUDS framework. 
Assume we are operating in an environment where historical data are continuously updated with new data streams upon the arrival of a new data batch.
A traditional forecasting approach would use the full-sample (i.e.\ all historical data including the new data stream) to re-estimate the model and produce forecasts, which we call full-sample forecasts.
This approach is visualized by the top four rectangles  in Figure \ref{fig:FFUDS_approach} and works well in stable data settings. 
However, to guarantee solid forecast performance in a platform data setting, we add two key features which consists of our main methodological novelty:  a forecast breakdown detection and a forecast combination component, as visualized in the bottom layer of Figure \ref{fig:FFUDS_approach}. Here, we first screen the streaming forecast losses, based on the forecast errors, 
for breakdowns. 
The forecast errors are the difference between the actuals and forecasts, as supplied via the two incoming paths into the forecast errors box.
Once a forecast breakdown is detected, 
we then combine the forecast based on the full-sample and the forecast based on the post-break sample to arrive at the final demand forecasts.
The forecast combination  is essential to
further monitor the forecast losses for breakdowns as it allows us to
simultaneously incorporate break information into the forecasts via the post-break sample and stabilize the forecasts via the full historical sample. The forecast loss stream  is then used for further forecast breakdown detection which, each time a break is detected, determines the starting date of the next post-break sample/estimation, hence the arrow to the post-break estimations box.

Given the model class we consider, this yields a forecasting framework that is accurate, fast-to-compute, and tailored towards the streaming data setting of on-demand service platforms. Indeed, first, the streaming data aspect is accounted for through efficient so called ``renewable" estimation upon arrival of each new data stream.
Second,  FFUDS  can be applied at large scale, in particular it can be separately implemented for each data stream. 
Third, estimation can be tailored towards the occurrence of sparse data points.
Moreover, forecast losses can be computed via different loss functions, thereby giving more or less weight to  sparse data points, as we will illustrate.
Fourth, the backbone of FFUDS, namely the forecast breakdown detection and corresponding forecast combination approach, is tailored towards the frequent changes the data streams are subject to and the prevention of performance drops.

The key backbone of FFUDS, the bottom layer set up to forecast unstable data streams, can be paired with any forecast method the practitioner prefers. In the next section, we provide details on  FFUDS and put forward a demand forecast method that delivers forecasts fast, which is crucial in environments where speed is vital, such as the ones where on-demand service platforms operate in.

\section{The FFUDS Framework}\label{Model}
The role of any data stream forecast system is to provide inputs for optimal decision making. 
We propose a novel framework that offers accurate forecasts when confronted with unstable data streams.
In Section \ref{subsec:stream_breakdowns} we discuss how 
we detect forecast breakdowns in data streams, in Section \ref{loss_functions} we motivate our choice to use different forecast loss functions to this end.
In Section \ref{baseline}, we introduce the baseline demand model that we use to produce forecasts in a minimal amount of time. 
These sections combined give rise to our Fast Forecasting of Unstable Data Streams (FFUDS) forecast framework, of which we give the algorithmic implementation in Section \ref{subsec:ffudsalgorithm}.

\subsection{Streaming Forecast Breakdowns} \label{subsec:stream_breakdowns}
We propose a procedure to detect \textit{instability} in forecast performance since a sudden increase  in forecast error loss signals the need to re-initiate the data set to estimate the model. 
In terms of notation, we denote the demand at deterministic frequency (e.g.\ minute, hourly, daily, etc.) for a specific 
data stream $i$ as $d_{it}$. 
Note that our procedure is designed to apply to all 
data streams (i.e. $i=1,\ldots,n$) simultaneously, though for notational simplicity we will refer to $d_t$ in the following discussion. Streaming demand observations $d_t$ become available in batches at lower frequency and this set of observations is denoted by $D_b$ with $b=1$ the initial data batch, and subsequent batches for $b=2,3,\dots$.  All data batches are of equal size $H$.

To test for a breakdown in  forecast performance at data batch frequency, define the forecast loss function for 
$b > 1$
and  horizon {$h=1,\ldots,H$ as $L_b^{(h)} = L(d_{H(b-1)+h}, d_{H(b-1)+h}^f)$ 
with  $d_{H(b-1)+h}^f$ denoting demand forecasts.\footnote{For ease of exposition, we keep the forecast horizon and batch size equal to $H$ but in practice these can differ, as is the case in our platform application below. 
Besides, we use separate notation for the sampling frequency of the original data ($h$ for hourly in our applications) and the data batch/loss frequency ($b$ for batches or business days in our applications) to denote time in the subindex. This makes the notation less standard but we need it to properly  define forecasts and their  conditioning information sets.
}
The time series $\hat L_b^{(h)}$,  having estimated $d_{H(b-1)+h}^f$ and which we denote $\hat d_{H(b-1)+h}^f$, is used for forecast breakdown detection.
While forecasts can be produced by any algorithm or model, for reasons outlined in the next section, we take $d_{H(b-1)+h}^f = \mbox{E} (d_{H(b-1)+h}\mid I_{H(b-1)})$, where $\mbox{E}( \cdot \mid \cdot )$ is the conditional expectation operator and $I_{H(b-1)}$ the information set at time $b$ since the last break, if any. 

We  develop a new forecast breakdown procedure tailored towards platform data streaming settings. In particular,
we compute the  time series $\hat L_{b} = \sum_{h=1}^{H} \hat L_b^{(h)}$ and build on the common approach in the change-point literature  of detecting breaks through minimizing a cost function over possible numbers and locations of break dates. We use the computationally efficient 
PELT algorithm of \cite{killick2012optimal}, available in the \verb|R| \citep{Rcoreteam} package \verb|changepoint| \citep{killick2014changepoint}, with a normal likelihood test that allows for changes in both mean and variance.\footnote{In Section \ref{subsec:ablation}, we consider a wide range of alternative change-point detection methods and compare them to PELT.}
Each time a break is detected at time $b$, we reset the ``post-break sample" of streaming data, thereby starting from the first available observation after the earliest detected break\footnote{When a change-point detection test returns more than one detected break, we recommend taking the first break date to define the post-break sample, see Section \ref{subsec:ablation}.} (i.e.\ the first hour of the next business day $b+1$ in our application). Similarly, we reset the forecast loss time series.

Once a forecast breakdown is detected, the question then arises how to adjust the demand forecast procedure. 
On the one hand,  ignoring the detected break is likely to lead to inferior forecast performance at some point in time. 
On the other hand, using post-break data to produce forecasts is not always directly possible due to lack of data available after the detected break to estimate model parameters. In addition, especially in case of small breaks, there is large uncertainty to  define the post-break period. 
As documented for macroeconomic data by \cite{MarcellinoStockWatson06} and \cite{Bauwens2013}, incorporating breaks in the model does not lead uniformly to forecast gains.\footnote{See \cite{Boot_Pick_2020} for a recent theoretical discussion.}
Given the uncertainty around the break dates and their size, using only data after an estimated break to estimate the model can yield worse forecasts than  ignoring the breaks.

In between these two extremes,  \textit{forecast combinations}  of full sample  and post-break forecasts can be considered. \cite{PESARAN2007134}  show that pooling forecasts from the same model but with different estimation windows,  i.e.\ using pre-break and post-break data, can improve out-of-sample forecast performance. 
To produce actual forecasts, \cite{PESARAN2007134}  consider both  simple average  and weighted forecast combinations. 
General results on optimal forecasts and weighting schemes in presence of breaks can be found in  \cite{PESARAN2013134}. 

We use a forecast combination approach of the full-sample and post-break forecasts.
The full sample model is 
deployed upon arrival of every new incoming data batch, is therefore  always ``on", and full-sample forecasts are simply computed from the estimated regression model on the full historical sample.
As soon as a break is detected, also the post-sample model is ``on" and we obtain post-sample forecasts from the  estimated regression model on the post-break sample only.\footnote{Note that if the post-break sample is too small to estimate the demand model, it can not be deployed. Then, the full historical sample-based forecast temporarily gets all the weight until sufficient new post-break data has become available, see Section \ref{subsec:ffudsalgorithm} for further details.}
We then combine both forecasts into a final demand forecast, and opt for equal weights.
 We opt for this  approach 
mainly because of its simplicity, ease in implementation and its established track-record of good performance in the forecast combination literature (e.g., \citealp{Elliott16}, Chapter 14).
In  fact, while procedures to  tune the forecast combination weights exist, \cite{CLAESKENS2016754}  provide a theoretical explanation for the stylized fact that forecast combinations with estimated optimal weights often perform poorly in applications.
Concretely, we estimate the model parameters with the full historical sample and with the post-break sample, produce multi-step ahead forecasts for both and then average these for each horizon.  

\subsection{Loss Functions} \label{loss_functions}
We use several types of loss functions $L$. This allows gaining insights on the sensitivity of our approach with respect to the loss function, and painting a broader picture when comparing its performance against benchmarks.
Figure \ref{fig:lossfunctions} illustrates the differences between the considered loss functions. We discuss the two statistical loss functions (Squared loss and Sape) and an economic loss function (Econ loss) which will be implemented in Section \ref{subsec:econ_impact} on economic impact.

\begin{figure}[t]
	\centering
	\begin{subfigure}[b]{0.25\textwidth}
		\vspace{0.2cm}
			\includegraphics[width=\textwidth]{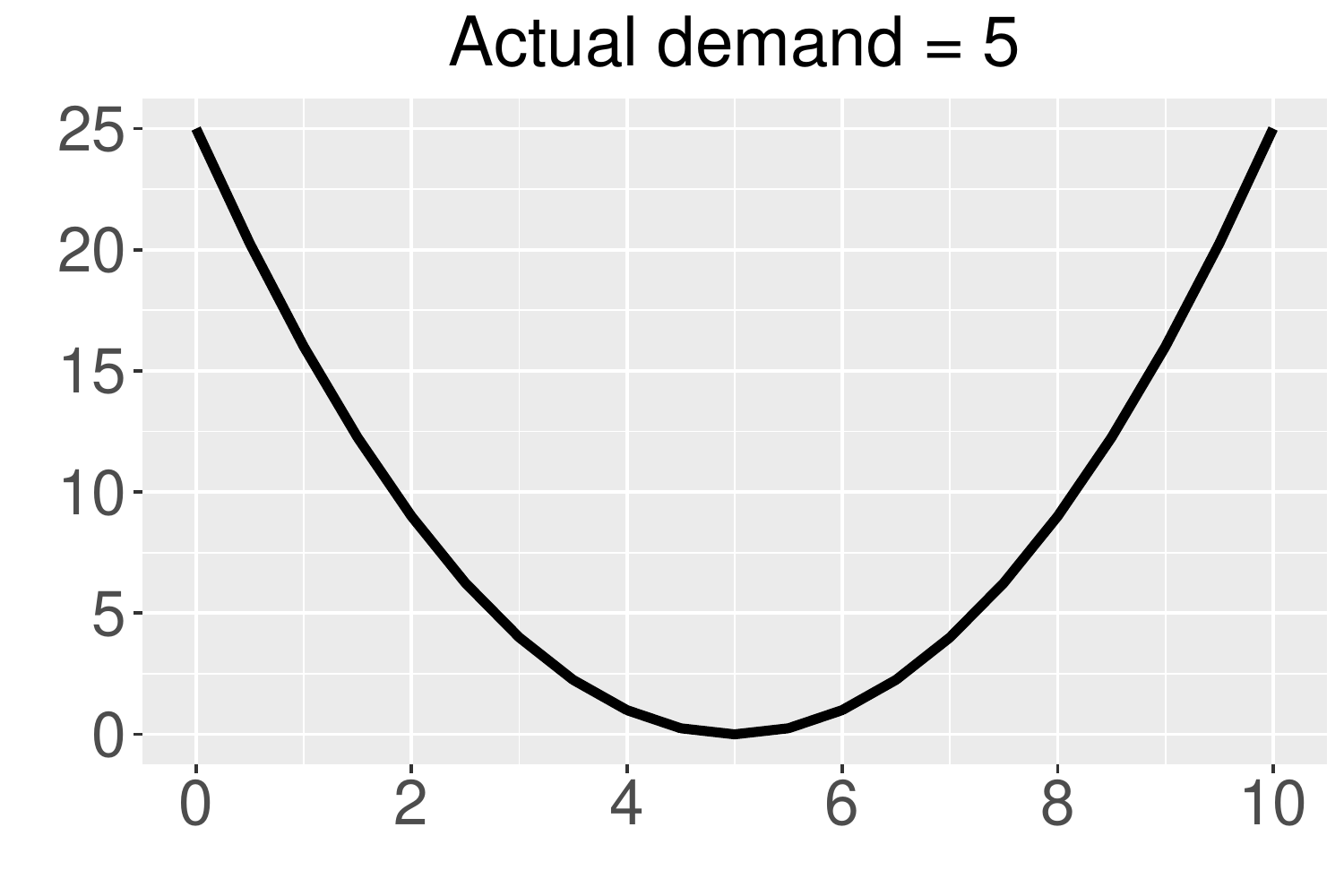}
		\caption{Squared loss} 	\label{squared_loss_5}
	\end{subfigure}
	\begin{subfigure}[b]{0.25\textwidth}
		\vspace{0.2cm}
		\includegraphics[width=\textwidth]{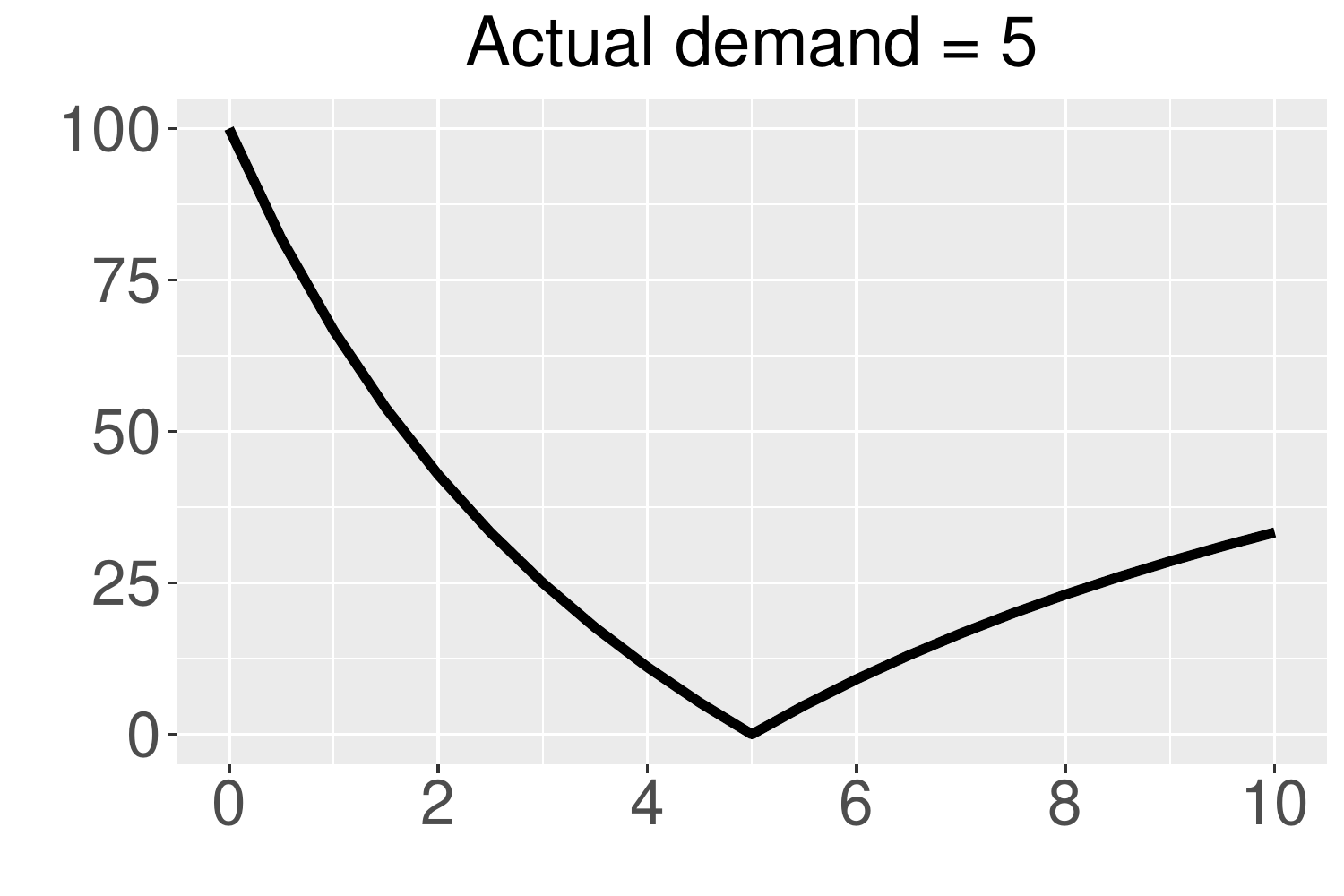}
		\caption{Sape loss} 
				\label{SMAPE_loss_5}
	\end{subfigure}
	\begin{subfigure}[b]{0.25\textwidth}
		\vspace{0.2cm}
		\includegraphics[width=\textwidth]{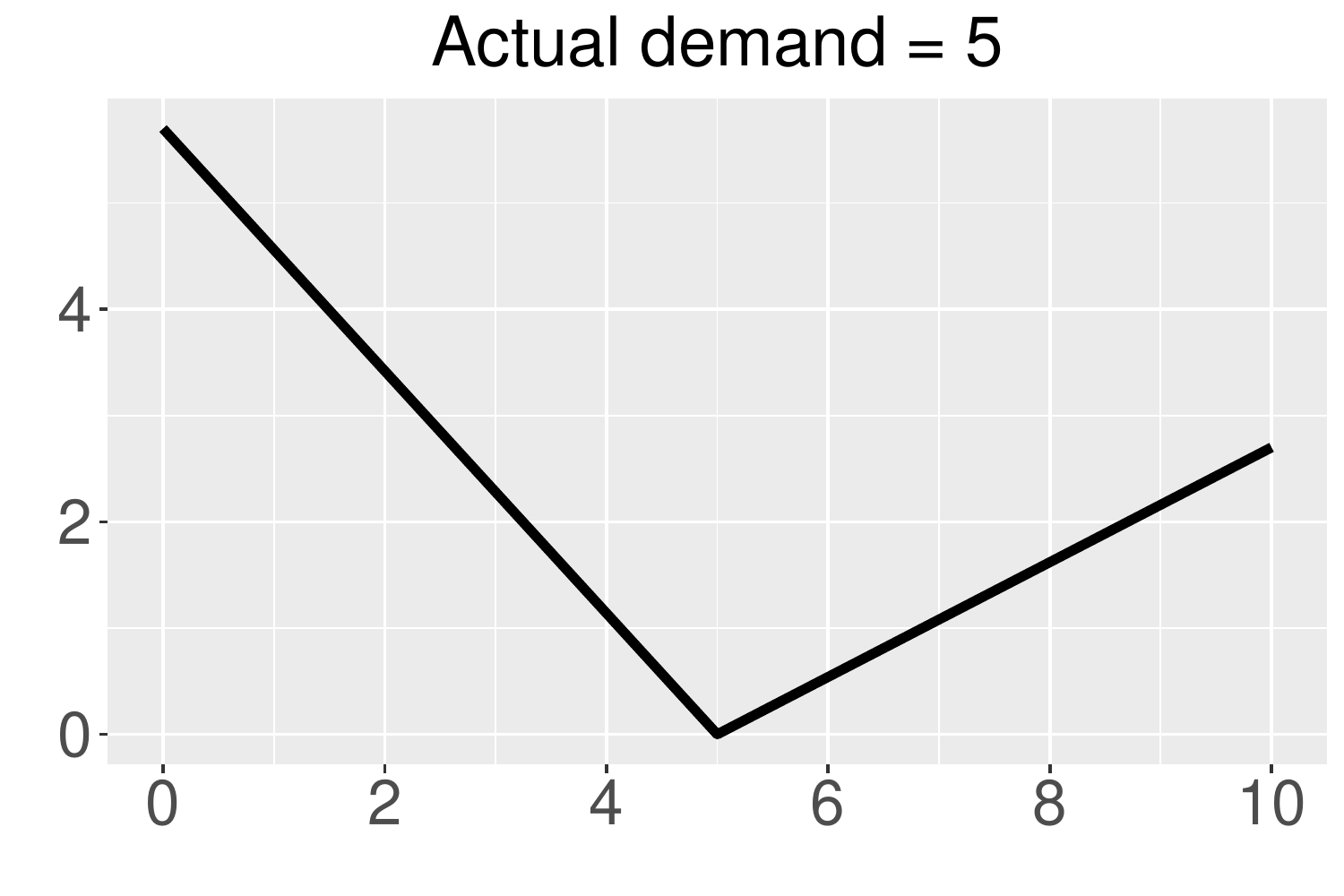}
		\caption{Econ loss} \label{econ_loss_5}
	\end{subfigure}
	\caption{Forecast error loss functions.} \label{fig:lossfunctions}

  \footnotesize \raggedright Notes:  Squared, sape and economic loss functions. On the x-axis are forecast values between 0 and 10. On the y-axis are the corresponding values of the respective loss functions for actual demands equal to five. \\[0.1cm]
\end{figure}

The first loss function is the classical squared error loss $L^{\mbox{sq}} (d_t,d_t^f)=(d_t - d_t^f)^2$ which is symmetric and quadratically increasing away from $d_t$, see Figure 
\ref{squared_loss_5}.  We use the root mean squared error (RMSE) which takes the square root of the average squared error losses when evaluating forecasts.
For evaluation 
of out-of-sample forecast accuracy,
we average over $L^{\mbox{sq}} (d_t,d_t^f)$ losses across the different out-of-sample time points, take the square root and call this the RMSE loss function.

Secondly, we study the symmetric absolute percentage error  loss  $L^{\mbox{sape}} (d_t,d_t^f)= 100 \mid d_t - d_t^f \mid / (|d_t| + |d_t^f|)$,  a popular  metric used at  platforms due to
its  relative nature, thereby facilitating comparisons across 
data streams, and the fact 
that demand $d_t$ for low-volume streams
can  be zero making the standard absolute percentage error loss $L^{\mbox{ape}} (d_t,d_t^f)= 100 \mid (d_t - d_t^f )/ d_t \mid $ not defined. However, the $L^{\mbox{sape}}$ comes with its own peculiarities  such as 
it being undefined when both actual and forecast demand are zero (i.e.\ perfect forecast setting), 
or its inability to differentiate forecast accuracy when actual demand is zero, then regardless of the value for $d_t^f$, a value of 100 is returned. 
When actual demand is non-zero, Figure \ref{SMAPE_loss_5}   highlights the asymmetric shape of $L^{\mbox{sape}}$ (for the case of $d_t=5$).
For out-of-sample forecast accuracy evaluation,
we average over $L^{\mbox{sape}} (d_t,d_t^f)$ losses across the different out-of-sample time points and call this the SMAPE loss.

The above discussed loss functions are traditional ``statistical" choices in the sense that they measure how far the forecast is from its realization. Rather than statistical loss, it might be more appealing for a platform business to measure forecast performance using an “economic" loss function by converting the forecasts into a money implied metric.
To assess the specific economic impact of forecast errors on the platform,  we define the following user-defined loss function 
$L^{\mbox{econ}} (d_t,d_t^f)= c_1 (d_t - d_t^f )  \1{d_t > d_t^f} + c_2 (d_t^f - d_t )  \1{d_t^f > d_t}$ where $\1{(\cdot)}$ denotes the indicator function.
The hyperparameter $c_1$ ($c_2$) represents the cost of under (over) forecasting demand. Asymmetry is introduced in this loss function when $c_1 \ne c_2$. Figure \ref{econ_loss_5} illustrates this for $c_1=1.14$ and $c_2=0.54$, which are the cost values provided by the platform in our application. We clearly see that too low forecasts are far more costly than forecasts exceeding actual demand.

\subsection{A Fast Baseline Demand Model} \label{baseline}
To employ our streaming forecast breakdown method, we require an approach to model and forecast demand data.
 Given our setup, a key feature that our demand model calls for is \textit{fast} estimation.
Based on the platform data characteristics from Section \ref{PlatformData}, we therefore propose a  demand model whose  parameter estimates can be obtained in closed-form, thus permitting real-time updates at large scale.
In particular, we use the following  parametric demand model:

\begin{eqnarray}
d_t = \alpha_0 + \sum_{j=1}^M  \alpha_j t^j   + \sum_{j=1}^J  \gamma_j S_{jt}  + \sum_{j=1}^{J}   \sum_{i=1}^P  \phi_{i,j}  Q_{jt}  d_{t-i} + \varepsilon_t,  \label{eq:generalmainmodel}
\end{eqnarray}
with trend term $t$, binary variables $S_{jt}$ taking value one at specific frequencies  to incorporate desired seasonality effects (e.g., minute, hour, day), and $Q_{jt}$  taking value one at times when different dynamics are connecting current demand $d_t$ with lagged (historical) demand $d_{t-i}$.
The unknown  parameters are the constant term $\alpha_0$, the trend parameters $\alpha_j$ ($j=1, \ldots, M$), the seasonality parameters $\gamma_j$ ($j=1, \ldots, J$) and the autoregressive parameters $\phi_{i,j}$ (for frequency $j$ at lag $i$); $\varepsilon_t$ is a zero mean white noise error term.
This general demand model allows for rich trend, seasonality, and lag configurations.

Demand model \eqref{eq:generalmainmodel} is, however, potentially highly parameterized, which may impact its forecasting performance. There are two ways to make the model more parsimonious. First, including domain expertise which directly sets some parameters 
 in equation \eqref{eq:generalmainmodel} to zero; we label this approach ``FFUDS-domain" in the remainder. 
 Second, employing a data-driven machine learning approach that identifies automatically the most relevant model parameters; we label this approach ``FFUDS-data" in the remainder.  The latter is useful in absence of domain expertise, but adds computational costs as can be seen from Section \ref{sec:implications} below.

Estimation of model \eqref{eq:generalmainmodel} can be done using ordinary least squares (OLS).  In fact, the model fits within the class of linear regression models and can be written in matrix form as $y =  X \beta + \varepsilon$ where 
$y$ contains the to be forecast demand,
$X$ collects all predictors namely the deterministic components (trend and seasonal dummies) as well as the lagged demand values, and 
$\beta$ is the vector of all model parameters $\alpha$, $\gamma$, and $\phi_{ij}$. 
 To estimate $\beta$, the OLS loss function minimizes the squared Euclidean distance between $y$ and $X \beta$, written shortly as  $||y - X \beta||^2$. This loss function has an analytically known gradient function $X'(y - X \beta)$ that is solvable in closed-form yielding the  estimator $\hat \beta = \left( X'X \right)^{-1} X'y$ without requiring a gradient descent type algorithm.

To accommodate our streaming data setting, model  \eqref{eq:generalmainmodel} needs to be 
continuously re-estimated as new demand data becomes available.  
To this end, we rely on  renewable OLS estimation \citep{Luo_2020_renewable}\footnote{\cite{Luo_2020_renewable} provide a more general framework for generalized linear models which require updating maximum likelihood estimators with more complex gradient functions, and that typically use numerical optimization methods. Apart from renewable estimation, the authors are also the first to provide incremental inference.} which requires some additional notation to define the online estimator $\tilde \beta$.
We denote $\tilde \beta_1 = \hat \beta_1$ as the OLS estimator for model \eqref{eq:generalmainmodel}  given the initial data batch $D_1$, and  $\hat \beta_b = \left( X_b'X_b \right)^{-1} X_b'y_b$ as the OLS estimator for the same model on a new  incoming data batch $D_b$. Then the renewable estimator based on historical data and incoming batches $b=2,3, \ldots$ can be written as
\begin{eqnarray}
\tilde \beta_b = \left( \tilde J_{b-1}   +  X_b'X_b  \right)^{-1} \left(  \tilde J_{b-1}  \tilde \beta_{b-1} +  X_b' X_b \hat \beta_b\right),   \label{eq:streamingols}
\end{eqnarray}
where $\tilde J_{b-1} = \sum_{j=1}^{b-1}  X_j'X_j$.
We see that for an updated estimator $\tilde \beta_b$, we only require to have the previously estimated $\tilde \beta_{b-1}$, $\tilde J_{b-1}$, the new $X_b$ and estimation $\hat \beta_b$. Hence, we avoid storing all historical data points and keep only summary statistics.
This  renewable estimator is also equivalent to the Bayesian posterior updating formula for linear regression models with the cumulated historical ($j=1,\ldots, b-1$) information  being the prior and $D_b$ the data. 
The size of $D_b$ can be different over time so that updated estimators can be computed independently from the frequency of $d_t$. 
Thanks to its recursive updating scheme, such a streaming estimator
reduces as much computing time as possible  and requires only minimal amounts of data to be stored. Both aspects have direct measurable financial impact for  platforms which typically operate via a cloud service provider such as Amazon Web Services (AWS) or Microsoft Azure. 
The renewable estimator above will, however, only work well in case of parameter stability and is therefore no substitute for  FFUDS.

\subsection{FFUDS Algorithm} \label{subsec:ffudsalgorithm}
We summarize the FFUDS procedure in Algorithm \ref{alg:FFUDS}. Lines 1 and 2 are initialization steps. The first time the model is fitted with the initial data batch for each of the 
data streams, we use the same parameter estimate to obtain full-sample and post-break sample forecasts, hence the first combined forecast is the same as the full-sample forecast. Line 3 requires selecting the forecast combination weight $w$ as hyper-parameter. Lines 4 and 5 run through each incoming data batch and each data stream. Line 6 reads the new available data batch. Line 7 uses this new batch to update the full-sample model estimate and produces new forecasts. Line 8 does the same for the post-break sample. Line 9 combines the forecasts of lines 7-8 using weight $w$. Line 10 uses the new data batch to compute forecast losses. The updated forecast loss stream is then used in line 11 as input for the breakdown test.
Lines 12 and 13 check for a forecast breakdown and if it occurs the post-break sample and forecast loss  stream are reset at the detected breakdown date. Note that if the length of the post-break sample is too short to estimate the model, due to insufficient data regarding the number of parameters to estimate in model \eqref{eq:generalmainmodel}, then we temporarily set $w=1$.  Alternative design choices are explored in Section \ref{subsec:ablation}. 
Finally, line 16 returns combined forecasts for all data streams, 
as input for optimal planning.

\begin{algorithm}
\caption{ Fast Forecasting of Unstable Data Streams (FFUDS)}
\label{alg:FFUDS}
\leading{15pt} 
\begin{algorithmic}[1]
\STATE Let $D_b^i$ be initial data batches for $b=1$ and $i=1,\ldots, n$ where $n$ is the number of data streams
\STATE Estimate  model \eqref{eq:generalmainmodel} on $D_1^i$, ($i=1,\ldots, n$),  compute initial full-sample and post-break forecasts from this model
\STATE 
Choose forecast combination weight $w \in (0,1)$
\FOR{$b = 2, 3, \ldots$}

	\FOR{$i = 1$ to $n$}
	  	\STATE Read data batch $D_b^i$ 
    	  	\STATE  Update full-sample model estimate including $D_b^i$, 
        compute full-sample forecasts $FULL$ 
    \STATE Obtain post-break model estimate including $D_b^i$, compute post-break forecasts $POST$ 
	  	\STATE Compute combined forecasts $\hat d^f = w FULL + (1-w) POST$
	  	\STATE Compute forecast losses $\hat L_b^{(h)} = L(d_{H(b-1)+h},\hat d_{H(b-1)+h}^f)$ for $h=1,\ldots,H$
	  	\STATE Test for forecast breakdown resulting in $TEST=1 (0)$ if reject (accept)
		\IF{$TEST=1$} 
			 \STATE Reset post-break sample and forecast loss stream $L_b$ 
		\ENDIF		  	
	\ENDFOR
\RETURN $n$ combined forecasts for decision making	
\ENDFOR
\end{algorithmic}
\end{algorithm}

\section{Platform Application} 
\label{sec:platform-application}
Stuart is a prime example of on-demand service platforms that matches providers and consumers of on-demand delivery services.
This platform thus faces the decision of how to match the dynamic supply and demand just like other two-sided on-demand service platforms.
We analyze UK demand data provided by this company at the hourly frequency from January 1, 2019 to March 31, 2021. 
Demand data are measured as the number of parcels to be delivered to 294 UK delivery areas  it serves.
To ease comparisons across areas (i.e.\ 
data streams), we assume that business operates everywhere on a daily basis between 9am and 11pm\footnote{Occasional overnight demand is integrated in the last hour of the business day.}, 
which implies $T=12,315$ observations.

In Section \ref{model-stuart}, we outline the business problem faced by  
this on-demand
service platform and present the details on the demand forecast model.
In Section \ref{stuart-data}, we detail  the data.
In Section \ref{benchmarks}, we discuss   relevant benchmark demand forecast models for our platform application.

\subsection{Global Platform Pricing Problem and Demand Forecast Model Specification} \label{model-stuart}
The platform faces a global pricing problem where one needs to maximize profits, defined as the difference between what the platforms receives for deliveries (revenue) and what it pays as wages to drivers (cost), subject to  driver participation constraints.
By solving the platform decision problem, the platform chooses the price for deliveries and wages for drivers, which is typically done in a separated way, see for example \cite{Garg_Nazerzadeh_2021}. 
Now, both the revenue and cost components of the optimization problem are a function of the demand for deliveries (they are positively impacted by it). Wages are then optimally set to ensure drivers are satisfying demand for deliveries at the right time and place. 
Having an accurate forecast of demand for deliveries is therefore a crucial input to reliably solve the platform decision problem. Besides, having access to accurate demand forecasts also reduces the need for surge pricing, i.e.\ sudden higher than usual wages so that drivers decide to work at times and in areas of high expected demand. 

To obtain area specific demand forecasts at the considered platform,  the local decision maker  faces the following  problem:  every night, the upcoming fleet of required couriers needs to be planned using hourly demand forecasts for the next seven days.
An automated procedure built by the global data science unit transfers these forecasts every night to all delivery areas. This procedure has to be quickly updated in case forecast performance deteriorates. 

We thus forecast hourly demand  $d_t$ (from 9am to 11pm) at the end of every day for the next seven days, implying forecast horizon $h=1,\ldots, 105$.  
This generates a daily stream of forecasts $d_t^f$ that can be evaluated against the actual demand 
the day after for $h=1,\ldots,15$, two days after for $h=16,\ldots, 30$, etc. The resulting forecast loss functions can be used for (cumulative) performance tracking and detection of forecast breakdowns as explained in Section \ref{subsec:stream_breakdowns}.
In particular, for forecast breakdown detection, we use the loss series computed at daily frequency as follows:   $\hat L_b = \sum_{h=1}^H L(d_{H(b-1)+h}, \hat d_{H(b-1)+h}^f)$ with $b=2, 3, \ldots$ and $H=15$.

 Given the hourly streaming data frequency, we use the following specific demand model
 \begin{eqnarray}
d_t = \alpha_0 + \sum_{j=1}^M  \alpha_j t^j   + \sum_{j=1}^7  \gamma_j D_{jt}  + \sum_{j=1}^{15} \left( \delta_j  +  \sum_{i=1}^P  \phi_{i,j} d_{t-i}   \right)H_{jt}  + \varepsilon_t,  \label{eq:mainmodel}
\end{eqnarray}
with variables $D_{jt} =1$ if $d_t$ is on Day $j$ and 0 otherwise, $H_{jt} =1$ if $d_t$ is on Hour $j$ and 0 otherwise. \\
For parameter identification purposes, we  set $ \alpha_0 =  \gamma_{1} = 0$. 
To keep the model parsimonious,
we take a linear trend ($M=1$) because it empirically dominates a quadratic trend ($M=2$), and set $P=105$.
We consider both a domain-knowledge informed and a data-driven version of FFUDS.

For the domain-version of FFUDS, we
 incorporate domain expertise from the platform to impose several restrictions.
Since 
mornings and evenings have different demand levels and therefore perhaps also dynamics, we set $\phi_{i,1}=\ldots=\phi_{i,4}$, $\phi_{i,5}=\ldots=\phi_{i,13}$ and $\phi_{i,14}=\phi_{i,15}$ 
to permit heterogeneity in the morning (9am-12pm), afternoon (1pm-9pm), and evening (10pm-11pm). We use the following simple dynamics:
for the morning hours, we  include the demand from the same hour of the previous week ($d_{t-105}$).
For the afternoon hours, we include the demand from the last hour ($d_{t-1}$),
the demand from the same hour of the previous day ($d_{t-15}$)  and the 
demand from the same hour of the previous week. 
For the evening hours, we include the demand from the last hour and the demand from the same hour of the previous week.
The autoregressive parameters corresponding to all other lags are set to zero. 
For the data-driven alternative of FFUDS, we include all lagged demand parameters  and the zero coefficients are learned in a data-driven way by using the lasso \citep{tibshirani1996regression},  a popular machine learning method available in the \texttt{R} package \texttt{glmnet} \citep{glmnet}.
Lastly, including specific holiday effects or external data such as weather can easily be handled in the model. For example, December 25 (Christmas) and January 1st (New Year) are typically known for very low demand, and we handle such effects  through dummy variables.

Finally, although demand from distinct areas share a similar overall pattern, they are bound to different effects  because of their heterogeneous size and geographical location. We therefore use the same model specification for all areas,  but allow for heterogeneity across areas by estimating the parameters separately.

\subsection{Data} \label{stuart-data} 

\begin{figure}[t]
	\centering
		\includegraphics[width=0.7\textwidth]{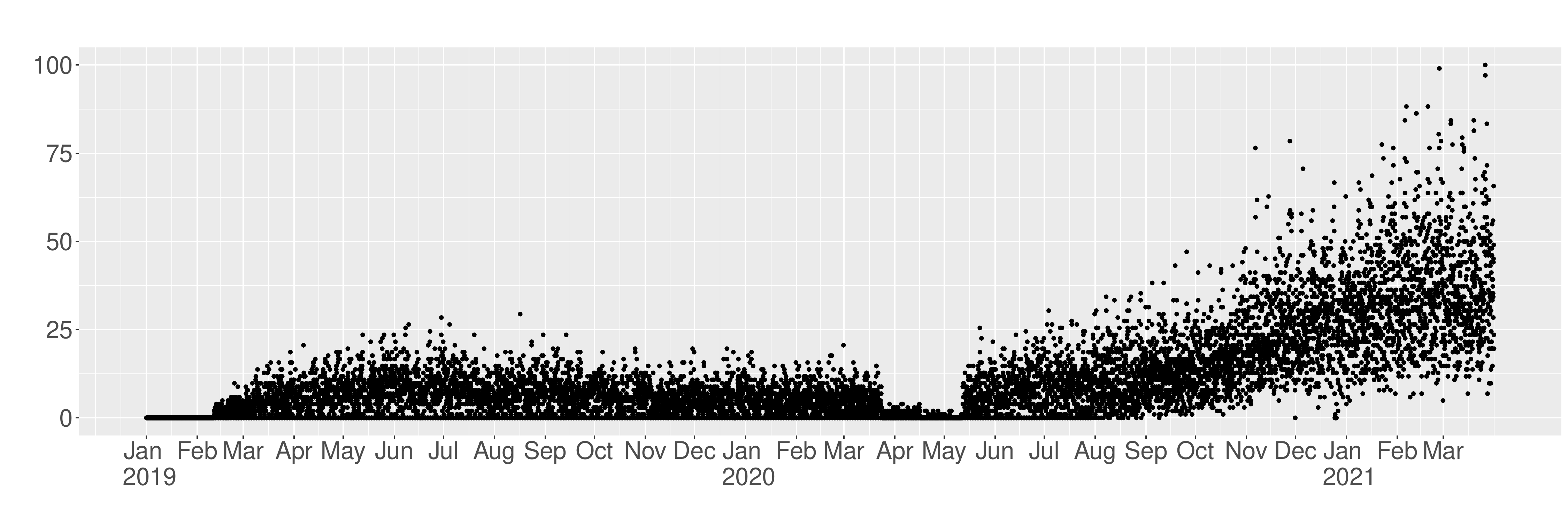}
	\caption{Hourly demand in Milton Keynes from January 1, 2019 until March 31, 2021. \label{fig:total_mk_demand}}
\end{figure}

The data from the on-demand 
service platform display all features typical of platform streaming data (Section \ref{subsec:platform_features}).
First, streaming data are observed at a high, namely hourly, frequency.
In 
Figure \ref{fig:total_mk_demand}, we visualize the  hourly demand for Milton Keynes Central as an example of a delivery area.\footnote{For data confidentiality reasons, raw demand values are scaled by a constant. The scaled demand values appear in figures and related text. Scaling is a commonly used technique and does not affect our forecasting framework.}
Second, high-dimensional streaming data are available, namely for 294 UK delivery areas.
Figures A3-A5 
(Appendix A) 
provide additional insights into the demand heterogeneity across delivery areas. Note that strong intraday and day of the week seasonal demand patterns arise (see Figure  A1).
As these seasonal patterns vary across delivery areas, it is important to forecast demand at the area level which we do.
Third, the range of the demand data varies greatly with sparse data points typically occurring both at re-current time points such as early morning (9am-12pm), as well as for specific extended time spells such as the first lockdown of the Covid-19 pandemic, see Figure A2. 
Fourth,
 demand is highly unstable. In the first few months of 2019, demand grows gradually and then 
flattens until March 2020. 
The Covid-19 pandemic and first lockdown result in a huge demand drop for several weeks, but then demand quickly starts to grow at an exponential rate until the end of our sample. 
The pandemic has accelerated the growth of online shopping to unprecedented and unexpected levels. For instance, the United Nations Conference on Trade and Development \citep{UNCTAD_2021} reports that global online retail sales' share as part of total retail sales rose from 16\% to 19\% already in 2020.
Hence, a modeling approach is needed that can swiftly adapt to these changes and their possibly unpredictable nature. 

We further investigate the instability in demand over the full sample period 
to gain insights in the occurrence of 
breaks.
To this end, 
we employ a popular change-point detection method for each delivery area's hourly demand series, namely the PELT nonparametric multiple change-point test \citep{killick2012optimal}.\footnote{We compared this hourly analysis with a daily one where we temporally aggregate the  data to daily demand series, 
and then detect change-points in the daily  series with PELT. Breaks remain present, in line with the hourly analysis.}
Although the tests are retrospectively scanning the entire demand series for breaks, we get insights on the frequency and commonality of break dates which 
motivate the need for a demand forecast method that remains accurate in presence of breaks.

\begin{figure}[t]
	\centering
	\begin{subfigure}[b]{\textwidth}
		\centering \includegraphics[width=0.7\textwidth]{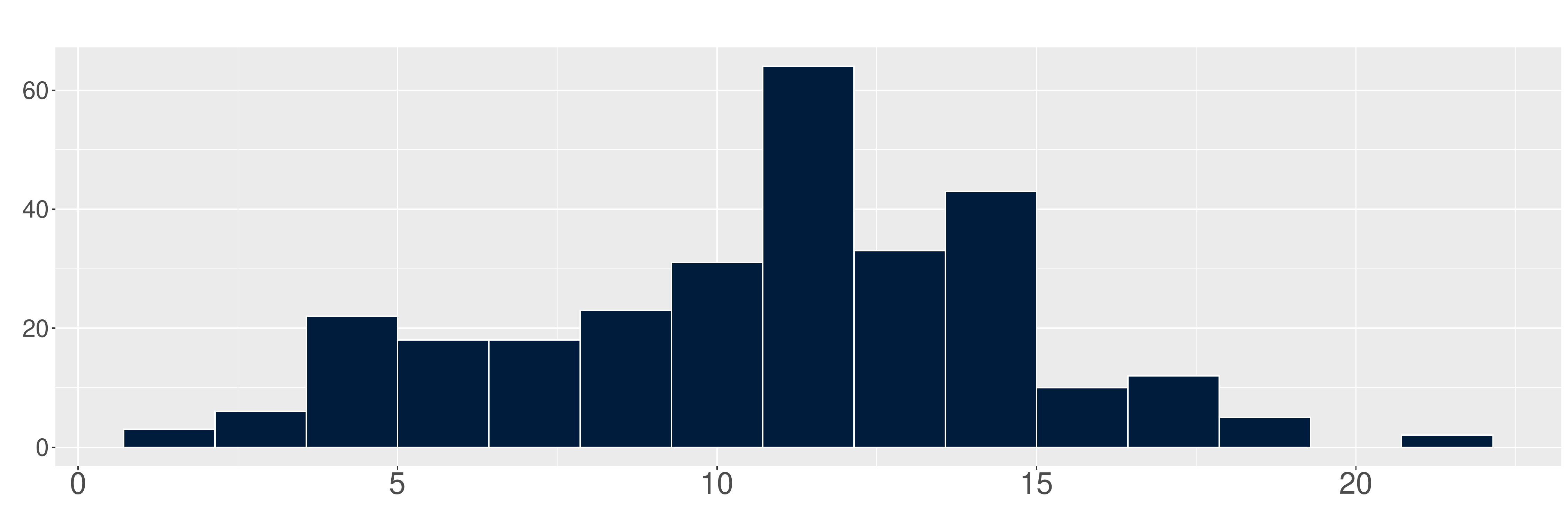} 	  \caption{Area count (vertical axis) of number of detected breaks (horizontal axis).}\label{fig:changepoints_histogram}
	\end{subfigure}
	\begin{subfigure}[b]{\textwidth}
		\centering \includegraphics[width=0.7\textwidth]{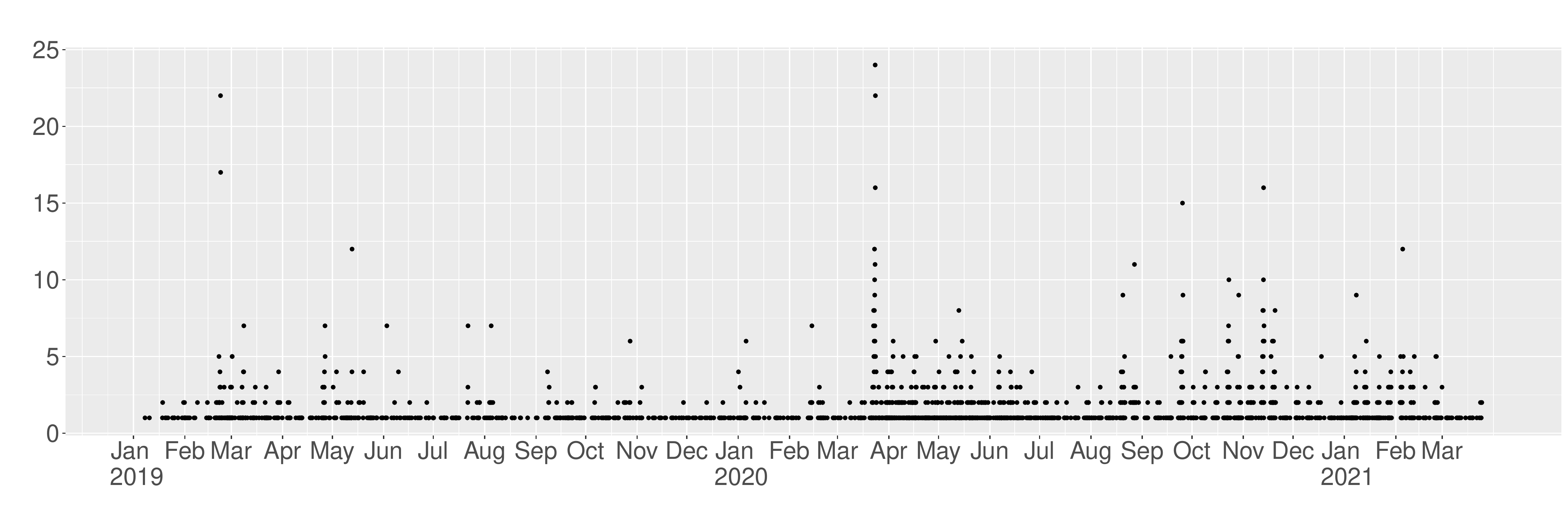}   
		\caption{Hourly time series plot of number of detected breaks summed over all  areas.}\label{fig:changepoints_scatter}
	\end{subfigure}
	\caption{Histogram (top) and time series (bottom) of the number of breaks (bottom) detected by  PELT.
	}    \label{fig:changepoints_hourlytimeseries}
\end{figure}        

Figure \ref{fig:changepoints_histogram} summarizes the number of detected breaks across delivery areas. On average about eleven breaks in each  area are detected, though the range is large with some areas having hardly any break and one area having even 22 breaks. 
We next investigate the timing of  the breaks  and check if they cluster among  areas. Figure \ref{fig:changepoints_scatter} plots the number of breaks detected across all areas over the full sample. 
We find at least one detected break in 15 percent of the day/hour date-stamps of the full sample. 
Breaks peak on March 22-23, 2020 where we count 181 breaks. 
This clearly points to the demand shifts related to the first Covid-19 lockdown. In addition, we find systematically more breaks between  April, 2020 and March, 2021  than in the pre-Covid period. In fact, in this period,  breaks affecting multiple delivery areas take place 35 percent of the time.

Demand  post-Covid differs considerably from that in the pre-Covid period. To illustrate this, we compute the daily average demand for the areas  active before March 2020 (Pre-Covid) and still in business after May 2020 (Post-Covid). 
The results are displayed in Figure \ref{fig:covid_scatter} with the 45 degree dashed line aiding to identify the directional change. 
Most areas see their demand being boosted to multiple times the pre-pandemic demand. 
A desirable property of a forecast model should therefore be a quick adaptation to a ``new" regime.

\begin{figure}[t]
			\centering
			{\includegraphics[width=0.7\textwidth]{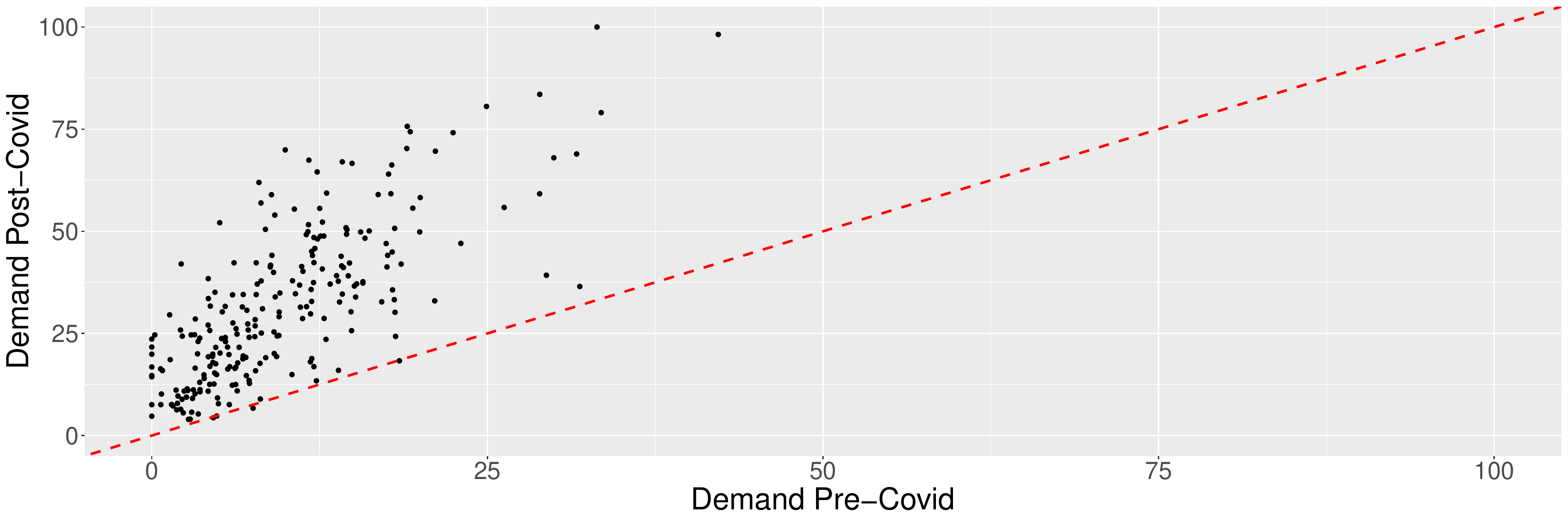} }			
			\caption{Scatterplot of pre- against post-Covid-19  average daily demand for all UK areas.				
			}
			\label{fig:covid_scatter}
\end{figure} 
  
\subsection{Benchmark Forecast Models} \label{benchmarks}
We compare the  performance of FFUDS against several benchmarks. 
The first benchmark is defined as $d_t =  d_{t-105}  + u_t $, with forecast $\hat d^f_t =  d_{t-105}$, which we call the ``Naive" model since  it implies that the forecast of a specific hour/day is the same value as observed in the previous week hour/day. An advantage of this specification is that it does not require parameter estimation, and it is expected to work well in the case of strong seasonality patterns as observed in our platform demand data.

The second benchmark is Facebook Prophet which implements automatic change-point detection in the parameters and is 
available in open-source software packages such as \verb|R| 
and Python. 
We use the standard default \texttt{prophet} package in \verb|R| \citep{prophetCRAN}.
As the third benchmark, we use the popular
Long-Short-Term-Memory (LSTM) network
that we re-train at monthly frequency with fixed hyper-parameters throughout the forecasting exercise.\footnote{Tuning hyper-parameters  daily is computationally too demanding.
Based on preliminary experimentation on a few areas, we choose as hyper-parameters one LSTM layer with ten units. We use ten epochs for training with a dropout rate of 20\% to avoid overfitting.}
We use the \texttt{keras} and \texttt{tensorflow} packages in \verb|R|  in our applications. For a more detailed description on LSTM and Prophet, we refer to Appendix B. 
Overall we note that, given the number of parameters involved and the  time for training them,  the deployment of deep learning models at scale is complex in the unstable environment of platform data streams. 
As  two final benchmarks, we use the popular, traditional time series models SARIMA and ETS. We use the default  implementations \texttt{auto.arima}  and \texttt{ets} in the  \texttt{R} package \texttt{forecast} (\citealp{forecastR_v2}, \citealp{forecastR_v1}) to respectively select the best SARIMA and ETS model according to the  Akaike Information Criterion corrected for sample size.

Note that the backbone of FFUDS-- its forecast breakdown detection and forecast combination approach --can be added to  any forecast method (Section \ref{subsec:FFUDS-framework}), including the benchmarks LSTM, SARIMA and ETS. 
We therefore not only report on their standard implementations, but we also investigate how their forecast performance is affected when adding breakdown detection.
Prophet, in contrast, already performs automatic change-point detection, but one can still tune the strength of its prior that regulates the flexibility of its change-point detection. We therefore also consider three versions of Prophet: its default setting (\texttt{changepoint.prior.scale=0.05}); a setting allowing
fewer (\texttt{changepoint.prior.scale=0.001}) as well as more change-points (\texttt{changepoint.prior.scale=0.5}).

We further highlight that, though simple at first sight, demand models should accurately capture trend and seasonal effects in the platform data. 
To illustrate this,
in a preliminary analysis we construct a relatively short 30-day window that we update every night after having observed fifteen new hours of demand. 
We have 792 such windows between January, 2019 and March, 2021. 
Using these segments, we fit model \eqref{eq:mainmodel} with $\phi_{ij}=0$, i.e. only  trend and seasonality,  for each of the delivery areas. Figure C1, panel (a), 
in Appendix C shows the resulting rolling window average $R^2$s. 
Strikingly, trend and seasonality explains around 75\% of average area demand variation in the beginning of 2019 but then increases oftentimes above  85\% until the start of the first Covid-19 lockdown where the average $R^2$s drop temporarily back to about 80\%. After this first lockdown, the average streaming $R^2$s gradually climb to new heights close to 95\%. 

If so much variation is explained by trend and seasonality, what is left in terms of dynamic properties in the demand series?
We investigate this by computing the amount of autoregressive dynamics after having filtered the trend and seasonality from the data. 
We compute a time series average, over the available areas, of first order autocorrelation coefficients. 
Figure C1, panel (b), 
reveals that, on average, the filtered demand series have overall low autoregressive dynamics. 
At the start of the sample the average streaming sample autocorrelation coefficient hover around 20\%, while they increase to just below 30\% at the end of the sample in 2021. 
Nonetheless, some episodes are characterized by streaming sample autocorrelation coefficients up to 42\%. 
It is therefore still worthwhile to study the presence of autoregressive dynamics in our streaming data setup. 
In the next section, we show that FFUDS achieves sizable forecast gains compared to the benchmarks.

\section{Results}\label{sec:forecast_results}

\subsection{Results for All UK Delivery Areas}
For all UK delivery areas, we implement a streaming exercise 
starting in February, 2019. Parameter estimates are updated, up to next-week-ahead forecasts are produced, forecast losses and next-business-day breakdown tests are implemented on a daily basis until March, 2021.
We synthesize the streaming aspect by splitting the  time span into pre-Covid and post-Covid periods, running respectively from February, 2019 to March, 2020 and April, 2020 to March 2021. In fact, the in-sample break detection tests reported in Section  \ref{stuart-data}  evinced that after March, 2021 substantially more breaks appear in the delivery areas, making possible comparing our approach in a break-poor and break-rich scenario. 

Table \ref{tabel:summar_forecast_smape}  summarizes over all delivery areas  the SMAPE performance of  FFUDS against the  benchmarks.\footnote{Similar insights are obtained for forecast performance in terms of squared loss, see Table C1 (Appendix C) 
for details.} For completeness we include both the domain-specific and data-driven version of FFUDS, but we reserve the discussion on their comparison to the ablation study in Section \ref{subsec:ablation}. We start by comparing FFUDS domain (for brevity ``FFUDS" in the following discussion) against the standard implementations of the benchmarks.
We zoom into the one-day-ahead forecast performance, i.e.\ one- to fifteen-hour-ahead forecasts, since they are in particularly used by the local area planners to decide on their next day operations, and  they enter our  forecast breakdown detection procedure.  
We display forecast errors in percentages relative to FFUDS. For instance, a relative forecast error of 120\% indicates that  FFUDS performs 20\% better than the benchmark.

\begin{table}
	\centering
	\caption{SMAPE forecast performance  \label{tabel:summar_forecast_smape}}
  \resizebox{0.85\textwidth}{!}{\begin{minipage}{\textwidth}	
	\begin{tabular}{lllllcccccccccc} \hline 
		  &&&    && \multicolumn{2}{c}{FFUDS} & Naive & Prophet & \multicolumn{2}{c}{LSTM} & \multicolumn{2}{c}{SARIMA} & \multicolumn{2}{c}{ETS} \\  	
 &&&    && domain & data && && && &\\
Breaks	  &&&    && $\checkmark$ &  $\checkmark$ &  & $\checkmark$ & & $\checkmark$  &  & $\checkmark$ & & $\checkmark$ \\  \hline 
	Pre-Covid &&&          && 100.00 & 100.11 & 125.20 &  103.68 &  111.38 & 106.72 &  114.97 & 110.94  & 106.78& 104.84  \\      
 Post-Covid &&&          && 100.00 & 94.17 & 120.28 &  115.53 &  121.14 & 106.20 &  104.92 & 101.47 & 101.56 & 99.01\\         
  \midrule
	\end{tabular}
 \end{minipage}}
\raggedright
\footnotesize

Notes: One-day-ahead SMAPE forecast performance of the benchmarks relative to FFUDS domain,  averaged across all areas.
Values above 100 indicate the percentage gain in forecast accuracy of FFUDS relative to the benchmark.  \\[0.1cm]
\end{table}

We  first examine the pre-Covid period, the relatively calmer period with few disruptions.
FFUDS considerably outperforms the Naive model, with an overall gain of 25\%. 
Even a relatively calm period for the platform data requires more than just capturing seasonality effects.
The more advanced Prophet method performs  better across all hours of the day, though FFUDS  dominates with an overall gain of 3\%. 
Finally, LSTM, SARIMA and ETS are outperformed by respectively 11\%, 14\% and 6\% by FFUDS in the pre-Covid period.

Post-Covid,  
the results are, overall, even more in favor of FFUDS. 
It strongly dominates  all benchmarks.\footnote{Note that the same numbers are displayed in Figure \ref{fig:2D_plot_errors_vs_time}.} In fact, Prophet is now  outperformed with 15\% compared to 3\% pre-Covid.
Standard LSTM, ignorant to the presence of breaks, becomes less competitive in the post-Covid period which is characterized by more disruptions and is therefore outperformed by 21\%. The traditional benchmarks SARIMA and ETS are, in contrast, more competitive post-Covid compared to pre-Covid.
Overall, we can conclude that FFUDS systematically yields  more accurate forecasts than the benchmarks. 

Next, we investigate how breakdown detection affects the performance of the benchmarks LSTM, SARIMA and ETS, see Table \ref{tabel:summar_forecast_smape} checked columns for breaks. 
Importantly, the performance of all benchmarks improves under breakdown detection; making the key backbone of FFUDS valuable for a wide range of forecasting methods. 
Still, using a method that delivers forecast fast, as we propose in Section \ref{baseline}, is crucial in environments where speed is vital. In fact, FFUDS remains the top performer  even when the benchmarks are augmented with forecast breakdown detection, apart from ETS with breakdown detection post-Covid. Furthermore, it considerably outperforms all benchmarks in terms of computing time, as we discuss in Section \ref{subsec:econ_impact}.
Finally, note that in Figure C2 of Appendix C, we report the SMAPE of Prophet relative to FFUDS under various settings of Prophet's build-in change-point detection.
We see that Prophet's recommended change-point  prior value delivers the most competitive performance vis-a-vis FFUDS.
In unstable data settings like ours, one should avoid small values of the change-point prior as they allow for too few change-points, which considerably hurts performance post-Covid where more breaks occur.

Figure C3 
provides disaggregated details about the forecast performance per hour over the seven-day horizon for respectively pre-Covid (column 1, SMAPE loss) and post-Covid (column 2, SMAPE loss). 
It turns out that forecast capacity depends more on the time of day rather than the day-ahead. For example, forecasting 4pm area demand on Day 1 is almost as easy as forecasting 4pm demand on Day 7. 
The same is true for other business hours. 
These intraday figures re-emphasize that, overall, FFUDS largely dominates the standard benchmarks pre- and post-Covid. They also show that in the pre-Covid period morning hours are  difficult to forecast as indicated by the orange lines. The main reason for this is the systematically recurring close to zero morning demand
across all delivery areas (Figure A2 
top panels). In such settings, the Naive method cannot be beaten. FFUDS  typically forecasts demand to be small, yet non-zero.
Due to the definition of the SMAPE, however, a non-zero forecast  of a zero actual value directly results in a 100\% forecast error, regardless of the forecast's magnitude. 
In such situations, the RMSE  better reflects the degree of over-prediction. From  Figure C3 
(third column), one can see that the Naive and LSTM forecasts are roughly 50\% better than the ones from FFUDS during the morning. 
While this appears to be a large difference at first sight, this number is driven by our normalization of the FFUDS forecast errors at 100\%. In practice, the difference is negligible since FFUDS over-predicts morning demand compared to these benchmarks by less than one package on average.
Post-Covid, morning demand considerably increases across all delivery areas (Figure A2, 
bottom panels). Then,
 FFUDS performs better during the morning hours. 
This can also be seen in the 
top panel  of Figure C3 
(column 2 vs column 1) which quantify for FFUDS the forecast  loss in absolute levels per hour over the seven-day horizon. 
The largest SMAPE values  decrease from 80\% (Pre-Covid) to 45\% (Post-Covid).
While in terms of SMAPE, the Naive method appears to still beat FFUDS, this only occurs due to its peculiar  definition at zero actual demand.  The RMSE metric reveals that FFUDS outperforms the Naive method by roughly 10\%. Finally, Figure C4 provides the same disaggregated forecast results  when adding breakdown detection to LSTM, SARIMA and ETS. Doing so greatly improves their forecast performance, as they now become more competitive to FFUDS especially during early morning hours and for short forecast horizons.

After studying all UK delivery areas together, we also focused on one representative  area in London, namely Wimbledon, to have a clearer view on  where FFUDS finds breaks and how it performs throughout the  2019-2021 sample period. 
A summary of this case study is included in Appendix D.

\subsection{Ablation Studies} \label{subsec:ablation}
We conduct a set of ablation studies to assess the contribution of the different components of FFUDS compared to its benchmarks. Detailed results are included in Appendix C. 

{\bf Predictors and Parameter Restrictions.}
FFUDS uses a demand model with trend \& seasonality and lagged demand dynamics  predictor sets.
Besides, zero parameter restrictions on the lagged demand dynamics are imposed to keep the model parsimonious either based on domain knowledge or in a data-driven way. 
In Table C2, 
we  investigate the incremental added value of each predictor set (i.e.\ trend \& seasonality; lags) and choice of coefficient restrictions (i.e.\ domain-expert vs data-driven). 
We find that
(i) the best performance is attained by including demand lags as predictors 
(pre-Covid and post-Covid). 
(ii) Imposing restrictions in a data-driven way 
does not improve forecast performance pre-Covid but does post-Covid; with an improvement of 6\%.
This improvement comes at a cost of additional computing time to compute the lasso, see Section \ref{subsec:econ_impact}.

{\bf Change-point Detection Method.}
FFUDS' first key component  concerns forecast breakdown detection. We therefore conduct a  dedicated evaluation to this component by redoing the forecast exercise where FFUDS makes use of 8 popular change-point detection alternatives to PELT, as listed in Table C3. We refer the interested reader to \cite{vandenburg2020evaluation} for a recent comparison of all the considered change-point methods.
Table C4 
reports the SMAPE  of FFUDS under these alternatives.
Its performance  is very robust to the choice of change-point detection method, with no significant differences in forecast performance.

Finally, a change-point detection method may return more than one detected break. In an additional sensitivity analysis, we compared forecast performance when either using the first or the last break date to define the post-break sample for those change-point detection methods where multiple change-points were detected simultaneously, see Table C5. Overall, using the first break date results in minor performance improvements, hence our recommendation to define the post-break sample accordingly.

{\bf Forecast Combination Weights.}
FFUDS' second key component concerns the forecast combination where we combine the full-sample and post-break forecast equally.
We now investigate its sensitivity to the choice of  forecast combination weights $w$ by redoing the forecast exercise with  weights $w=0, 0.25, 0.5, 0.75$ and 1 on the full-sample.
Figure C5 
summarizes the SMAPE  performance for these different choices of $w$.
The equal-weight approach results in the most stable forecast performance across pre- and post-Covid periods. 
In the calmer pre-Covid period, putting more weight on the full-sample is beneficial; whereas in the post-Covid period with more breaks, the opposite occurs. 
The above results can only be obtained ex-post when having observed the entire data stream. Equal forecast combination weights appear  to yield good and stable performance, in line with the forecasting literature.

{\bf Post-break Sample.}
When the post-break sample is too small to estimate the demand model, the full-sample forecast gets all weight in the forecast combination. In our application, this occurs the first seven days after each break is detected, since we use up to 105 (=7days$\times$15hours) lags in the model. Afterwards the weight between both is split equally again.
We consider two alternative approaches that allow us to re-estimate the demand model as of the first day after a detected break. 
The first approach puts all coefficients corresponding to lags larger than 1 in  model \eqref{eq:mainmodel} to zero, and estimates the model with the lasso, which maintains its good performance on small samples. The second approach takes the Naive model which does not require estimation. In both cases, we find virtually no difference in forecast performance (i.e. less than 1\%) , which is to be expected since this change only affects a minority of days in the out-of-sample period, and hence has a negligible impact on future streaming series and the overall performance of our approach. 
 
\section{Implications \label{sec:implications}}
The natural question to ask next is how the enhanced forecast performance of our approach translates in economic gains (Section \ref{subsec:econ_impact}) and managerial insights (Section \ref{subsec:managerial_insights}). 

\subsection{Economic Impacts} \label{subsec:econ_impact}
We investigate the economic impact of FFUDS' superior forecast performance.
First, we quantify monetary impact, using the economic loss function (Section \ref{loss_functions})  with hyper-parameters $c_1=1.14$ and $c_2=0.54$. 
When forecast/planned demand is lower than actual demand, independent couriers are not always available to handle the extra deliveries and business opportunities are missed. A relatively high cost of 1.14 is associated with this circumstance.
In the opposite situation, when forecast/planned demand exceeds actual demand, couriers are potentially paid without having to deliver. The platform evaluates this situation as less  costly than the former. 
Table \ref{tabel:summar_forecast_econ} is the economic analogue of Table \ref{tabel:summar_forecast_smape} summarizing economic forecast performance when using the economic loss function for all methods. Figures C6-C7 
are the economic heat map analogues of Figures C3-C4. 

Also in terms of economic forecast performance, both versions of FFUDS largely dominate the other benchmarks.
To summarize monetary impact, we sum
these hour/day specific area  monetary losses over 2021-Q1, the last available quarter. Doing this for FFUDS domain and comparing to 
the benchmarks gives the yearly gains (in UK pounds) displayed in the bottom row of Table \ref{tabel:summar_forecast_econ}.
Compared to the best performing benchmark, namely 
LSTM with breakdown detection, we still obtain a striking 
\pounds18,404
gain on a yearly basis. 
The gains over the other benchmarks are even larger, even more so when compared to FFUDS data which results in an additional yearly gain of around \pounds500,000 compared to FFUDS domain.
Hence, while  the statistically measured relative forecast gain against the best benchmark is only a few percentage points, which might appear small  at first sight,  the economic gains add up considerably.

\begin{table}
	\centering
	\caption{Economic forecast performance \label{tabel:summar_forecast_econ}}
  \resizebox{0.72\textwidth}{!}{\begin{minipage}{\textwidth}	
	\begin{tabular}{lccccccccccccccccc} \hline 
 && \multicolumn{3}{c}{FFUDS} && Naive && Prophet && \multicolumn{2}{c}{LSTM} & \multicolumn{2}{c}{SARIMA} & \multicolumn{2}{c}{ETS}  \\ 
&& domain && data && && && &&&&\\
Breaks && $\checkmark$ && $\checkmark$ &&  && $\checkmark$  && & $\checkmark$  & & $\checkmark$  & & $\checkmark$   \\ \hline 	
Pre-Covid 	&&  100.00 && 103.45 &&  130.15 &&  120.87 &&  116.46  & 107.26 & 121.60   & 118.33 & 111.53 & 108.64 \\
Post-Covid 	&&  100.00 && 97.10 &&  127.07 &&  130.60 & &  123.53  & 101.71 &  113.77 & 107.05 & 108.89   & 102.81 \\ \hline
Yearly gain (\textsterling)   &&   -  && -501,324 && 1,604,324 && 3,070,548 & & 1,583,956  & 18,404 & 1,334,076 & 784,100 & 622,640 & 155,740 \\  \midrule 
	\end{tabular}
 \end{minipage} }
 \raggedright
\footnotesize

Notes: One-day-ahead economic forecast performance of the benchmarks relative to FFUDS domain,  averaged across all areas.
Values above 100 indicate the percentage gain in forecast accuracy of FFUDS relative to the benchmark. The bottom row summarizes the yearly gain of FFUDS over the benchmarks in UK Pounds.  \\[0.1cm]
\end{table}

Second, we investigate computational cost of  both versions of FFUDS compared to its benchmarks. 
We omit the Naive model from this comparison as it does not require any estimation, only retrieval of a demand value from last week, hence its computing time (less than a second) and costs are negligible. 
We measure computing time as the number of seconds it takes to do one sequential daily update of estimating and then forecasting a full week-ahead of hourly demand for  all UK areas. Because the sample size is largely different in the beginning of 2019 than at the end, 
we measure computing time on both. 
Table \ref{tabel:computing} reports computing times in seconds.\footnote{For the benchmark methods LSTM, SARIMA, ETS, computing times are only reported for their standard implementations, adding breakdown detection to them hardly increases computing time as can be seen from the vertical axis in Figure \ref{fig:2D_plot_errors_vs_time} where the numbers for the short sample are displayed.} 

For the short sample, FFUDS domain is roughly 
45 (compared to ETS) to 5,500 (compared to LSTM) times faster. 
For the long sample, FFUDS domain is a stunning 
60 to 27,000 times faster compared to respectively ETS and LSTM-- the most extreme benchmarks in terms of computing time.
LSTM requires over 200 hours computing time across all areas, hence forecasting over-night  is then simply impossible in practice (unless parallel programming is used).
Note that it is the computation time dimension where FFUDS with domain expertise gains most of its importance over the data-based version of FFUDS and other benchmarks, still also FFUDS data largely beats all other benchmarks in terms of computing time.

These results imply that higher frequency than daily forecast updates, which are becoming increasingly required for on-demand platforms, can become compromisingly slow for these benchmarks, or other packages based on sampling methods. In contrast, FFUDS domain estimates and forecasts all areas in less than a minute with minimal data storage. 
This speed is also strategically important because platforms are nowadays challenged on their carbon footprints due to the running of computationally intense algorithms.   
Importantly, FFUDS'  drastic reduction in computing time does not come at a cost in terms of estimation accuracy.
Indeed, Figure \ref{fig:2D_plot_errors_vs_time} summarizes the computing time (vertical axis) and forecast errors (SMAPE, horizontal axis) of the considered methods. FFUDS (domain and data) is the only method placed at the bottom left, thereby combining low computing time with high prediction accuracy.

\begin{table}
	\centering
	\caption{Computational metrics \label{tabel:computing}}
 \resizebox{0.68\textwidth}{!}{\begin{minipage}{\textwidth}	
	\begin{tabular}{lllcccccccccccccc} \hline 
 & &&  \multicolumn{6}{c}{Computing Time (seconds)} &&& \multicolumn{6}{c}{Computing Costs (US dollars)} \\
&  && \multicolumn{2}{c}{FFUDS} & Prophet & LSTM & SARIMA & ETS &&& \multicolumn{2}{c}{FFUDS} & Prophet & LSTM & SARIMA & ETS \\ 
& && domain & data & & & & &&& domain& data&& & & \\ \hline
 & Short sample && 2.94 & 15.11 & 346.87 & 16,287.73  & 1,300.49 & 137.20  &&& 0.11 & 0.66 & 15.05 & 706.80 & 56.43 & 5.95 \\
 & Long sample && 26.31 & 246.98 & 5,912.03 & 725,209.1 & 4,868.02 & 1,680.90  &&& 1.14 & 10.72 & 256.55 & 31,470.05 & 211.25 & 72.94 \\ \midrule 
\end{tabular}
\end{minipage} }
 \raggedright
\footnotesize

Notes: Computing time (in seconds on a Microsoft Windows 11 Pro with 16-core AMD Ryzen Threadripper PRO i5 4,3 GHz processor) and costs (in US dollars) for  FFUDS and benchmarks.  \\[0.1cm]
\end{table}

Finally, beyond the strategically important new services that can be offered to customers and drivers, the speed of our procedure has also direct financial consequences. To illustrate this, we take the example of cloud computing costs. An AWS m5.2xlarge EC2 instance costs at the time of writing approximately 0.428 US Dollar (\$)  per hour. 
Table \ref{tabel:computing} summarizes the yearly costs (in US Dollars) of FFUDS and its benchmarks for forecasts updates done all year around. 
Yearly costs are drastically lower for FFUDS than for the benchmarks, especially LSTM. In case of the long estimation sample, for instance, the difference becomes a striking \$1.14 for FFUDS domain compared to \$31,470 for LSTM.
Note that these costs increase linearly with the number of delivery areas within a given country, and that most on-demand platforms operate internationally.

\subsection{Managerial Insights} \label{subsec:managerial_insights}
Opening new delivery areas, merging or splitting existing ones, is together with expected market growth of strategic importance for any on-demand delivery platform. In fact, local demand forecasting accuracy impacts the supply of drivers and their compensation, speed of delivery and therefore customer satisfaction. Figure \ref{fig:BCG} sheds light on the relationship between market share, market size, market growth for FFUDS' (domain) SMAPE loss  at the delivery area level.\footnote{SMAPE is a relative loss measured in percentages and therefore appropriate when comparing areas of different sizes. The RMSE and ECON functions measure loss in levels and therefore automatically increase for larger areas.}
Market growth is defined as the ratio of a delivery area's demand from the last available quarter relative to demand from the first available quarter.
Market share of a delivery area is defined as the ratio of total demand of the area relative to total UK demand in 2021-Q1. 

\begin{figure}[t]
	\centering
	\begin{subfigure}[b]{0.45\textwidth}
		\includegraphics[width=0.8\textwidth]{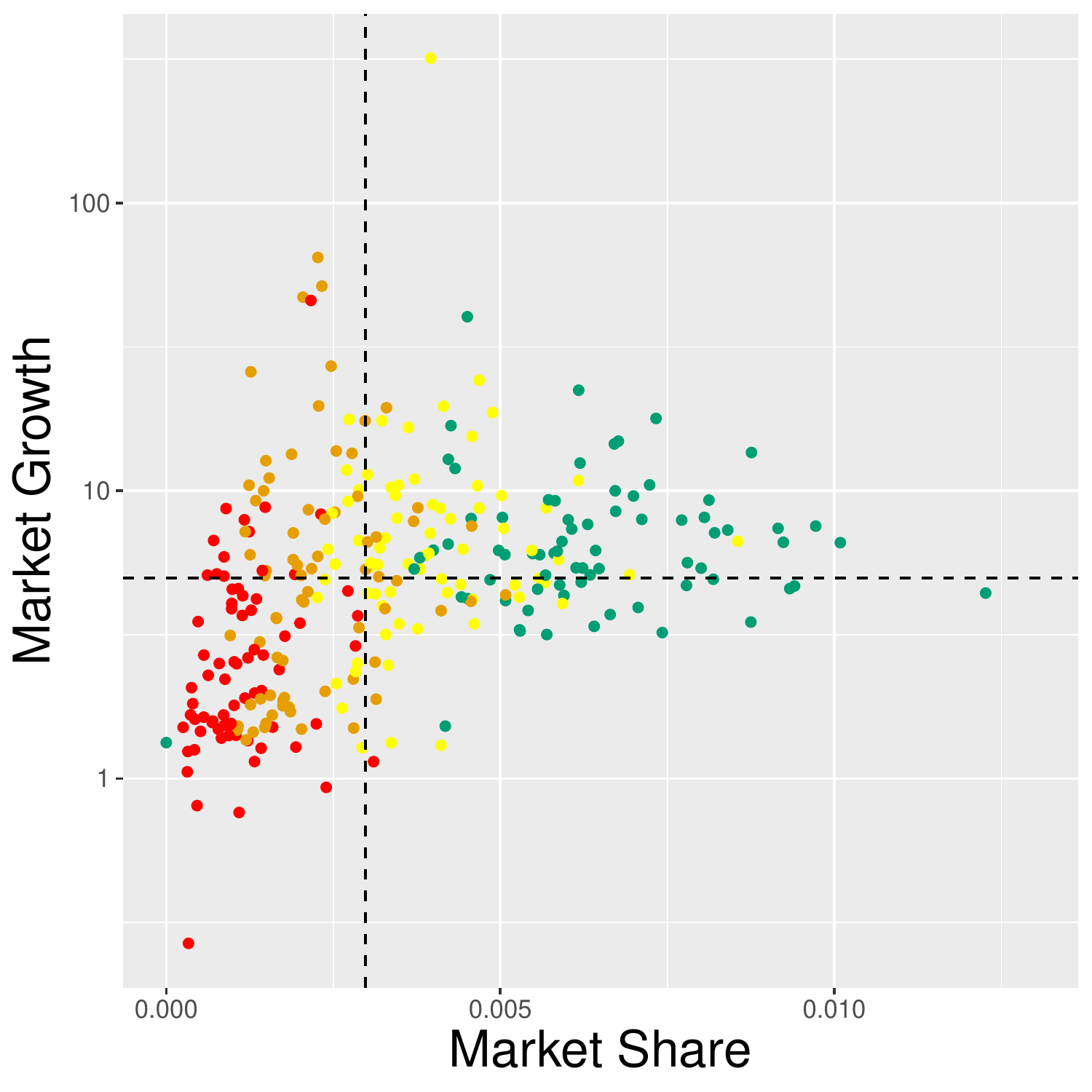}
		\caption{Market share - Market growth\label{fig:Marketshare_Marketgrowth}}
	\end{subfigure}
	\begin{subfigure}[b]{0.45\textwidth}
		\includegraphics[width=0.8\textwidth]{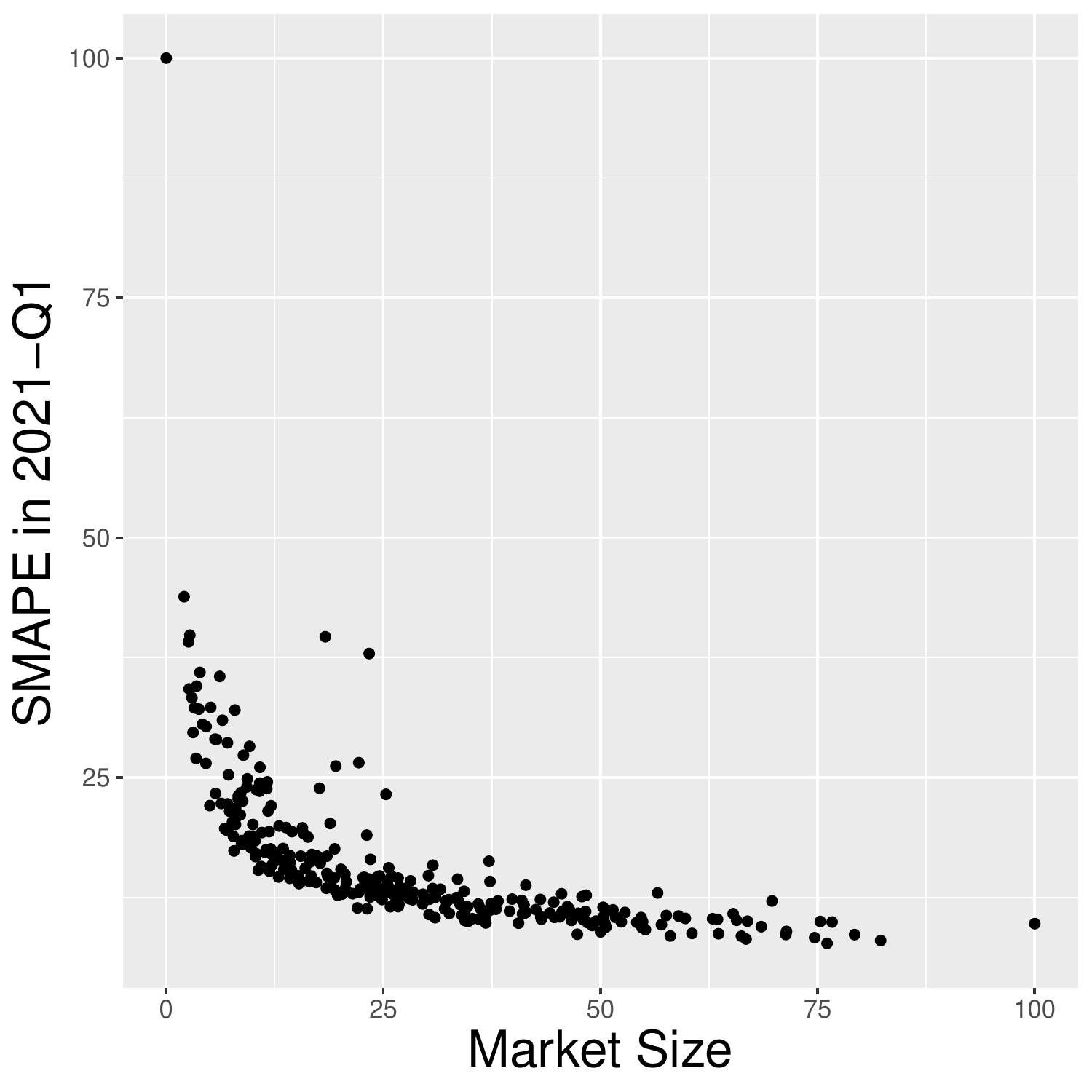}
		\caption{Market size - Forecast accuracy \label{fig:Marketsize_Forecastaccuracy}}
	\end{subfigure}
	\caption{Performance of the new approach related to market characteristics.  
		\label{fig:BCG}}
\end{figure}

Figure \ref{fig:BCG} panel (a) displays on a market share - market growth log-scale the SMAPE  performance of all areas. The dashed lines indicate the median levels of market share and growth, and we group by quartiles the SMAPE losses decreasing from red (worst), orange, yellow to green (best). The left lower quadrant, containing relatively small and no growth areas turn out to be the most complicated to forecast demand for. Performance improves somewhat for small areas with large market growth, see the left top quadrant. The picture turns yellow and mostly green in the two  right  quadrants, which constitute larger areas that typically grow between four and twenty percent. 

Figure \ref{fig:BCG} panel (b) emphasizes the effect of delivery area size on  forecast performance. Smaller areas are difficult to forecast relative to larger ones. However, this relationship is not linear and the gains of increasing the size of the areas taper off after levels of 30  in terms of total Q1-2021 demand. 
This finding is important because very large areas are less straightforward to manage in terms of supply and pricing strategies. Areas that grow too large can therefore be re-organized without loss of demand forecast accuracy. On the other hand, forecasting-wise  opening new areas that suffer from initial low demand makes sense only if rapid growth is expected.

Another strategically important issue for delivery platforms is scale, requiring wide market coverage to have a large enough volume to cover fixed costs, but also for example to strike deals with nationwide chains. Consequently, understanding the geographical differences in terms of operational efficiency allows prioritizing where to expand. We analyze the forecast performance of FFUDS for all UK delivery areas and look into differences between small and large city regions. Overall, we find that smaller cities with a small amount of  areas are more complicated to forecast than larger cities. Figures C8 and C9 (Appendix C) 
illustrate this for London and Newcastle. The size of the circles reflect the size of the SMAPE in 2021-Q1. The London region with many areas outperforms Newcastle on average, though one can also see areas in the center of London that are not that easy to forecast either which might be due to fierce local competition or volatile demand. 

\section{An Additional Application} \label{app:citibike}
To illustrate the generality of the FFUDS framework, we provide an additional application to a bicycle sharing system.  The data comes from Citi Bike in New York City. Citi Bike is owned by Lyft, an American leading mobility service provider, and operates in New York City since 2013.

Certain types of  Citi Bike system data is made  available to the public on the website https://s3. \linebreak amazonaws.com/tripdata/index.html. We download monthly data files from January 2019 (201901-citibike-tripdata.csv) to May 2023 (202305-citibike-tripdata.csv). 
Each dataset contains information about the station name, pick up and drop off geolocation, driver age, gender, etc, and is provided in irregularly spaced time frequency as the system operates 24 hours per day and records each trip.

For our analysis, we transform the data into hourly (0am to 11pm) demand  for bicycles by counting the number of departures at pick up stations. We further geographically aggregate the data into H3 cells which are based on the Hexagonal Hierarchical Spatial Index of Uber who uses it to optimize their ride-hailing taxi service platform. We use the  \texttt{R} package \texttt{h3jsr}  (\citealp{H3cells_v1}) to transform geolocations into H3 cells.
For our chosen granularity, this results in 16 pick up areas, for which we have 38,664 hourly observations each.

To logistically manage Citi Bike's platform, a forecast system  is required to dispatch enough bicycles at geolocations to satisfy demand.
The FFUDS framework is applied given an initial data batch from January 2019, and subsequently we produce at daily frequency 24 hour-ahead forecasts starting in February 2019 for which we compute SMAPE forecast loss streams. The demand for bicycles model is specified without domain knowledge. In particular, we use regression model (1) with a linear trend, hour of the day, day of the week and monthly seasonality dummies, and 168 demand lags, thereby considering a history of one entire business week ($24\times7 = 168$ lags). The regression parameters are estimated by the lasso to automatically identify the most relevant model parameters, so we only consider ``FFUDS-data" as we do not have access to domain expertise.

\begin{table}[t]
	\centering
	\caption{Citi Bike Application: SMAPE forecast performance.  \label{tabel:summar_forecast_smape_citibike}}
	  \resizebox{0.9\textwidth}{!}{\begin{minipage}{\textwidth}	
	\begin{tabular}{lllllccccccccc} \hline 
	Period	  &&&    && FFUDS & Naive & Prophet & \multicolumn{2}{c}{LSTM} & \multicolumn{2}{c}{SARIMA} & \multicolumn{2}{c}{ETS} \\ 
 Breaks	  &&&    && $\checkmark$ &  & $\checkmark$ & & $\checkmark$ & & $\checkmark$ & & $\checkmark$ \\  \hline 	
Pre-Covid &&&          && 100.00 &  104.69  & 146.88   & 108.29 & 100.27 & 109.35 & 105.51& 123.13 & 110.99\\      
Post-Covid &&&          && 100.00 & 109.91  & 177.81  & 111.02& 104.20 &  113.29& 105.99 & 152.10 & 125.10\\           
  \midrule
	\end{tabular}
\end{minipage}}
\raggedright
\footnotesize

Notes: One-day-ahead SMAPE forecast performance of the benchmarks relative to  FFUDS,  averaged across all H3 cells.
Values above 100 indicate the percentage gain in forecast accuracy of FFUDS relative to the benchmark.\\[0.1cm]
\end{table}

Table \ref{tabel:summar_forecast_smape_citibike} summarizes the results of FFUDS against the same benchmarks we considered in the platform application. FFUDS outperforms the considered benchmarks both pre- and post-Covid. Its relative performance is better post-Covid which can be explained by  the fact that 83\% of the detected breaks (on average around 28 breaks) are detected in this period. 
The importance of forecast breakdown detection also becomes apparent when considering the benchmarks LSTM, SARIMA and ETS as their performance considerably improves with breakdown detection, especially so for LSTM which then becomes the most competitive benchmark in terms of forecasts accuracy compared to FFUDS.
Note also that the Naive approach dominates all other standard benchmarks, which illustrates that the changing complex mix of seasonality, trends, and dynamics in the Citi Bike data is difficult to capture by direct application of standard available demand forecast tools.

Finally, Figure \ref{fig:2D_plot_errors_vs_time_citibike} displays a visual trade-off between the forecast performance and computing time of FFUDS versus its benchmarks. FFUDS is the only method that, relative to the Naive one, entails negligible computing time  and combines excellent forecast accuracy with low computing time. This makes FFUDS a framework that is scalable in a production environment.

 \begin{figure}[t]
\centering
\includegraphics[width = 0.45\textwidth]{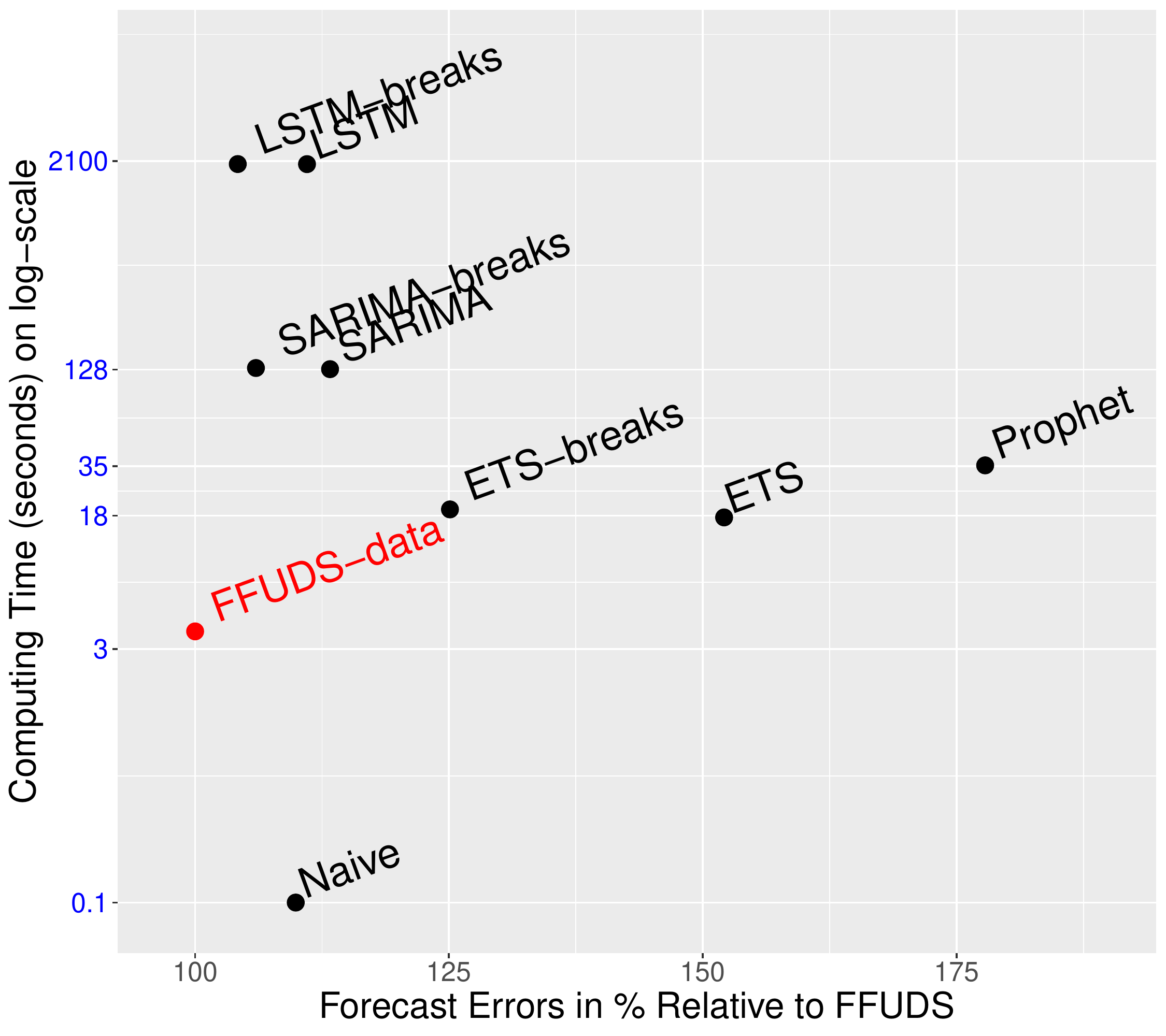} 
\caption{Citi Bike Application: Performance of the new method} 
\label{fig:2D_plot_errors_vs_time_citibike}
\footnotesize \raggedright  Notes:  Forecast errors in percentages versus computing time in seconds (in blue) for the forecasting methods: FFUDS (ours in red),LSTM (standard and with breaks), Prophet, SARIMA (standard and with breaks), ETS (standard and with breaks) and the Naive method.\\
\end{figure}

\section{Conclusion} \label{Conclusion}
On-demand service platforms are subject to irregular growth. Keeping them operationally efficient is challenging because such data-driven platforms require forecasting at a large scale for decision making. A successful forecast framework must be adaptive to rapidly changing environments, yet because of scale, can typically use own historical data only. Since platform data is streaming and forecasts need to be produced at high-frequency, the approach has to be computationally fast. The proposed FFUDS architecture offers such a forecasting framework.

A careful examination of platform data characteristics, reveals that, while strong trend and seasonality effects are present, the data streams are heavily impacted by episodes of instability. Therefore, we build Fast Forecasting of Unstable Data Streams (FFUDS), an automated framework that uses historical data and rests on three key components.
First, the use of a break detection procedure on streaming forecast losses, which allows dynamically adapting the data weighting scheme used for forecasting. 
Second, the use of a forecast combination approach where we simultaneously incorporate break information and stabilize the forecasts.
Third, our proposed general demand model
incorporates seasonality, trends, and nonlinear dynamic effects, that is simple enough to estimate in closed-form thereby permitting a fast streaming data setting.
We empirically validate our approach using demand from all UK delivery areas from an on-demand delivery service platform, find it to outperform several industry benchmarks, and analyze the strong positive economic and managerial implications of such enhanced forecast performance. Our approach is computationally fast so that it permits efficient forecast updates when new data becomes available, and it is applicable in the large-scale multi-country setting in which platforms typically operate.
There are two immediate future research extensions regarding our approach. First, in our application, it turns out that  FFUDS  performs less well in some morning hours of the business day. Indeed, in the setting of sparse data it is difficult-- for all forecast methods --to  deliver accurate forecasts. It would be interesting to develop tailored forecast approaches for sparse data settings. Second, we recommended to use FFUDS with fixed equally weighted forecast combinations, but this may be suboptimal in practice. The development of a data-driven rule for the forecast weighting scheme could further improve the performance of FFUDS.

In our platform application, the logistics services span 294 delivery areas. By forecasting demand for each area separately, we  implicitly assume the demand forecasting tasks to be independent across areas. In a sensitivity analysis, we investigate whether forecast performance can be improved by expanding regression model \eqref{eq:mainmodel} with past demand data of neighboring areas, thereby allowing for demand spillovers. Yet, such a multivariate forecasting approach did not result in improved forecast accuracy. 
While local operational planning units can thus construct demand forecast for each delivery area independently, 
driver wages are not necessarily determined independently per area given that drivers can move between areas to maximize their revenues, see \cite{Bimpikis_OR_2019}. Therefore, conditional on demand forecasts, driver wages are ideally optimized by the platform per marketplace of which the delivery areas are part of, see \cite{Besbes_mansci_2021} for a discussion on this complex optimization problem, which is currently still an open research question.

Future research should also be devoted to 
the inclusion of supply data of the independent couriers to handle further automation of prescriptive decision making regarding pricing.
Of particular interest is surge pricing which platforms use  as a signaling device by increasing a worker's compensation at a given delivery area above the regular price of the entire market place  to balance demand and supply in  areas with a shortage of workers compared to the demand, see  e.g., \cite{Guda_Subramanian_mansci2019} and \cite{Garg_Nazerzadeh_2021} for more details. 
Another topic on our agenda when considering both demand and supply data is  explicitly incorporating  the objective function and constraints regarding the optimal decision problem into the estimation problem, along the lines recently provided by  \cite{Elmachtoub_2022}.

The framework we develop has been successfully applied to an on-demand delivery service platform operating in the UK, but generalizes to any other business setting that encounters unstable data streams as illustrated via the bike share data analysis. 
FFUDS has further potential in  a wide range of other application fields. For instance, 
for offline stores such as those in large shopping malls, 
high-frequency consumer traffic forecasts are  needed but may face instability due to  marketing campaigns of the focal or neighboring stores in the mall.
Also online market places such as Walmart or Amazon require fast forecasts  of streaming web traffic data (collected via URL access logs) at the order of seconds to minutes which typically display bursts or sudden drops. 
A final example consists of hospitality platforms that deliver hotel occupation rate forecasts to their clients for the optimization of hotel personnel, food and leisure activities, but that in recent years face various instabilities, amongst others, due to issues of suddenly changing COVID-19 restrictions, changing weather conditions due to climate change, or geopolitical events.

\noindent
\section*{Acknowledgments}
We thank the senior editor, associate editor and referees for their constructive and detailed comments which substantially improved the quality of the manuscript.
We are very grateful to Benjamin Wolter for expert advice, to Arnaud Dufays, Sarah Gelper, Harris Kyriakou, Olivier Scaillet for  comments and checks provided on earlier versions of the paper and
to the participants at the Symposium on Statistical Challenges in Electronic Commerce Research (SCECR 2022),
the Conference on Data Science, Statistics \& Visualisation (DSSV 2022), the  International Symposium on Forecasting (ISF 2022),
and the Workshop on Information Technologies and Systems (WITS 2022) for helpful discussions. IW was financially supported by the Dutch Research Council (NWO) under grant number VI.Vidi.211.032.

\bibliographystyle{apalike}
\bibliography{bibfile}

\clearpage
\newpage

\begin{appendices}
	
	\renewcommand{\thefigure}{A\arabic{figure}}
	\renewcommand{\thetable}{A\arabic{table}}
	\setcounter{figure}{0}
	\setcounter{table}{0}
	
	\section{Platform Application: Additional Data Insights} \label{app:data}
	We provide additional insights into the
	(i) typical demand patterns of the data, and
	(ii) demand heterogeneity across delivery areas.
	
	Figure \ref{fig:total_uk_intraday}  shows intraday UK demand.
	While demand is aggregated over all product categories (food, fashion, health, etc.), 
	the large majority of deliveries consists of food. This explains why demand starts low at 9am, is moderately high around lunchtime (12-2pm),  goes down only mildly between 2-5pm, sharply increases afterwards and peaks around 8pm, after which it   decreases again to levels comparable to late afternoon. 
	Apart from a strong intraday demand pattern, pronounced day of the week fluctuations are present  as well, see Figure \ref{fig:total_uk_weekday}. Average demand is lowest on Mondays, steadily increases until Thursdays, jumps on Fridays and Saturdays, and lowers on Sundays. 
	These intra-day and weekly demand patterns also arise for the delivery areas albeit some areas have no reduction in late afternoon demand, or peak later than 8pm.

	\begin{figure}[h]
		\centering
		\begin{subfigure}[b]{0.49\textwidth}
			\includegraphics[width=\textwidth]{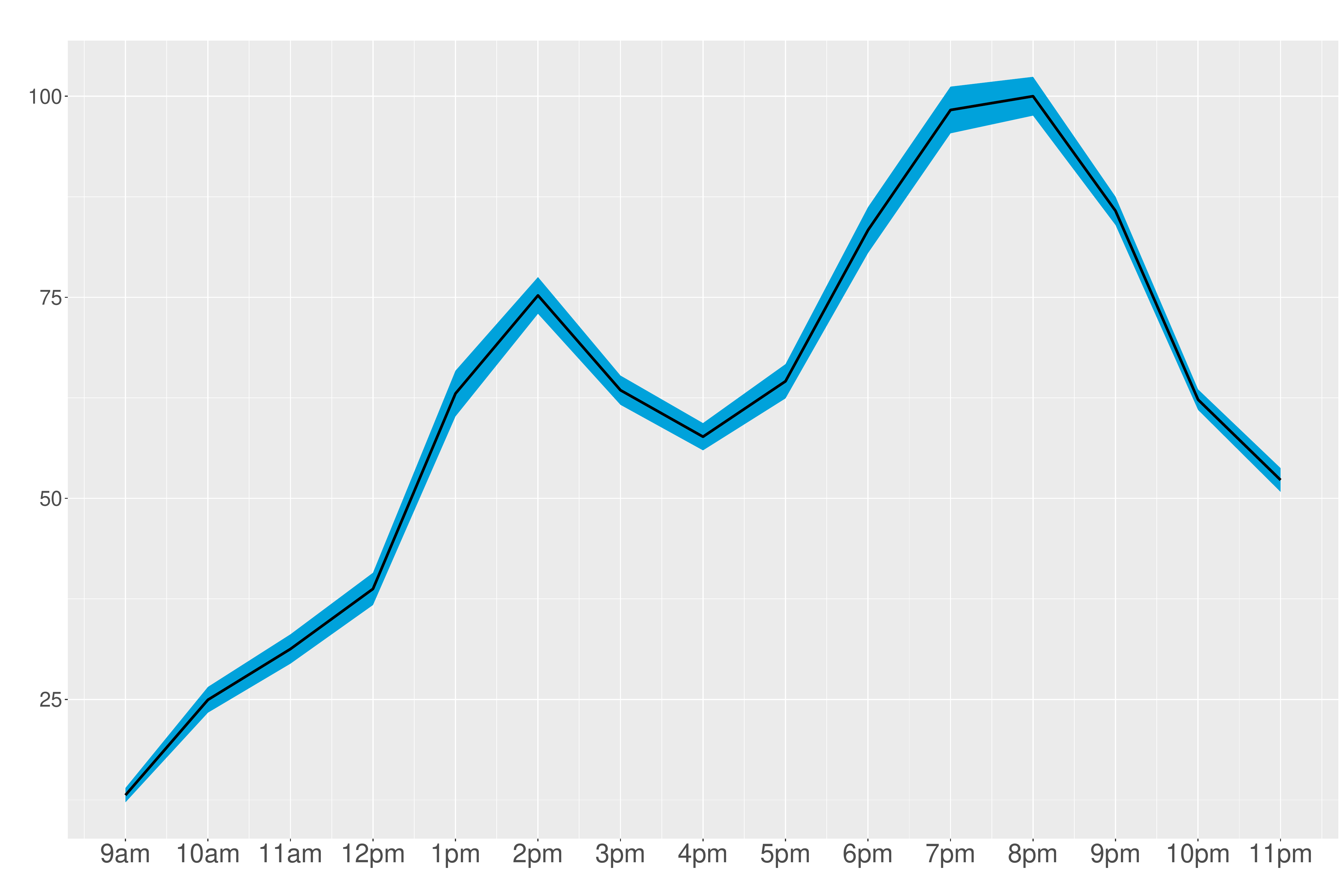}
			\caption{Intraday Demand \label{fig:total_uk_intraday}}
		\end{subfigure}
		\begin{subfigure}[b]{0.49\textwidth}
			\includegraphics[width=\textwidth]{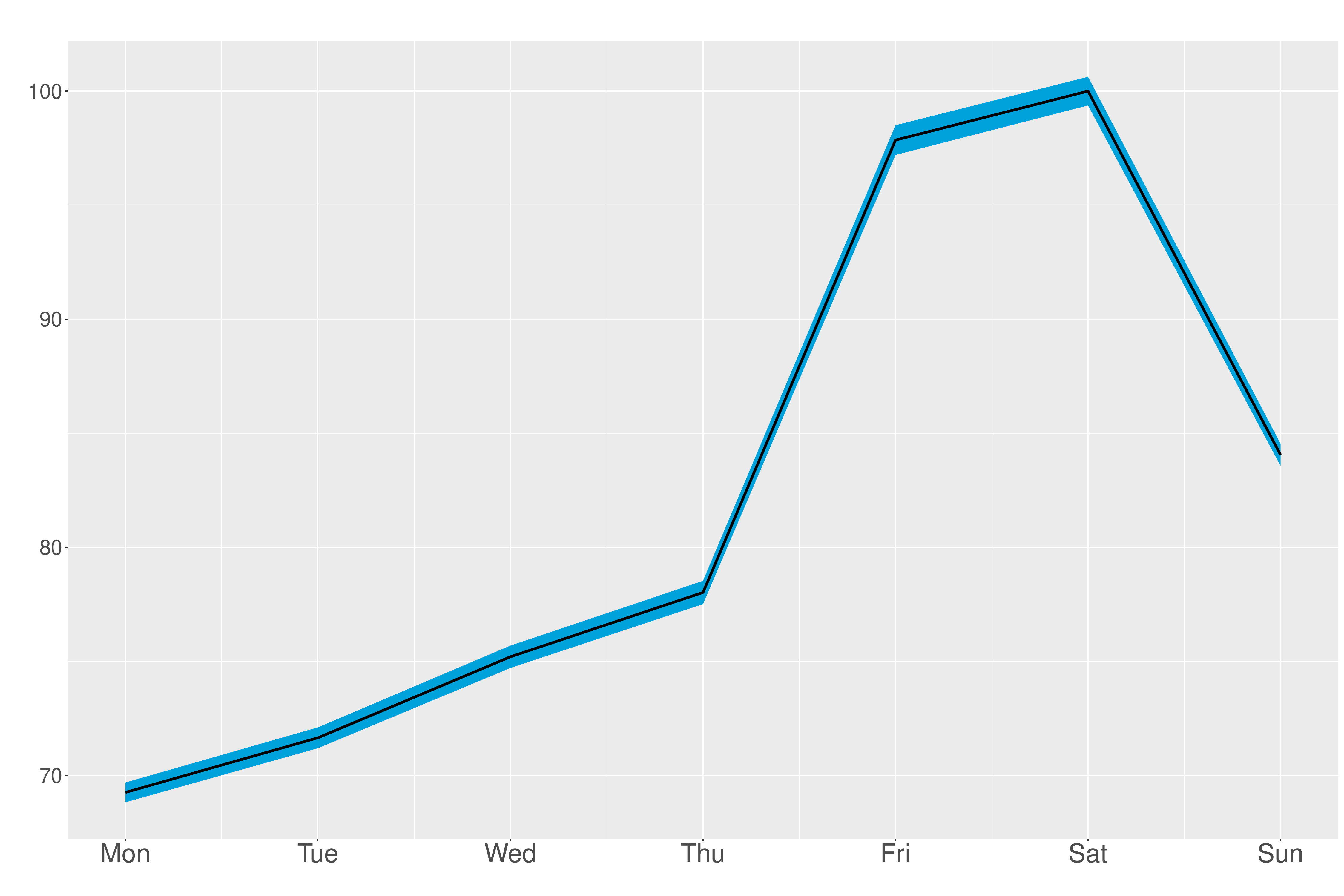}
			\caption{Day of Week Demand \label{fig:total_uk_weekday}}
		\end{subfigure}
		\caption{Intraday hourly demand and day of the week average demand for UK. One standard error   bands are displayed in blue.  
			\label{fig:total_uk_intraday_weekday}}
	\end{figure}
	
	Figure \ref{fig:low_morning_demand} further zooms into the  sparsity of the platform data, which mainly manifests itself early morning (9am to 12pm). We display histograms for average morning  demand in each delivery area of the UK, Pre-Covid (top) and Post-Covid (bottom).
	Pre-Covid, morning demand is for the majority of delivery areas zero.
	Post-Covid, morning demand considerably increases across all delivery areas due to changing changing customer habits.

	\begin{figure}[h]
		\centering
		\includegraphics[width=\textwidth]{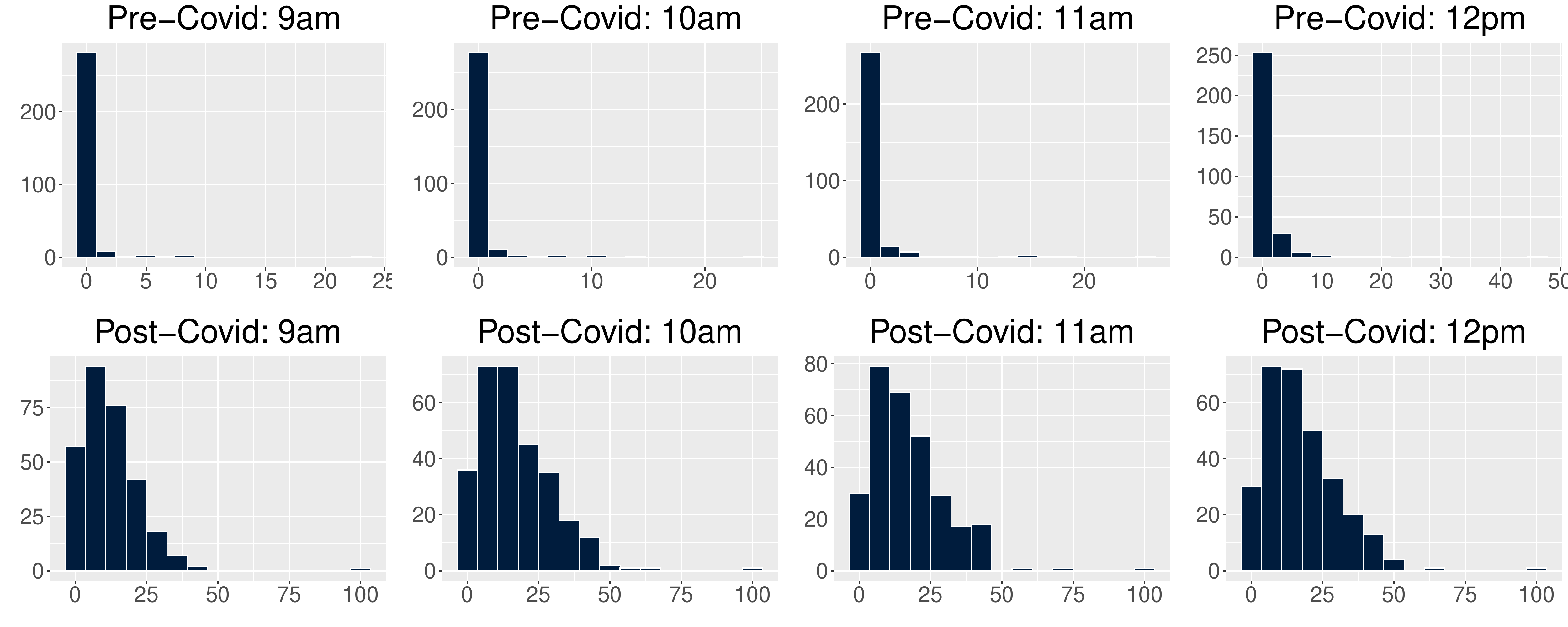}
		\caption{Histograms for average demand in morning hours in each delivery area of the UK, Pre-Covid (top) and Post-Covid (bottom). \label{fig:low_morning_demand}
		}
	\end{figure}
	
	\begin{figure}
		\centering
		{\includegraphics[width=16cm,height=8cm]{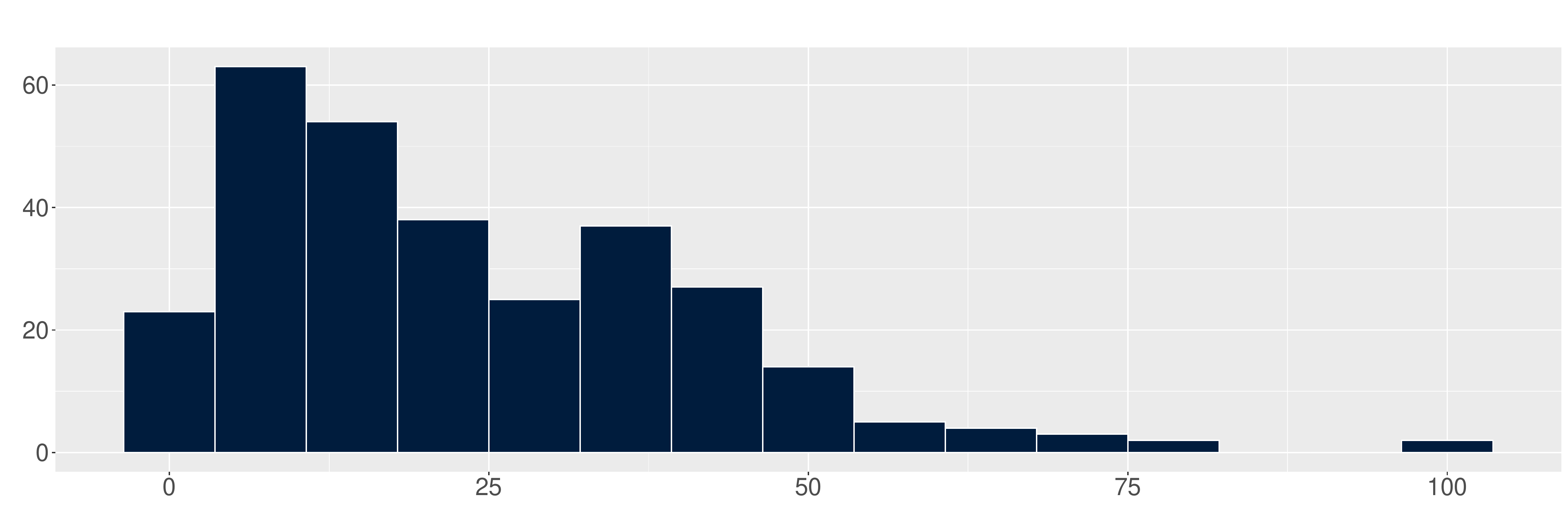} }	
		\caption{Histogram for aggregated demand in each delivery area of the UK.
		}
		\label{fig:total_areas_demand}
	\end{figure}

	Figure \ref{fig:total_areas_demand}  shows the histogram of total demand for the delivery areas over the entire sample period.  An interesting tri-modal pattern emerges. The first and largest group has total demand of less than 25, the second largest group up to 50. A small third group has high demand of almost 100, the largest delivery area being Bethnal Green in the London region. 
	Figure \ref{fig:map_UK} visualizes the demand on a UK map to highlight the geographical coverage of the  delivery areas. Each circle is placed at the center of one out of the 294 areas, giving by its latitude and longitude, and its diameter is proportional to the area's total demand. Unsurprisingly, most cities are covered with multiple delivery areas, the number of which is directly related to population size. Figure  \ref{fig:map_London} zooms in on the London area and reveals that the delivery area total demand is not uniformly distributed.  The main reason for this is that fleet planning is organized with respect to drivers' availability rather than clients' demand, thereby giving rise to substantial area-demand  heterogeneity ranging from housing over business towards tourist areas.
	
	\begin{figure}[t] 
		\centering
		{\includegraphics[width=10cm,height=8cm]{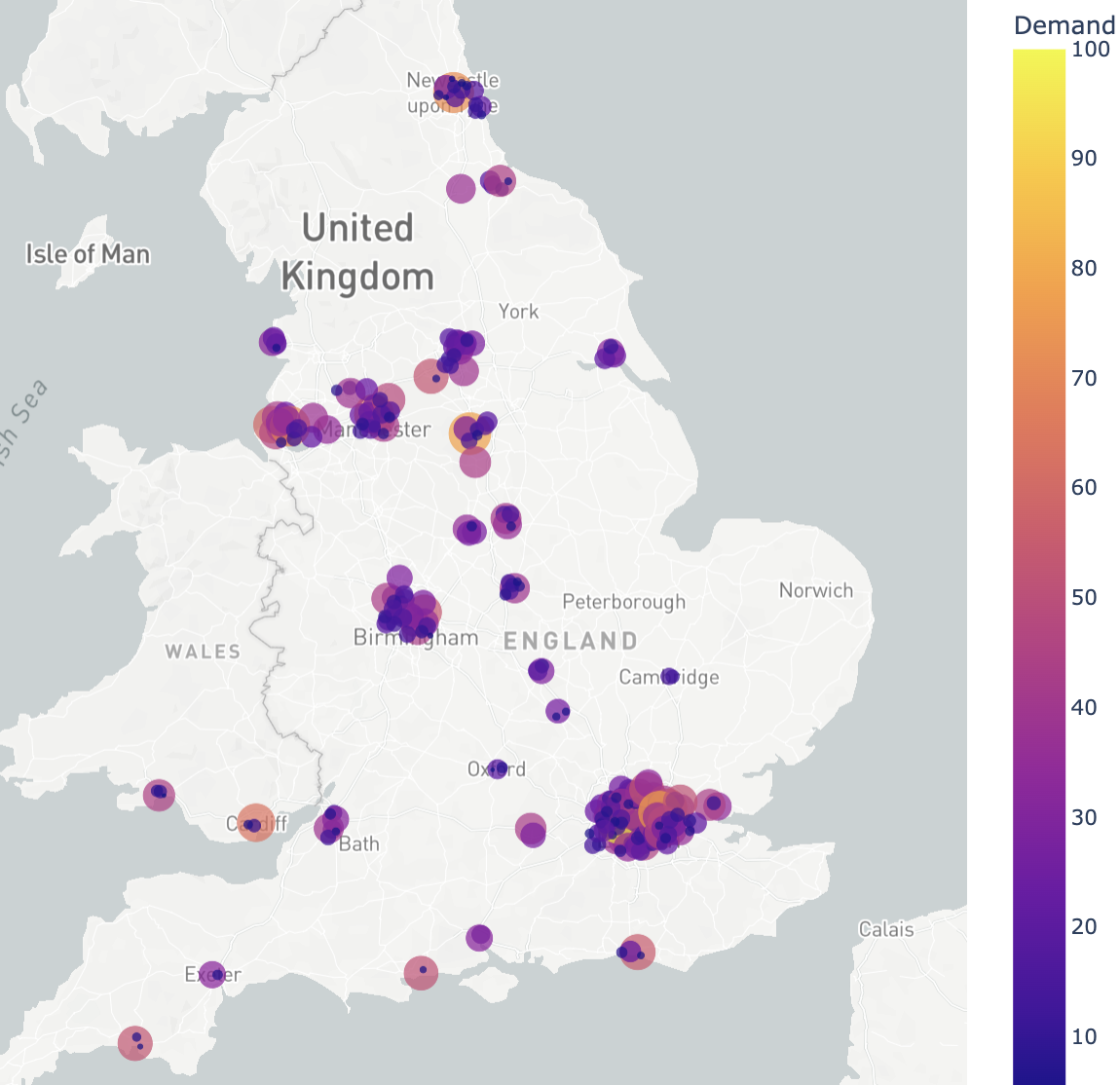} }	
		\caption{Aggregated demand in each delivery area in the UK. 
		}%
		\label{fig:map_UK}
	\end{figure} 
	
	\begin{figure}
		\centering
		{\includegraphics[width=10cm,height=6cm]{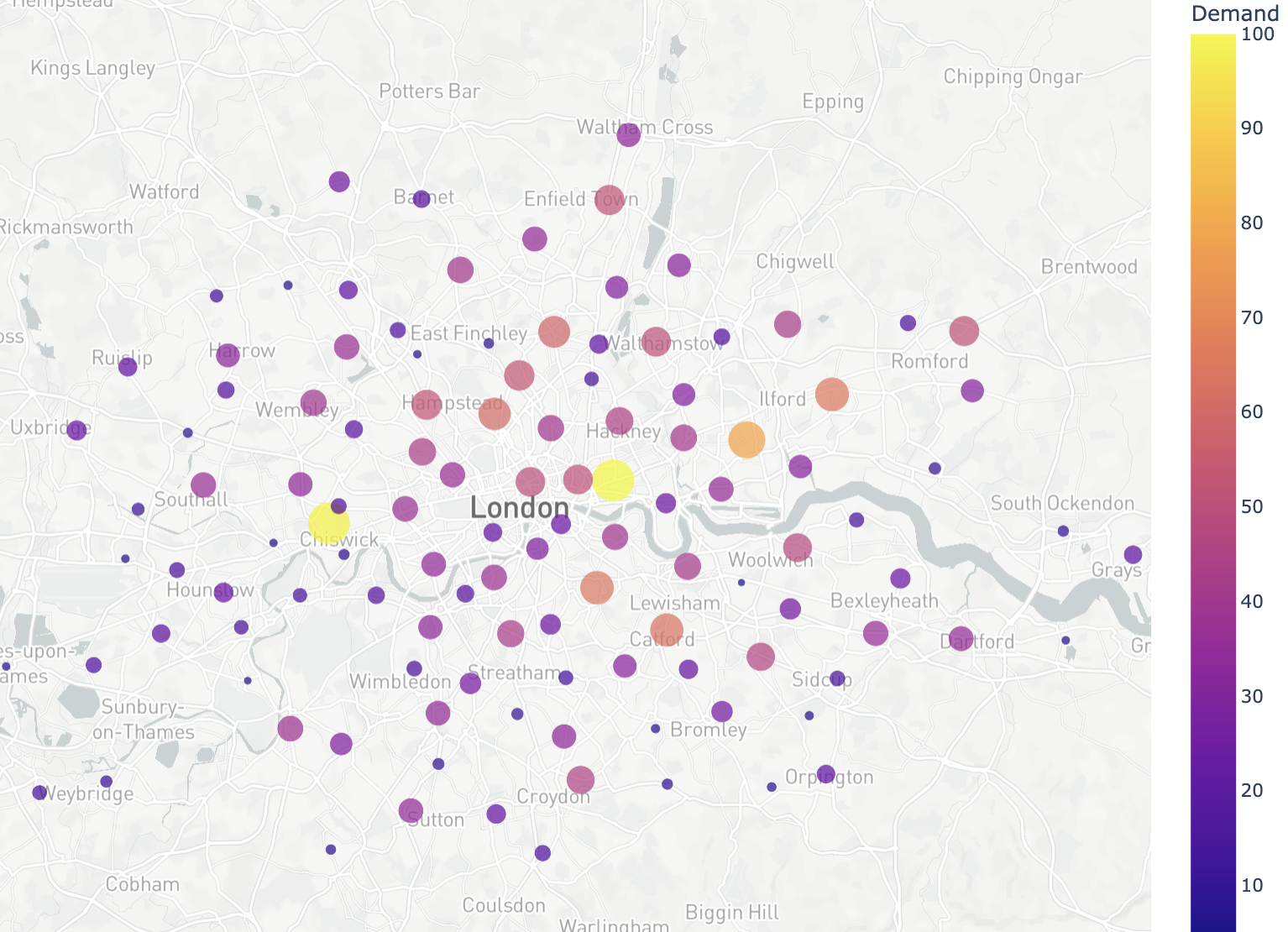} }
		\caption{Aggregated demand in each delivery area in  the London region. 
		}%
		\label{fig:map_London}
	\end{figure} 
	
	\clearpage
	\newpage
	
	\section{Benchmarks Methods} \label{app:prophet_lstm}
	In Sections \ref{app:prophet} and \ref{app:LSTM}, we provide a brief description of the Prophet and LSTM forecasting approaches. We denote the demand at hourly frequency for one specific delivery area as $d_t$. Both approaches rely on historical information up to time $t$ to forecast $d_{t+h}$ where $h$ is the forecast horizon.

	\subsection{Prophet}  \label{app:prophet}
	The main idea behind Prophet is decomposing $d_t$ into the following three deterministic parts:
	\begin{eqnarray}
		d_t = g_t + s_t + v_t + \varepsilon_t \label{eq:prophet}
	\end{eqnarray}
	with $g_t$ the trend, $s_t$ and $v_t$ the seasonal and holiday parts respectively. 
	The zero mean error term $\varepsilon_t$ represents idiosyncratic changes not captured by the three parts. 
	The advantage of this deterministic approach is that the forecast for $d_{t+h}$ can be directly computed as $g_{t+h} + s_{t+h} + v_{t+h}$.
	The trend part $g_t$ is specified by a flexible piece-wise logistic growth model accounting for change-points in the time series at known (specified by the user) or unknown dates (determined by the data only). Growth rate, capacity and offset parameters drive this trend specification.
	The seasonal part $s_t$ is captured by a standard Fourier series yielding smooth effects, and for which parameter coefficients need to be estimated. Finally, holiday effects $v_t$ are typically unsmooth and are therefore integrated in the model using indicator functions each of which are multiplied by a parameter representing the specific average effect. Assuming a law for $\varepsilon_t$, e.g.\ Gaussian, the likelihood  of model \eqref{eq:prophet} 
	is known and is then combined with uninformative priors for posterior inference on the parameters. This allows integrating model parameter uncertainty and can take care of out-of-sample forecast uncertainty, but these features require additional computation time.  See \cite{prophet_2018} for more detailed explanations.
	
	\subsection{LSTM Neural Network}  \label{app:LSTM}
	Long-Short-Term-Memory (LSTM) networks were first introduced by \cite{Hochreiter_1997}, and have become increasingly popular for time series forecasting. 
	They are part of recurrent neural networks but they do not suffer from the vanishing/exploding gradient problem that arises when  minimising the forecast error loss function.
	An LSTM filters information through an input ($i_t$), forget ($f_t$), output ($o_t$) gate. The forget gate ensures that the input and output gates consider only the important information of the new input and previous time period respectively. A cell state $c_t$ introduces some memory to the LSTM in order to remember the past. More formally, the structure of an LSTM  to forecast $d_{t+h}$ is a follows:
	\begin{eqnarray}
		i_t &=& \sigma \left( W_i X_t + U_i h_{t-1} + b_i  \right) \nonumber \\
		f_t &=& \sigma \left( W_f X_t + U_f h_{t-1} + b_f  \right)  \nonumber\\
		o_t &=& \sigma \left( W_o X_t + U_o h_{t-1} + b_0  \right)  \nonumber\\
		c_t &=&   f_t \odot  c_{t-1}  +  i_t \odot \tanh  \left(  W_c X_t + U_c  h_{t-1} +b_c \right) \nonumber\\
		h_t &=& o_t \odot \tanh (c_t) \nonumber\\
		\hat d_{t+h} &=& W_d h_t + b_d\nonumber
	\end{eqnarray}
	where the weight matrices $W$, $U$,  and bias vectors $b$ contain the parameters to be estimated, $X_t$ is the historical demand data, $\sigma$ is the sigmoid function, $\odot$ is the element-wise (or Hadamard) product operator. The number of units in the LSTM is for example reflected by the dimension of the input function $i_t$.  The initial cell state $c_0$ and hidden state $h_0$ are set to zero. Stochastic gradient descend methods are employed to minimise the squared loss $\Sigma_t (d_{t+h} - \hat d_{t+h})^2$ summed over the available demand data.

	\renewcommand{\thefigure}{C\arabic{figure}}
	\renewcommand{\thetable}{C\arabic{table}}
	\setcounter{figure}{0}
	\setcounter{table}{0}
	
	\newpage
	\clearpage
	\section{Platform Application: Additional Results} \label{app:other_loss}

	\begin{table}[ht]
		\caption{RMSE forecast performance.   \label{tabel:summar_forecast}}
		\resizebox{0.86\textwidth}{!}{\begin{minipage}{\textwidth}	
				\begin{tabular}{lllllcccccccccc} \hline 
					&&&    && \multicolumn{2}{c}{FFUDS} & Naive & Prophet & \multicolumn{2}{c}{LSTM} & \multicolumn{2}{c}{SARIMA}  & \multicolumn{2}{c}{ETS} \\ 
					&&& && domain & data & & & & & & & \\
					Breaks	  &&&    && $\checkmark$ & $\checkmark$ &  & $\checkmark$ & & $\checkmark$  &   & $\checkmark$ &  & $\checkmark$ \\  \hline 	
					Pre-Covid &&&          && 100.00 & 100.71  & 128.54 &  107.74 & 110.99 & 101.69 & {120.48} & 115.89 & {110.56} & 107.03 \\      
					Post-Covid &&&          && 100.00 & 101.21 &  119.07 &  138.68 &  116.23 & 100.16 & {111.01} & 106.41 &  {106.25} & 102.70\\         
					\midrule
				\end{tabular}
		\end{minipage}}
		
		\raggedright
		\footnotesize
		Notes: One-day-ahead RMSE forecast performance of the benchmarks relative to FFUDS domain,  averaged across all areas.
		Values above 100 indicate the percentage gain in forecast accuracy of FFUDS relative to the benchmark.  \\ [0.1cm]
	\end{table}

	\begin{table}[ht]
		\caption{SMAPE forecast performance for different choices of predictor sets and parameter restrictions. \label{FFUDS-components-ablation}}
		\centering
		\begin{tabular}{llcccccc} \hline 
			\multicolumn{2}{l}{Components to Vary}   &  \multicolumn{6}{c}{Variations of FFUDS} \\ 
			& & (1) & (2) & (3) & (4) & (5) & (6) \\\hline 
			Predictors & Trend \& seasonality & \checkmark & \checkmark & & \checkmark &  & \checkmark \\
			& Lagged demand dynamics & \checkmark & & \checkmark & & \checkmark &   \checkmark\\
			Restrictions  & Domain-expertise & \checkmark & \checkmark & \checkmark & &  &  \\
			& Data-driven & & & & \checkmark & \checkmark  & \checkmark \\ \hline 
			Pre-Covid            & & 100.00 & 104.75 & 102.06 & 104.96 &  100.95 & 100.11\\ 
			Post-Covid           & & 100.00 & 114.39 & 97.72 & 114.48 & 94.02 &  94.17 \\ \hline 
		\end{tabular}
	\end{table}
	
	\begin{table}[ht]
		\caption{Overview of considered change-point detection methods.}
		\resizebox{0.75\textwidth}{!}{\begin{minipage}{\textwidth}
				\begin{tabular}{lllll} \hline 
					Abbreviation & Method & Reference & \texttt{R}-package & Package Reference \\ \hline 
					PELT & Pruned Exact Linear Time & \cite{killick2012optimal} & \texttt{changepoint} & \cite{killick2014changepoint} \\  
					AMOC    & At Most One Change & \cite{hinkley1970inference} & \texttt{changepoint} & \cite{killick2014changepoint} \\  
					BINSEG  & Binary Segmentation & \cite{scott1974cluster} & \texttt{changepoint} & \cite{killick2014changepoint} \\  
					BOCPD   & Bayesian Online Changepoint Detection & \cite{adams2007bayesian} & \texttt{ocp} & \cite{ocp} \\
					CPNP    & Nonparametric Change Point Detection & \cite{haynes2017computationally} & \texttt{changepoint.np} & \cite{changepointnp} \\
					EPC     & Energy Change Point & \cite{matteson2014nonparametric} & \texttt{ecp} & \cite{ecp-article} \\
					KCPA    & Kernel Change-Point Analysis & \cite{harchaoui2008kernel} & \texttt{ecp} & \cite{ecp-article} \\
					SEGNEIGH& Segment Neighborhoods & \cite{auger1989algorithms} & \texttt{changepoint} & \cite{killick2014changepoint} \\
					WBS     & Wild Binary Segmentation & \cite{fryzlewicz2014} & \texttt{wbsts} & \cite{wbts} \\  \hline 
				\end{tabular}
		\end{minipage}}
	\end{table}

	\begin{table}[ht]
		\caption{SMAPE forecast performance for different change-point detection methods. \label{FFUDS-cpt-methods}}
		\centering
		\begin{tabular}{lcccccc}
			\hline
			FFUDS with &&& Pre-covid &&& Post-covid \\ 
			\hline
			PELT &&& 100.00 &&& 100.00 \\ 
			AMOC &&& 99.98 &&& 99.90 \\ 
			BINSEG &&& 99.96 &&& 99.81 \\ 
			BOCPD &&& 100.41 &&& 99.72 \\ 
			CPNP &&& 100.13 &&& 99.81 \\ 
			EPC &&& 100.75 &&& 102.66 \\ 
			KCPA &&& 100.75 &&& 102.66 \\ 
			SEGNEIGH &&& 100.01 &&& 99.39 \\
			WBS &&& 100.02 &&& 99.48 \\ 
			\hline
		\end{tabular}
	\end{table}

	\begin{table}[ht]
		\caption{SMAPE forecast performance for the change-point detection methods with multiple detected change-points, either defining the post-break sample after the earliest or the latest detected break. } \label{multiple-breaks}
		\centering
		\begin{tabular}{lcccccc}
			\hline
			FFUDS with  &&& Pre-covid &&& Post-covid \\ 
			\hline
			PELT &&& 100.00 &&& 100.00 \\
			SEGNEIGH, earliest break &&& 100.01 &&& 99.39 \\ 
			SEGNEIGH, latest break &&& 100.01 &&& 99.40 \\ 
			BOCPD, earliest break &&& 100.41 &&& 99.72\\ 
			BOCPD, latest break &&& 100.41 &&& 99.73 \\ 
			CPNP, earliest break &&& 100.13 &&& 99.81 \\ 
			CPNP, latest break &&& 100.14 &&& 99.82 \\ 
			\hline
		\end{tabular}
	\end{table}

	\begin{figure}[ht]
		\centering
		\begin{subfigure}[b]{0.45\textwidth} 
			\includegraphics[width=\textwidth,height=7cm]{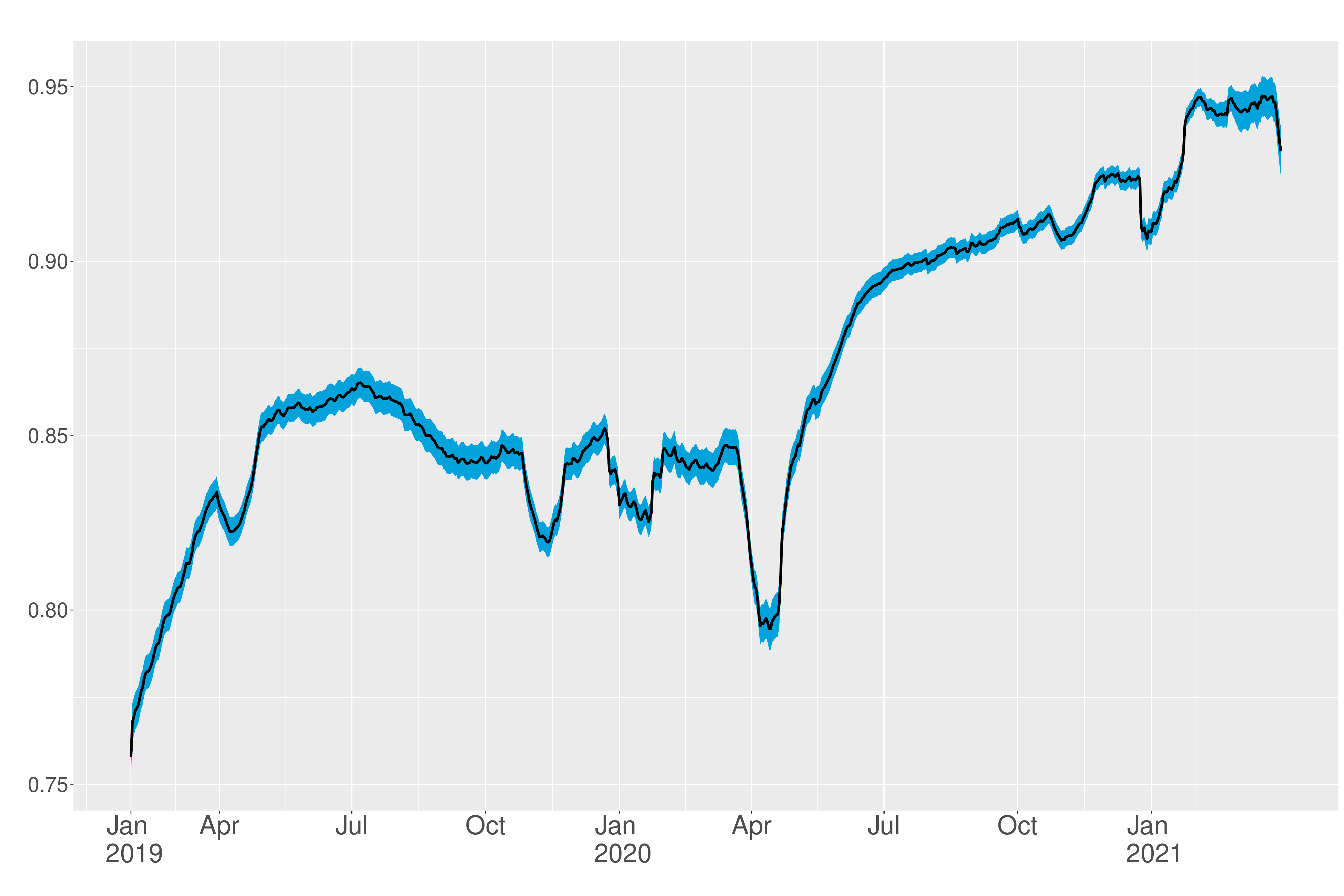}
			\caption{R-squared\label{fig:Rsquared_trend_season}}
		\end{subfigure}
		\begin{subfigure}[b]{0.45\textwidth}
			\includegraphics[width=\textwidth,height=7cm]{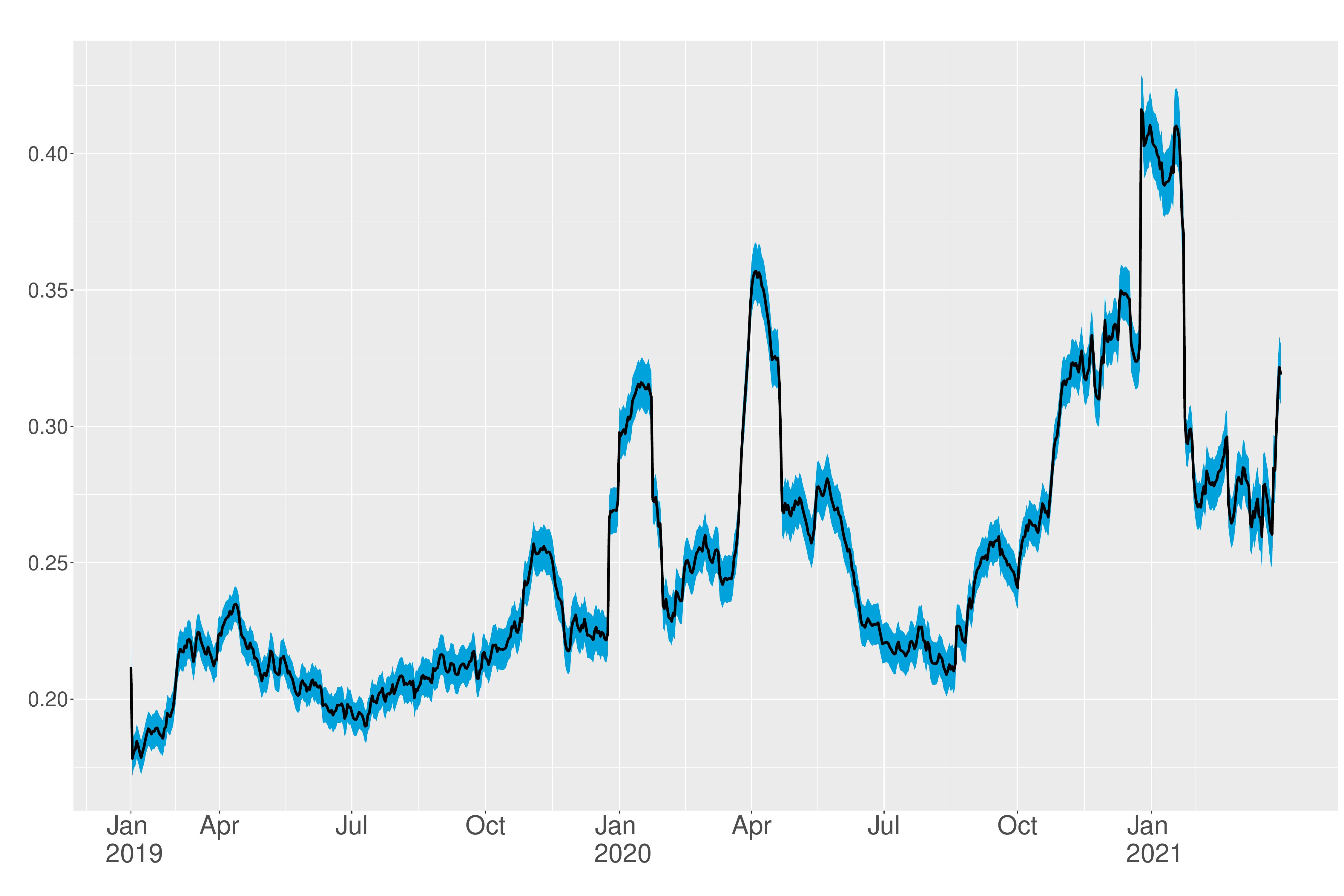}
			\caption{First Order AutoCorrelation Coefficient\label{fig:ACF_trend_season}}
		\end{subfigure}
		\caption{Streaming average $R^2$ (a) and First Order AutoCorrelation Coefficient (b) for the trend and seasonality model fitted to the delivery areas. One  standard error bands are displayed in blue.  
			\label{fig:Rsquared_ACF_streams}}
	\end{figure}

	\begin{figure}[ht]
		\centering
		\includegraphics[width=0.49\textwidth]{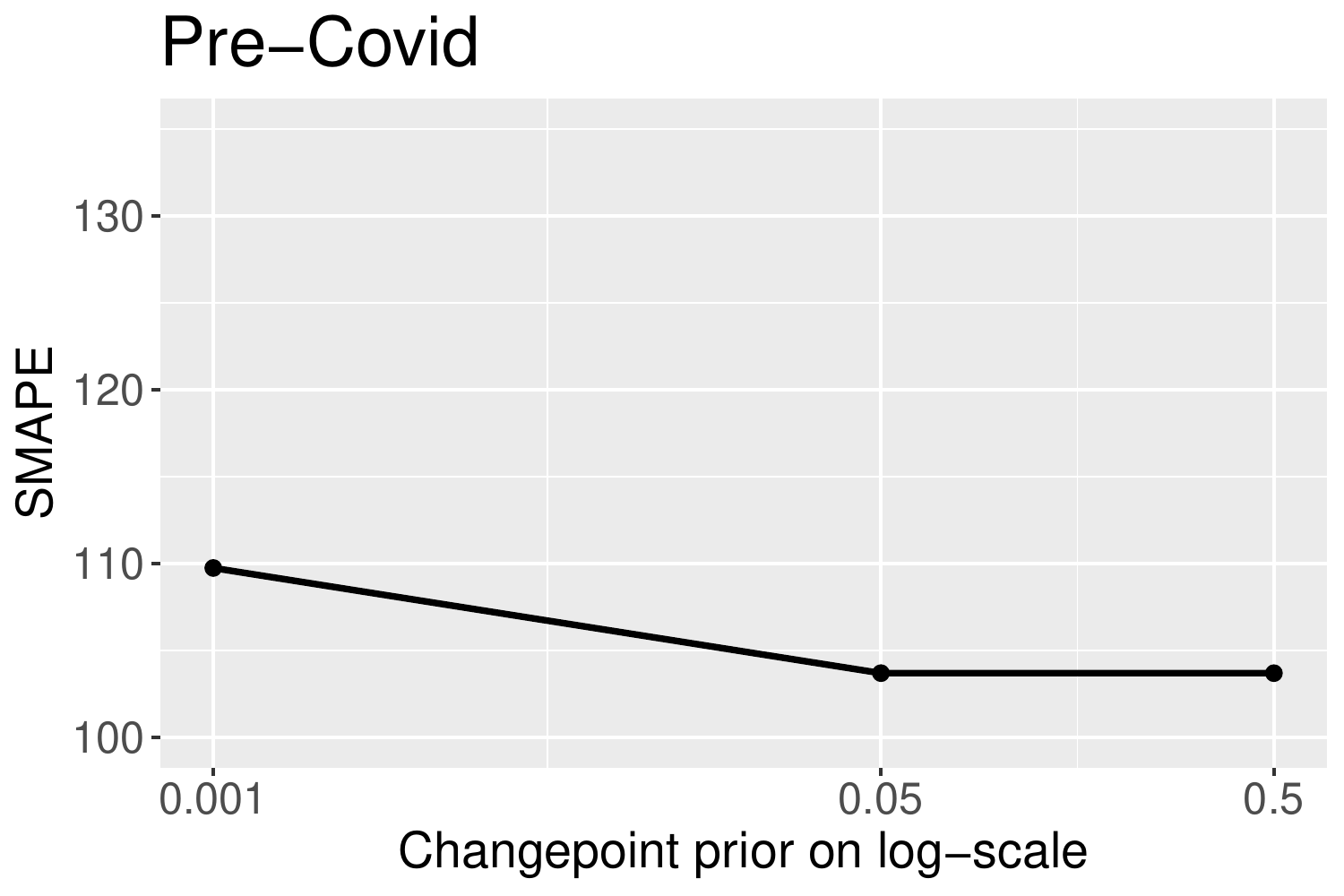}
		\includegraphics[width=0.49\textwidth]{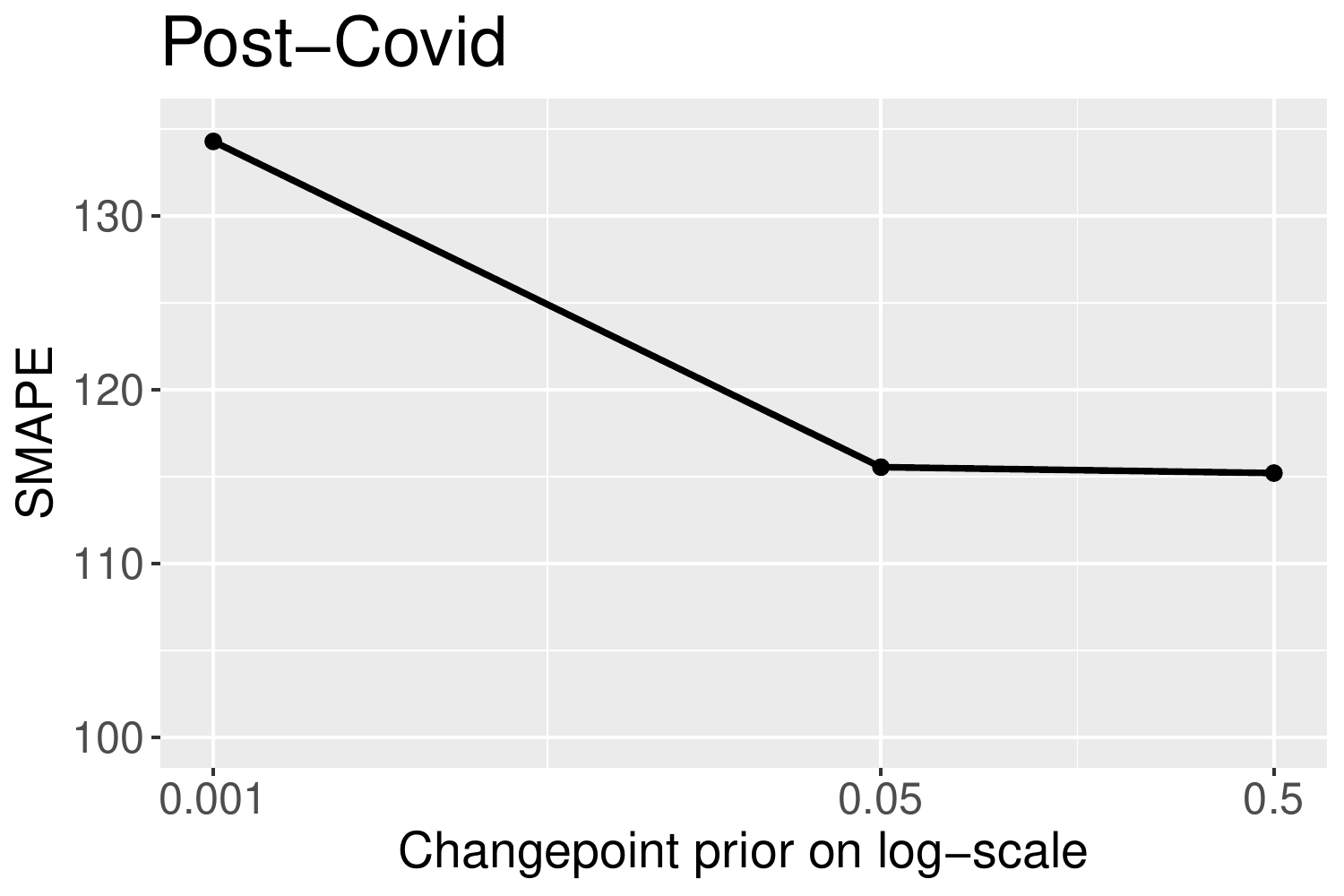}
		\caption{SMAPE forecast performance of Prophet for different change-point priors relative to FFUDS. \label{prophet_sensitivity}}
	\end{figure}
	\color{black}
	
	\begin{figure}[t]
		\begin{center}
			\includegraphics[width=0.23\textwidth]{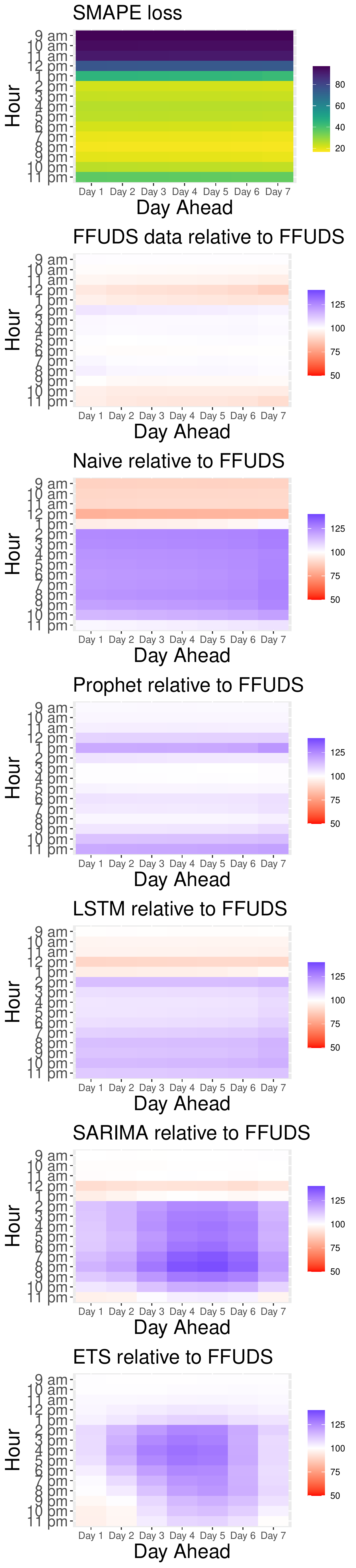}
			\includegraphics[width=0.23\textwidth]{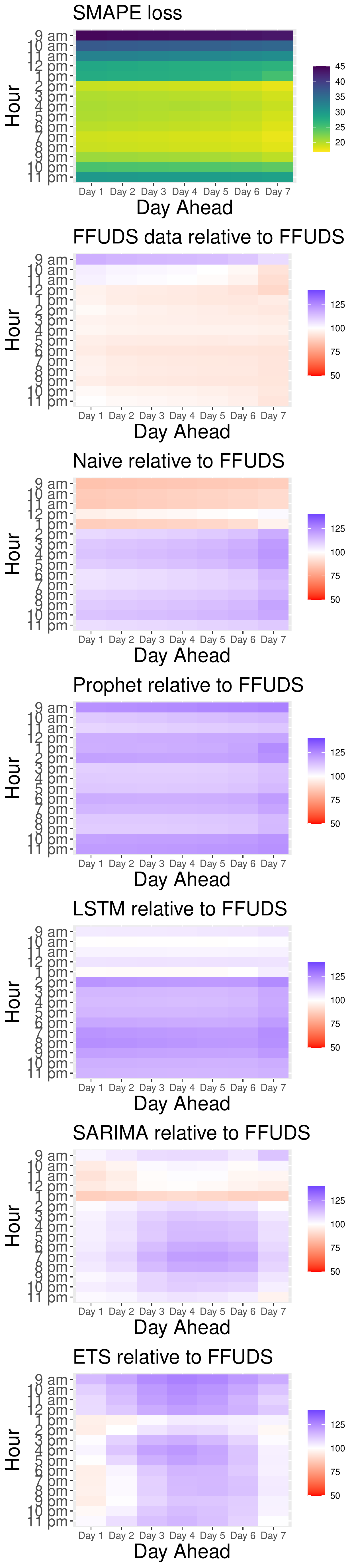}	\includegraphics[width=0.23\textwidth]{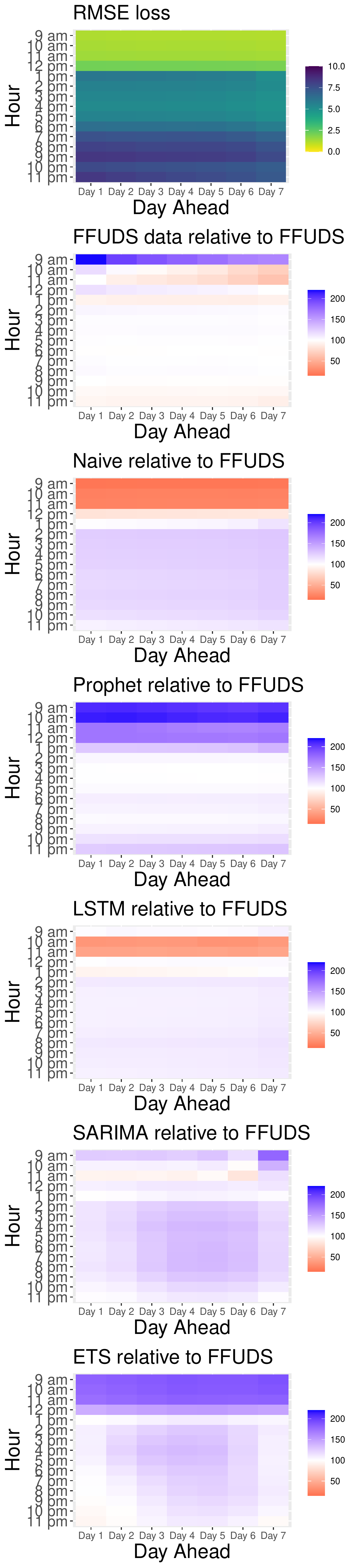}
			\includegraphics[width=0.23\textwidth]{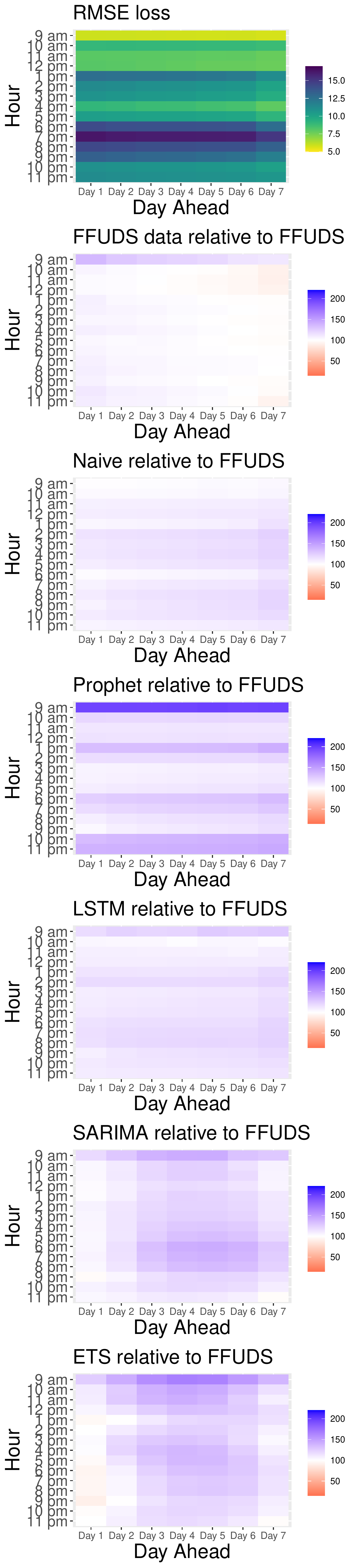}
			\caption{SMAPE and RMSE forecast performance heat maps.}   \label{fig:heatmap:smapermse}
		\end{center}
		{\footnotesize \raggedright 
			Notes:  Heat maps of SMAPE (columns 1-2: Pre- and Post-Covid) and RMSE (columns 3-4: Pre- and Post-Covid) forecast performance averaged across delivery areas for different hours of day (vertical axis) and days-ahead (horizontal axis).
			Top: Loss for FFUDS domain.
			Bottom rows: Respectively FFUDS data, Naive, Prophet, LSTM, SARIMA and ETS relative to FFUDS domain.
			Values above (below) 100\% visualized in blue (red) indicate better (worse) 
			performance of FFUDS (domain) compared to the benchmark. Equal performance is visualized in white. } \\[00.5cm]
	\end{figure}
	
	\begin{figure}[t]
		\begin{center}
			\includegraphics[width=0.23\textwidth]{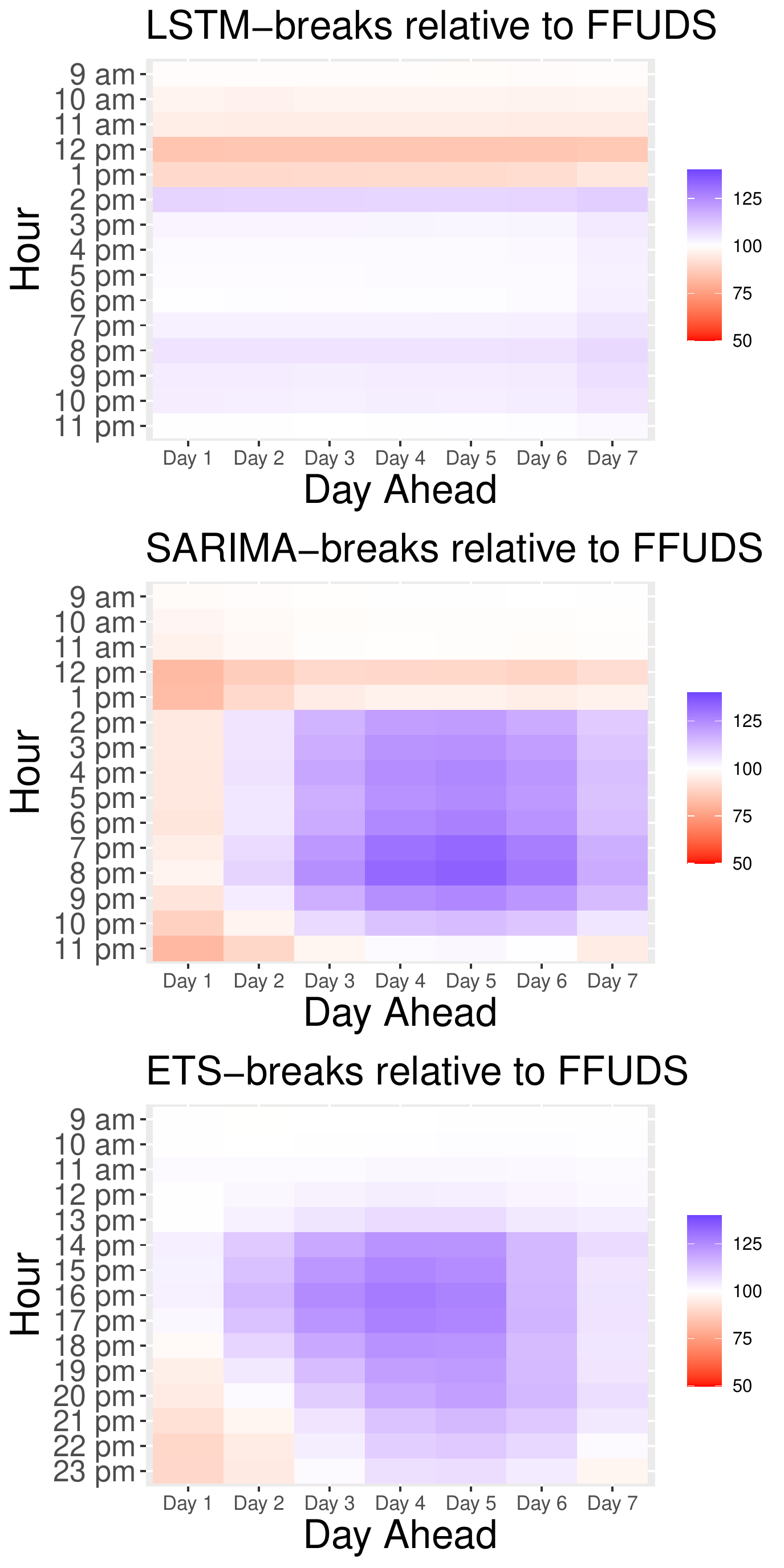}
			\includegraphics[width=0.23\textwidth]{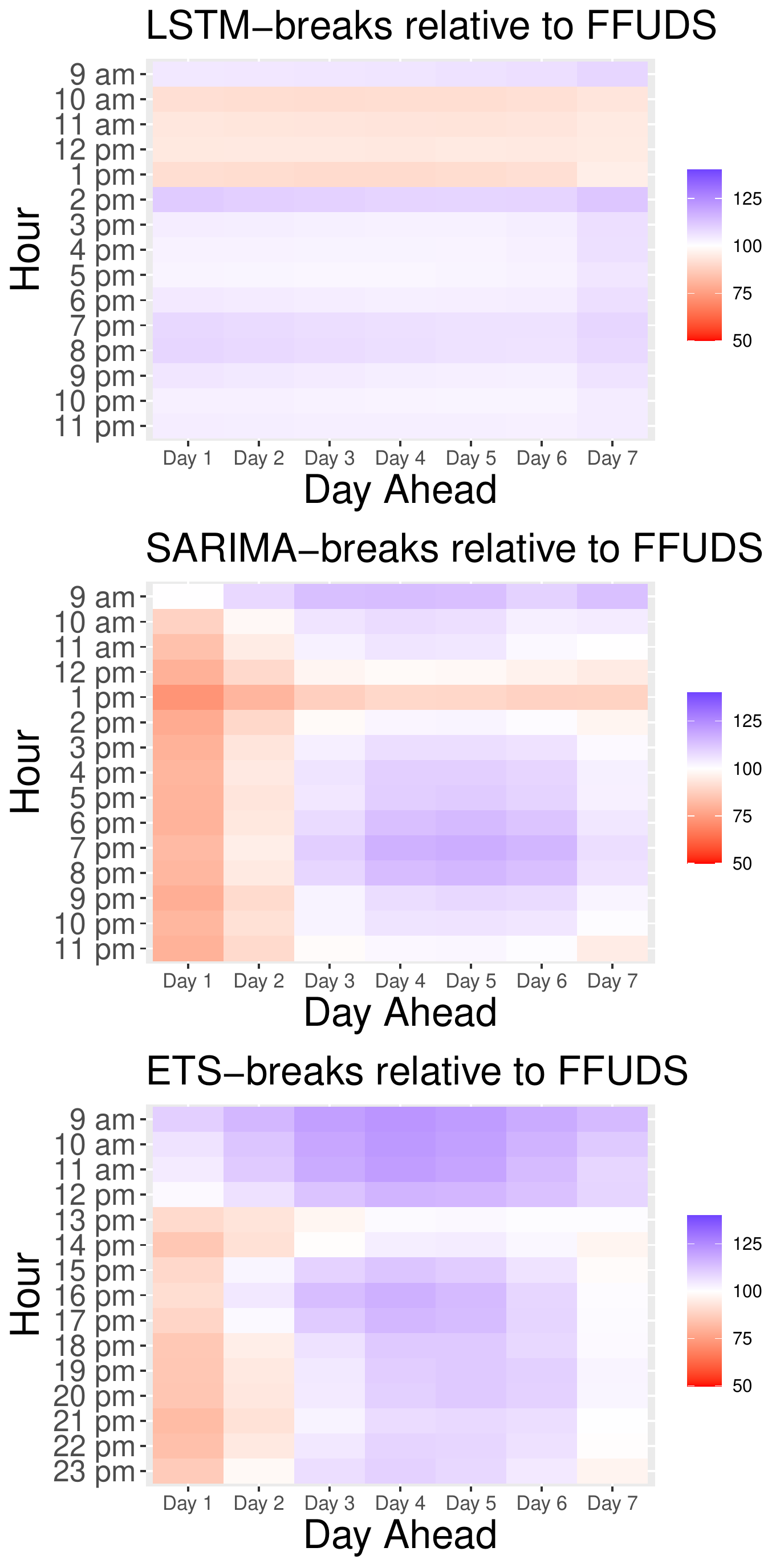}	\includegraphics[width=0.23\textwidth]{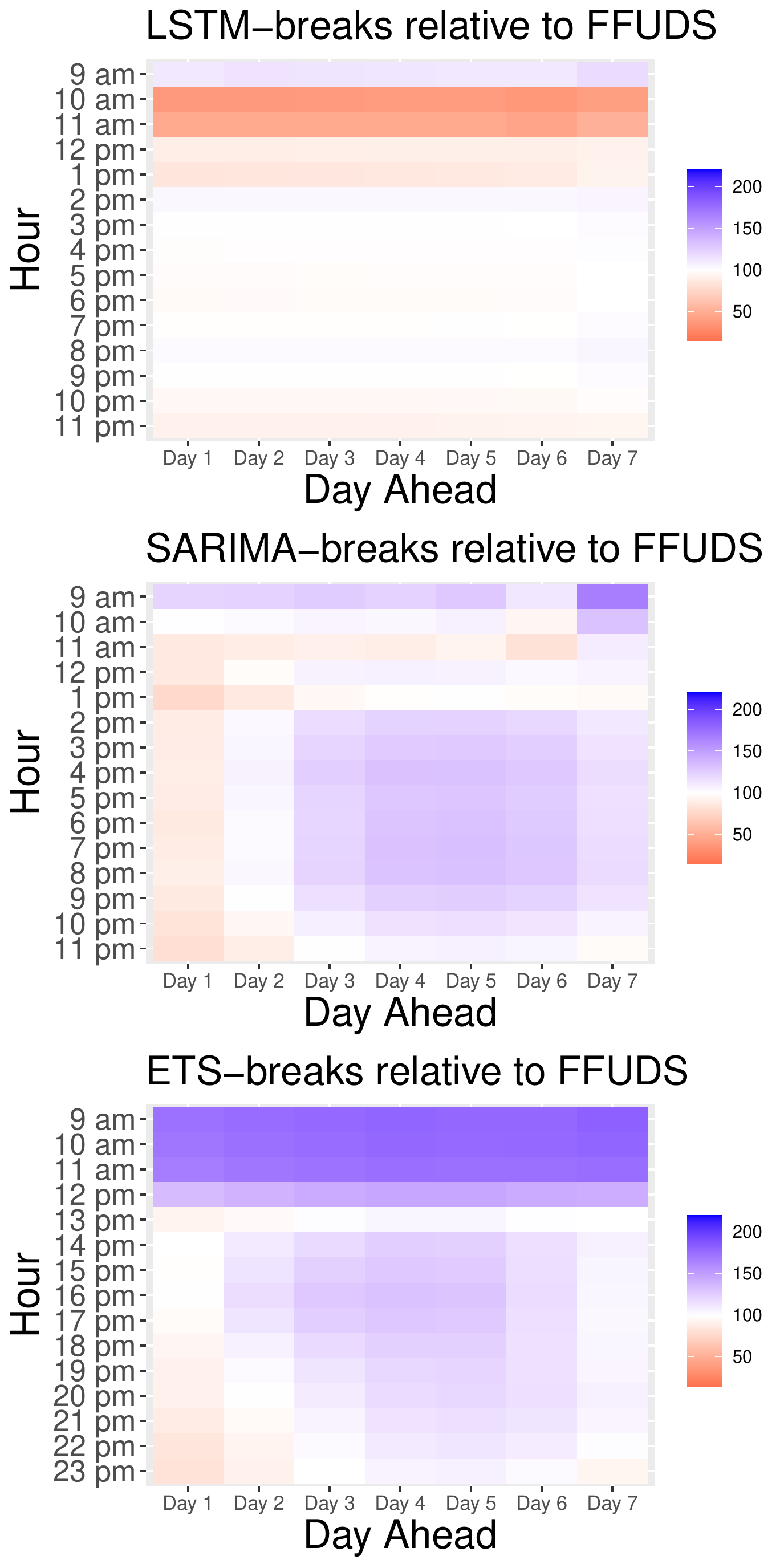}
			\includegraphics[width=0.23\textwidth]{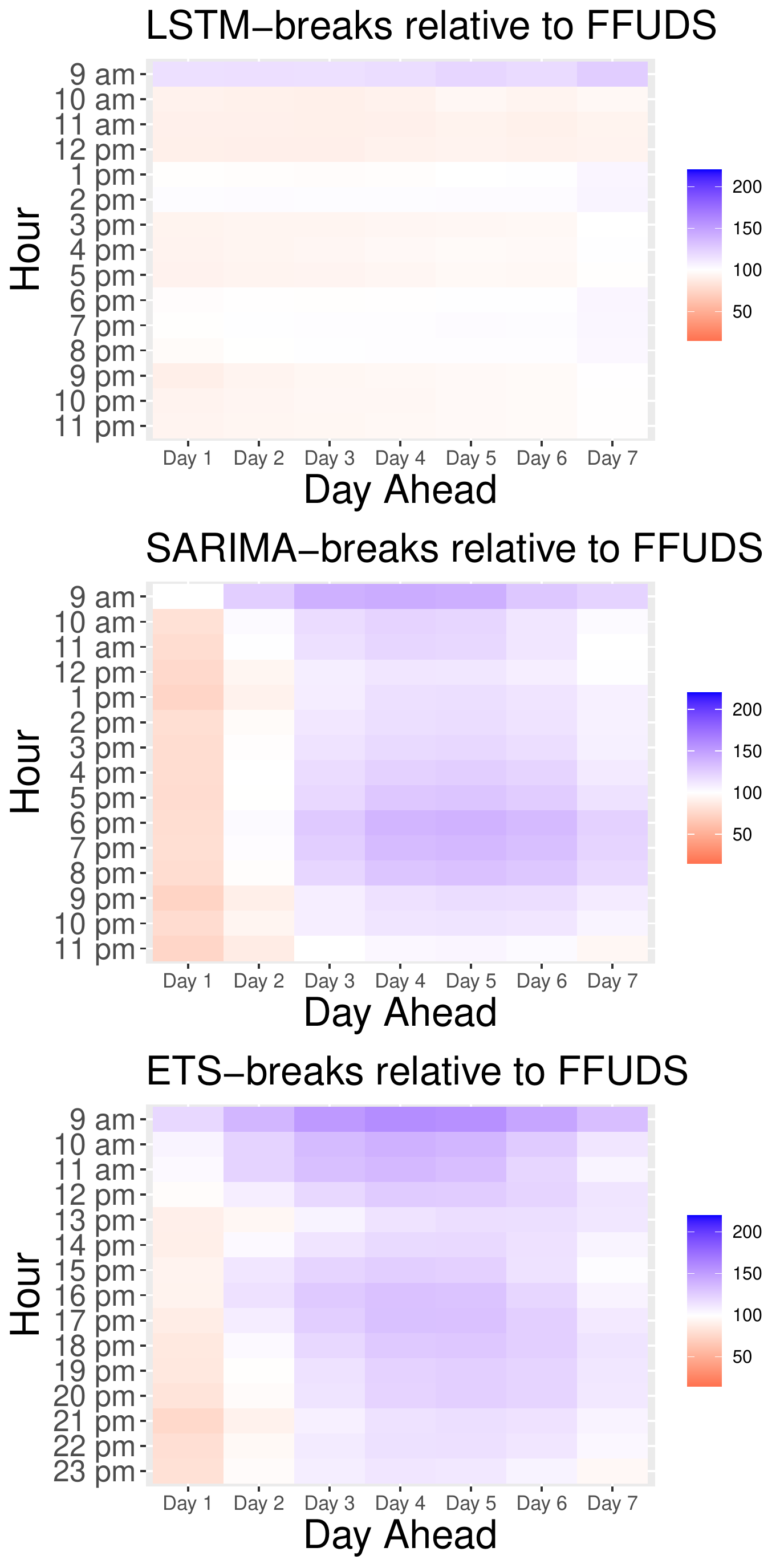}
			\caption{SMAPE and RMSE forecast performance heat maps for benchmarks with breakdown detection.}   \label{fig:heatmap:smapermse}
		\end{center}
		{\footnotesize \raggedright 
			Notes:  Heat maps of SMAPE (columns 1-2: Pre- and Post-Covid) and RMSE (columns 3-4: Pre- and Post-Covid) forecast performance averaged across delivery areas for different hours of day (vertical axis) and days-ahead (horizontal axis).
			Rows: Respectively LSTM, SARIMA and ETS all with breakdown detection relative to FFUDS (domain).
			Values above (below) 100\% visualized in blue (red) indicate better (worse) 
			performance of FFUDS (domain) compared to the benchmark with breakdown detection. Equal performance is visualized in white. } \\[00.5cm]
	\end{figure}

	\begin{figure}[h]
		\centering
		\includegraphics[width=0.49\textwidth]{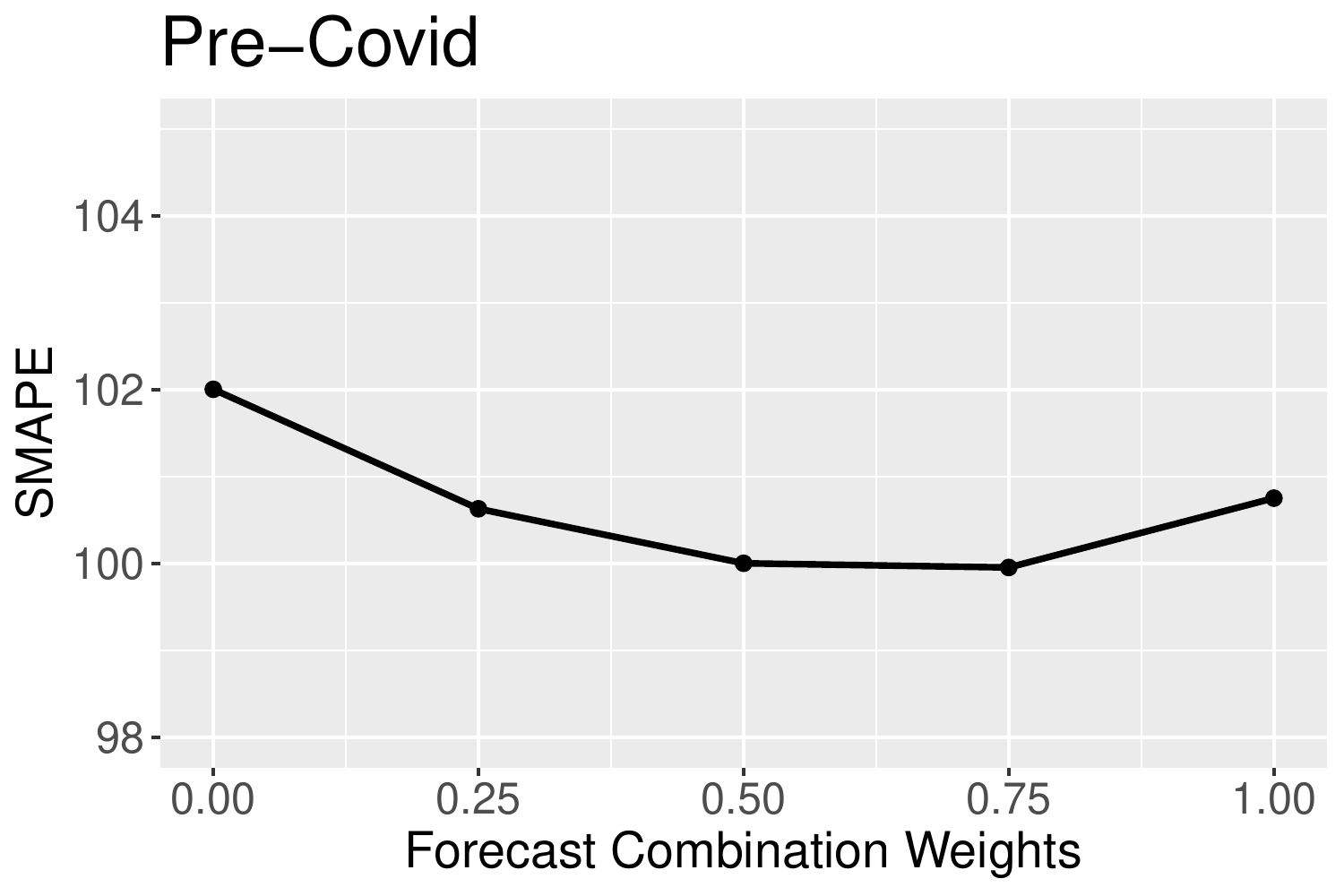}
		\includegraphics[width=0.49\textwidth]{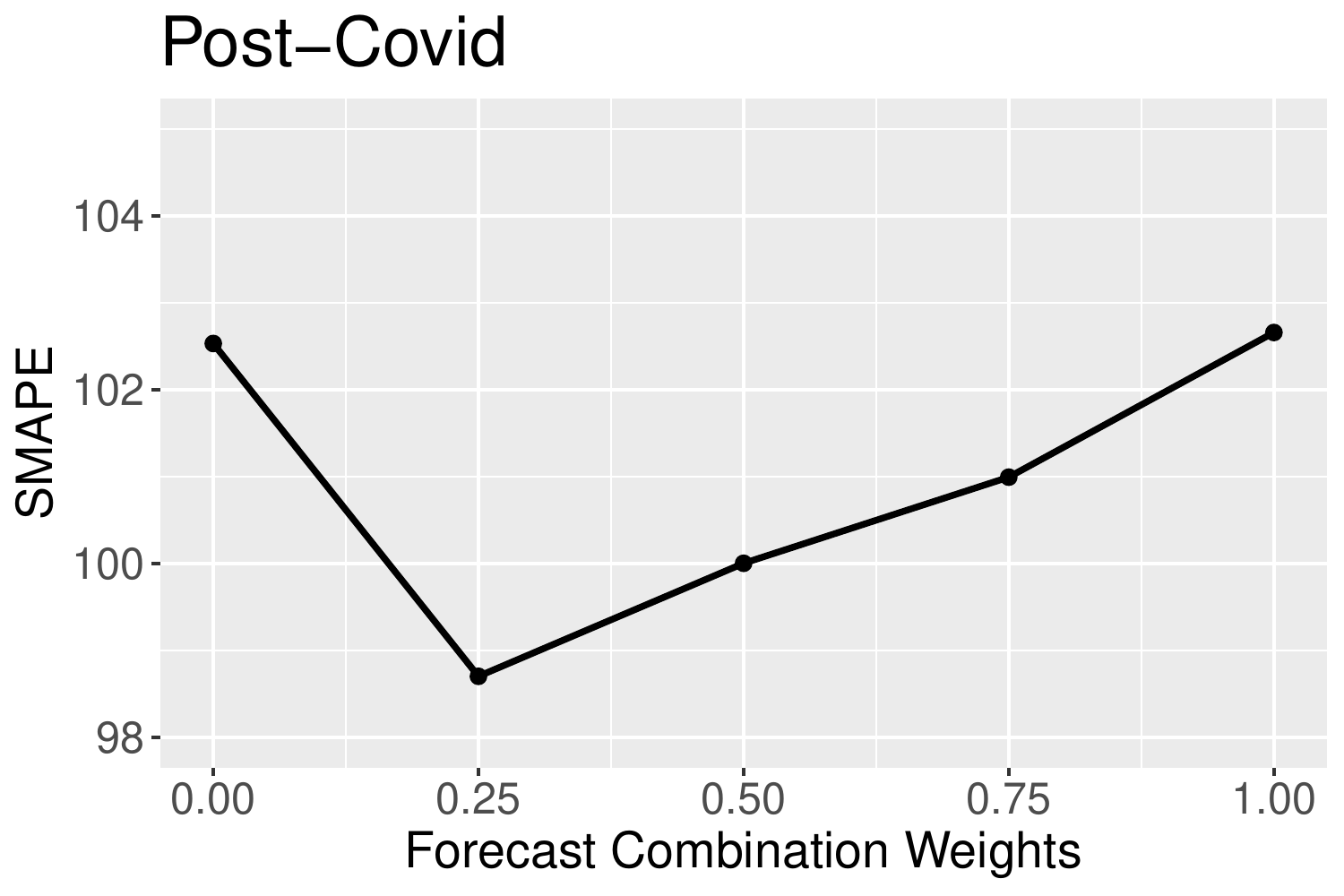}
		\caption{SMAPE forecast performance for different choices of forecast combination weight on the full-sample.} \label{FFUDS-forecast-weights}
	\end{figure}

	\begin{figure}[t]
		\begin{center}
			\includegraphics[width=0.23\textwidth]{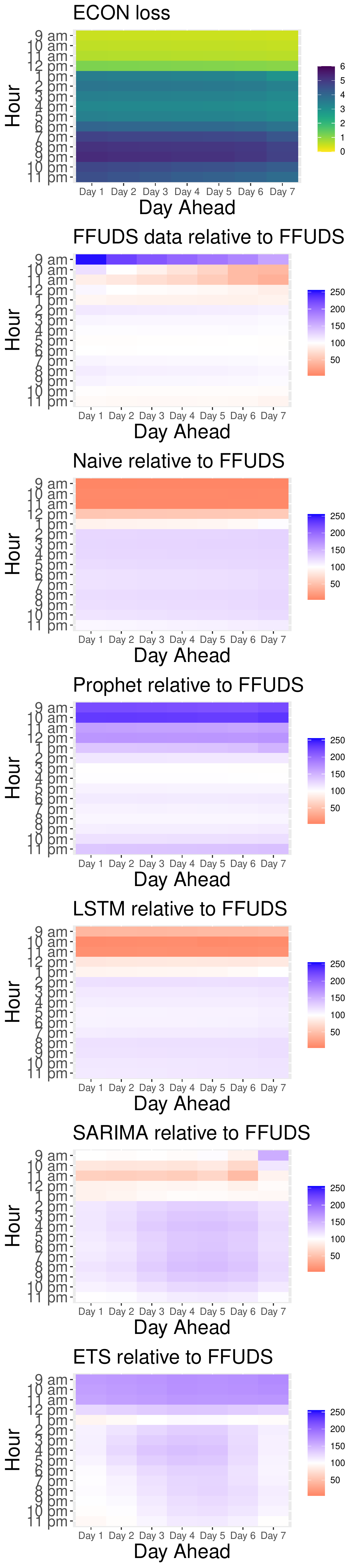}
			\includegraphics[width=0.23\textwidth]{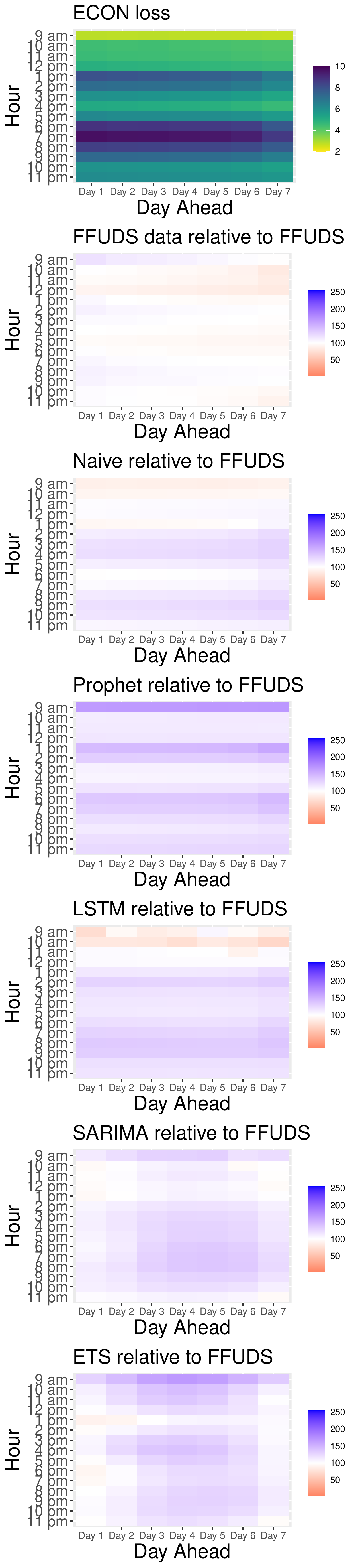}
			\caption{ECON forecast performance heat maps. } 
			\label{fig:heatmap:econ}
		\end{center}
		{\footnotesize \raggedright Notes:  Heat maps of ECON (columns 1-2: Pre- and Post-Covid) forecast performance averaged across delivery areas for different hours of day (vertical axis) and days-ahead (horizontal axis).
			Top: Loss for FFUDS domain.
			Bottom rows: Respectively Naive, FFUDS data, Prophet,  LSTM, SARIMA and ETS relative to FFUDS domain.
			Values above (below) 100\% visualized in blue (red) indicate better (worse) 
			performance of  FFUDS (domain) compared to the benchmark. Equal performance is visualized in white.} \\[0.1cm]
	\end{figure}

	\begin{figure}[t]
		\begin{center}
			\includegraphics[width=0.23\textwidth]{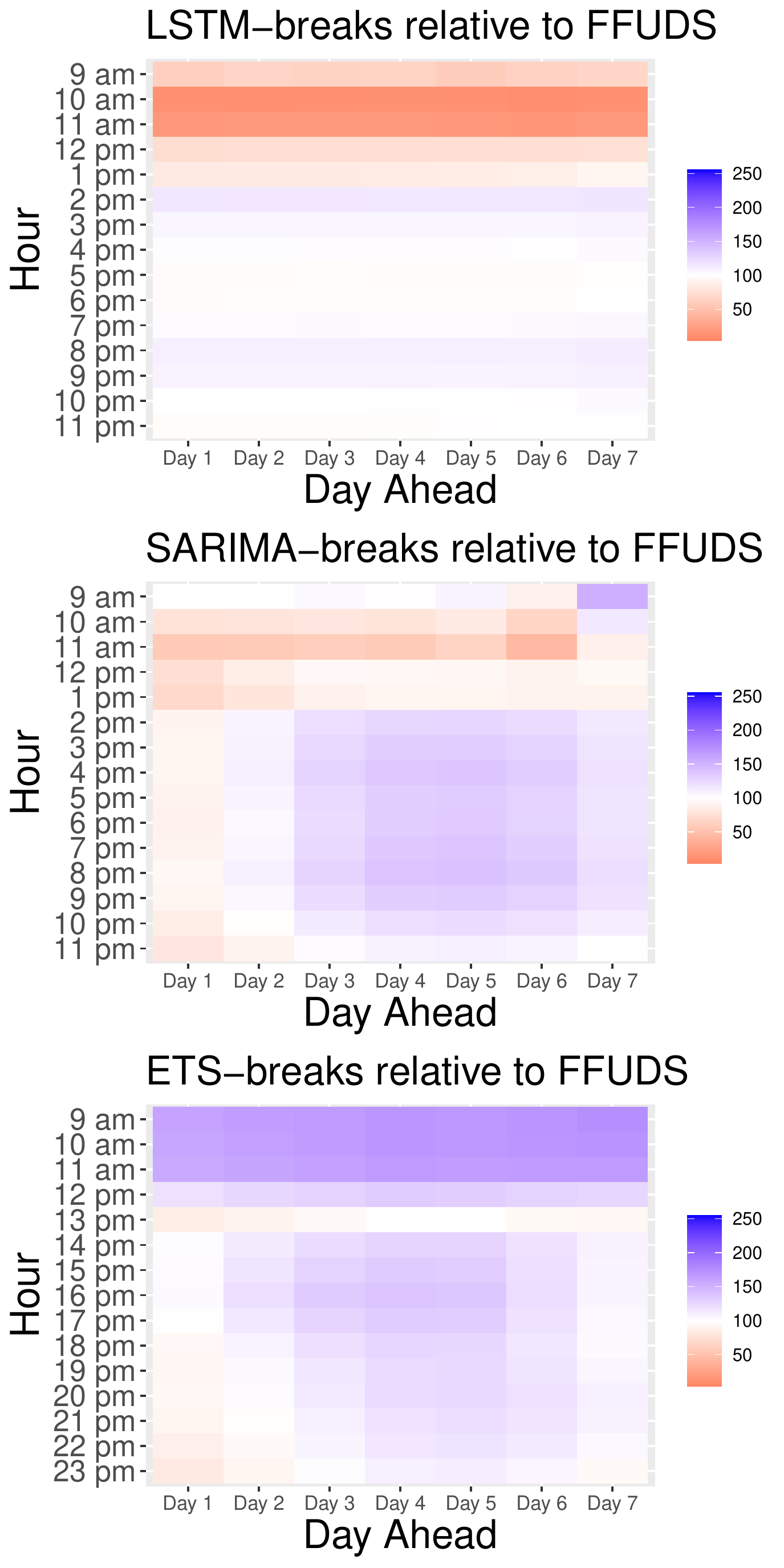}
			\includegraphics[width=0.23\textwidth]{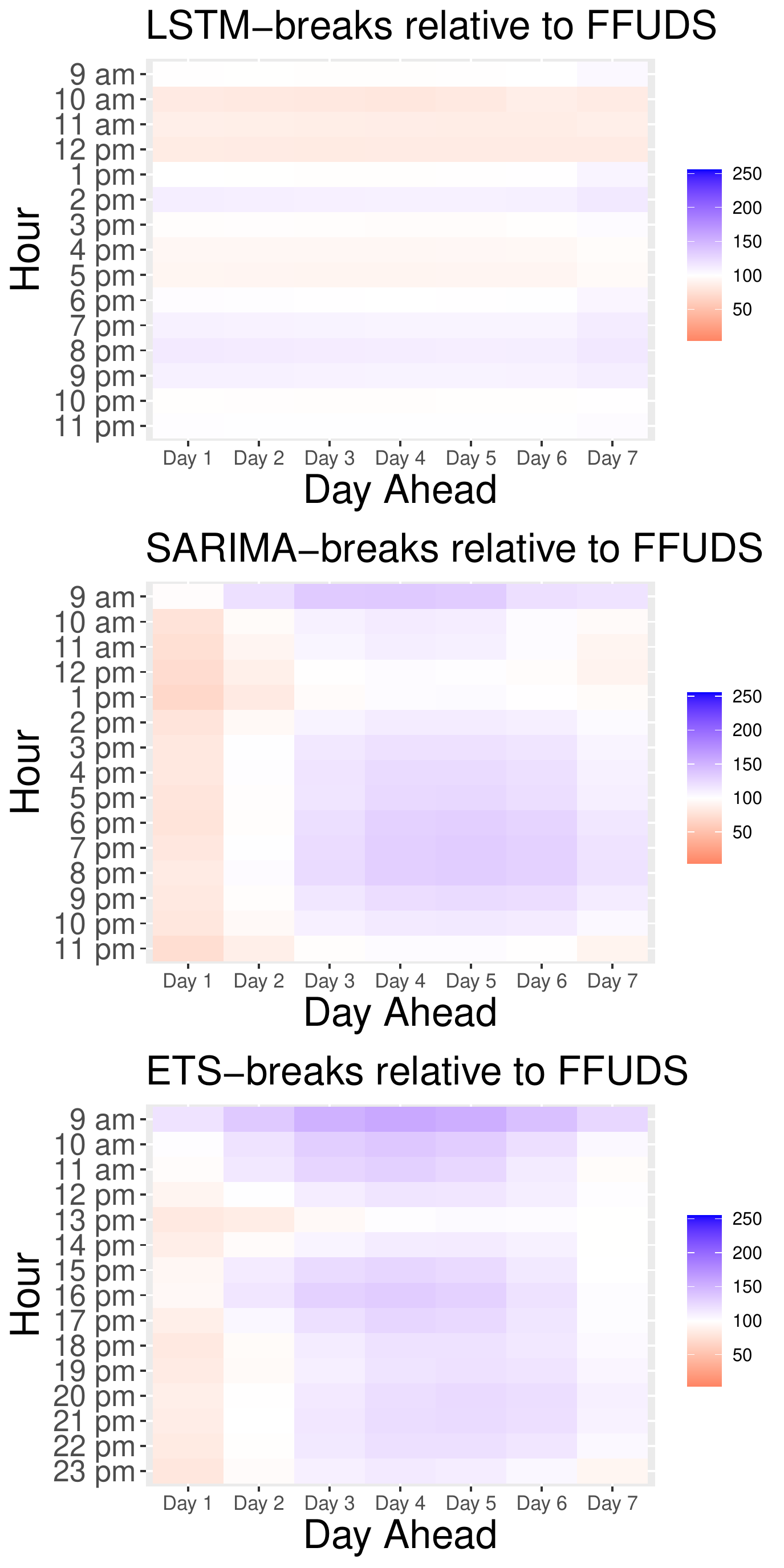}	
			\caption{ECON forecast performance heat maps for benchmarks with breakdown detection.}   \label{fig:heatmap:smapermse}
		\end{center}
		{\footnotesize \raggedright 
			Notes:  Heat maps of ECON  forecast performance averaged across delivery areas for different hours of day (vertical axis) and days-ahead (horizontal axis).
			Rows: Respectively LSTM, SARIMA and ETS all with breakdown detection relative to FFUDS (domain).
			Values above (below) 100\% visualized in blue (red) indicate better (worse) 
			performance of FFUDS (domain) compared to the benchmark with breakdown detection. Equal performance is visualized in white. } \\[00.5cm]
	\end{figure}

	\begin{figure}[htbp]
		\centering
		{\includegraphics[width=16cm,height=9cm]{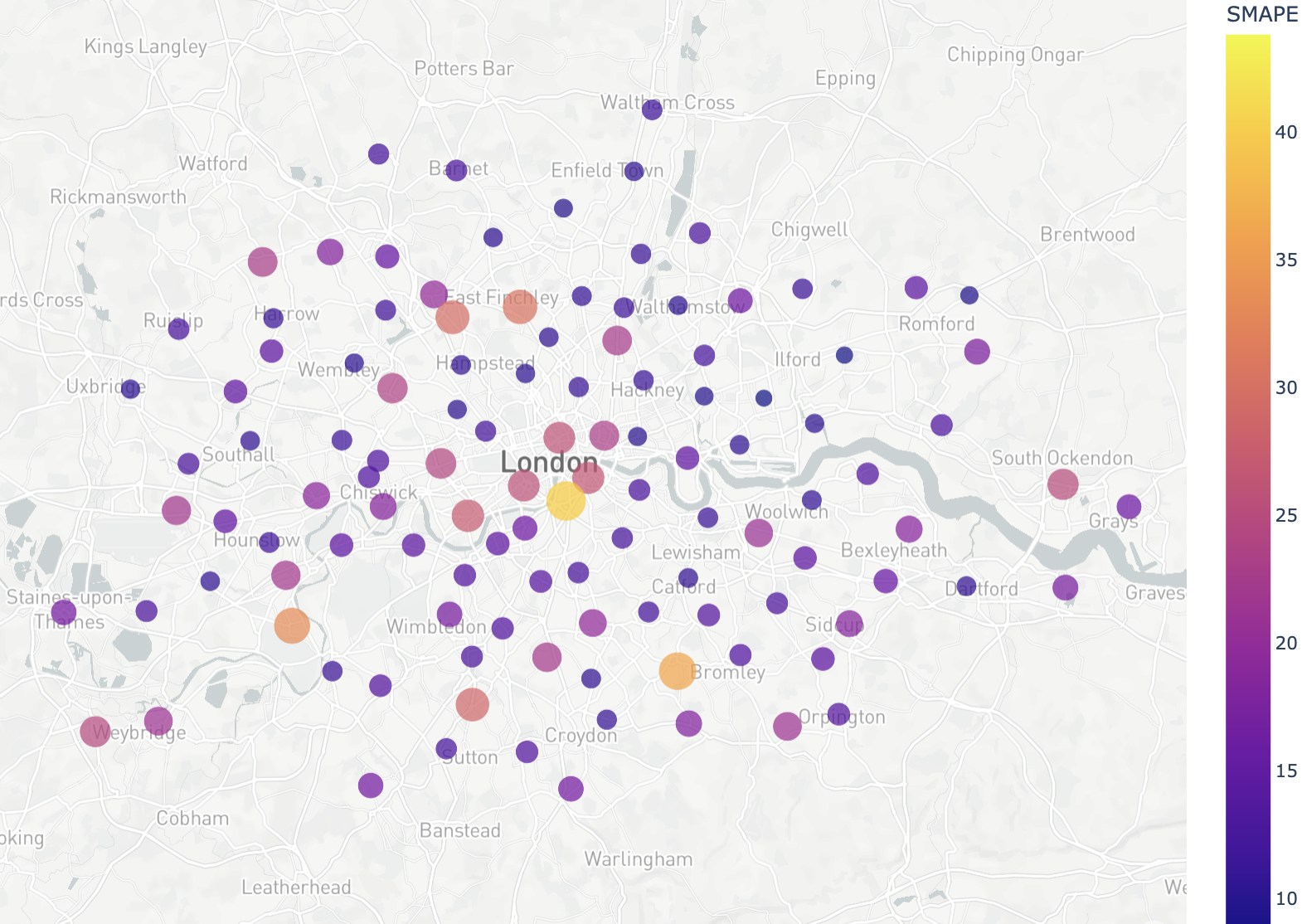} }	
		\caption{London region SMAPE loss; the larger the circle the larger the loss.}  \label{fig:map_SMAPE_london}
	\end{figure} 
	
	\begin{figure}[htbp]
		\centering
		{\includegraphics[width=16cm,height=7cm]{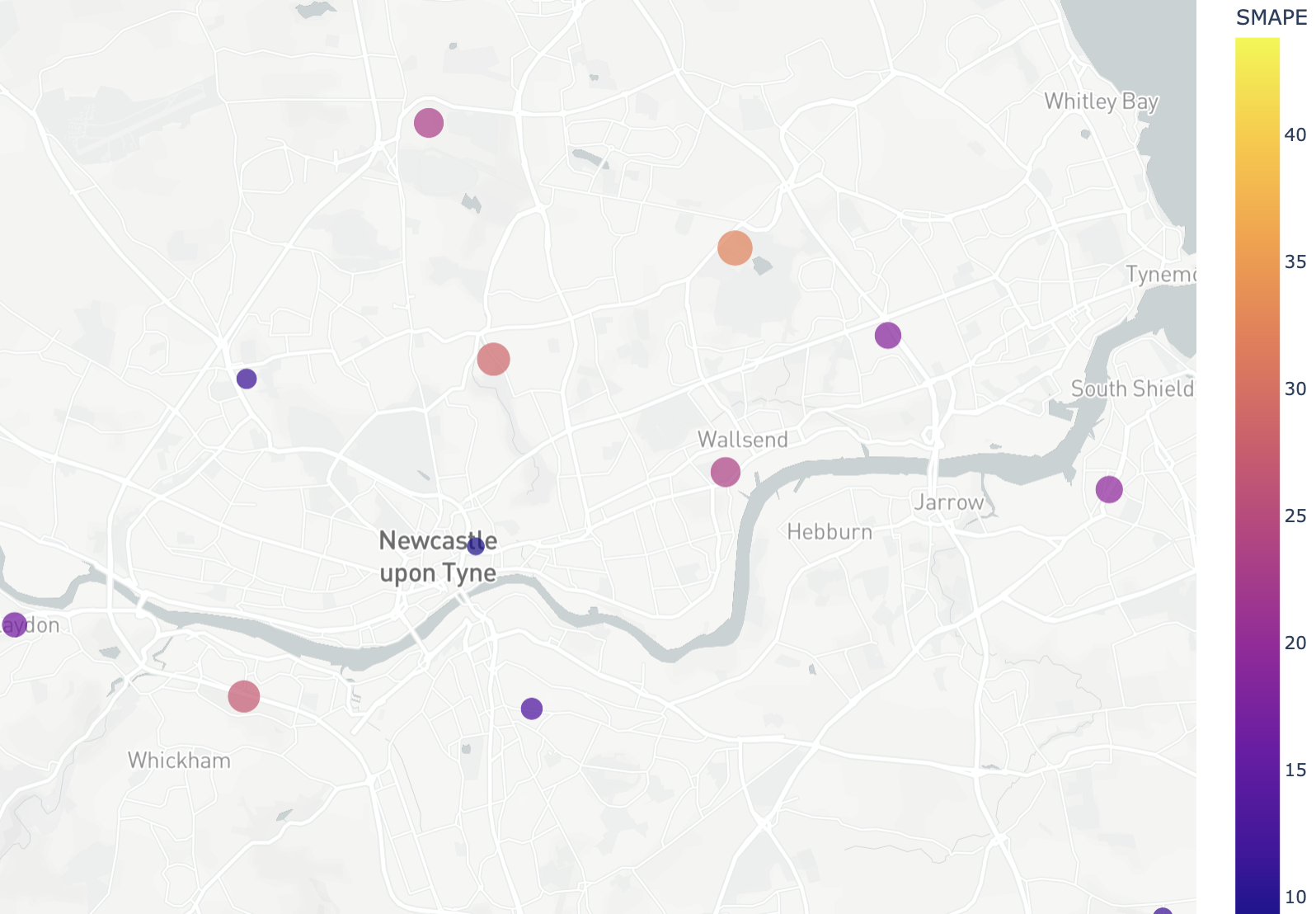} }		
		\caption{Newcastle region SMAPE loss; the larger the circle the larger the loss.}
		\label{fig:map_SMAPE_newcastle}
	\end{figure} 
	
	\newpage
	\clearpage
	
	\renewcommand{\thefigure}{D\arabic{figure}}
	\renewcommand{\thetable}{D\arabic{table}}
	\setcounter{figure}{0}
	\setcounter{table}{0}

	\section{Case Study of the Platform Application: Wimbledon}
	In this supplementary appendix, we focus on one representative  area in London, namely Wimbledon, to have a clearer view on  where FFUDS finds breaks and how it performs throughout the  2019-2021 sample period. 
	
	Figure  
	\ref{fig:wimbledon} 
	presents the streaming results for  Wimbledon. 
	Panel (a) 
	shows demand aggregated to a daily frequency. We clearly see the changes in the local growth patterns, the impact of the Covid-19 first lockdown and the subsequent strong and almost linear growth. The vertical dashed blue lines indicate the nine breaks detected by FFUDS (domain) on the streaming SMAPE  as displayed in
	panel (b). The implied break dates from SMAPE loss exhibit clear changes in forecast loss for most changes, especially towards the end of the sample where SMAPE reaches levels between 10-30\% compared to 20-60\% before Covid-19. Panel (c) 
	highlights the streaming relative performance of FFUDS compared to Prophet, the platform's current forecast method, by simply dividing the respective daily losses over the sample period.  In favor of FFUDS, the ratio is  above 100\% 
	(red line)
	in extended periods, especially since July 2020. For some specific days, Prophet dominates.

	\begin{figure}
		\centering
		\begin{subfigure}[b]{0.45\textwidth}
			\includegraphics[width=\textwidth,height=7cm]{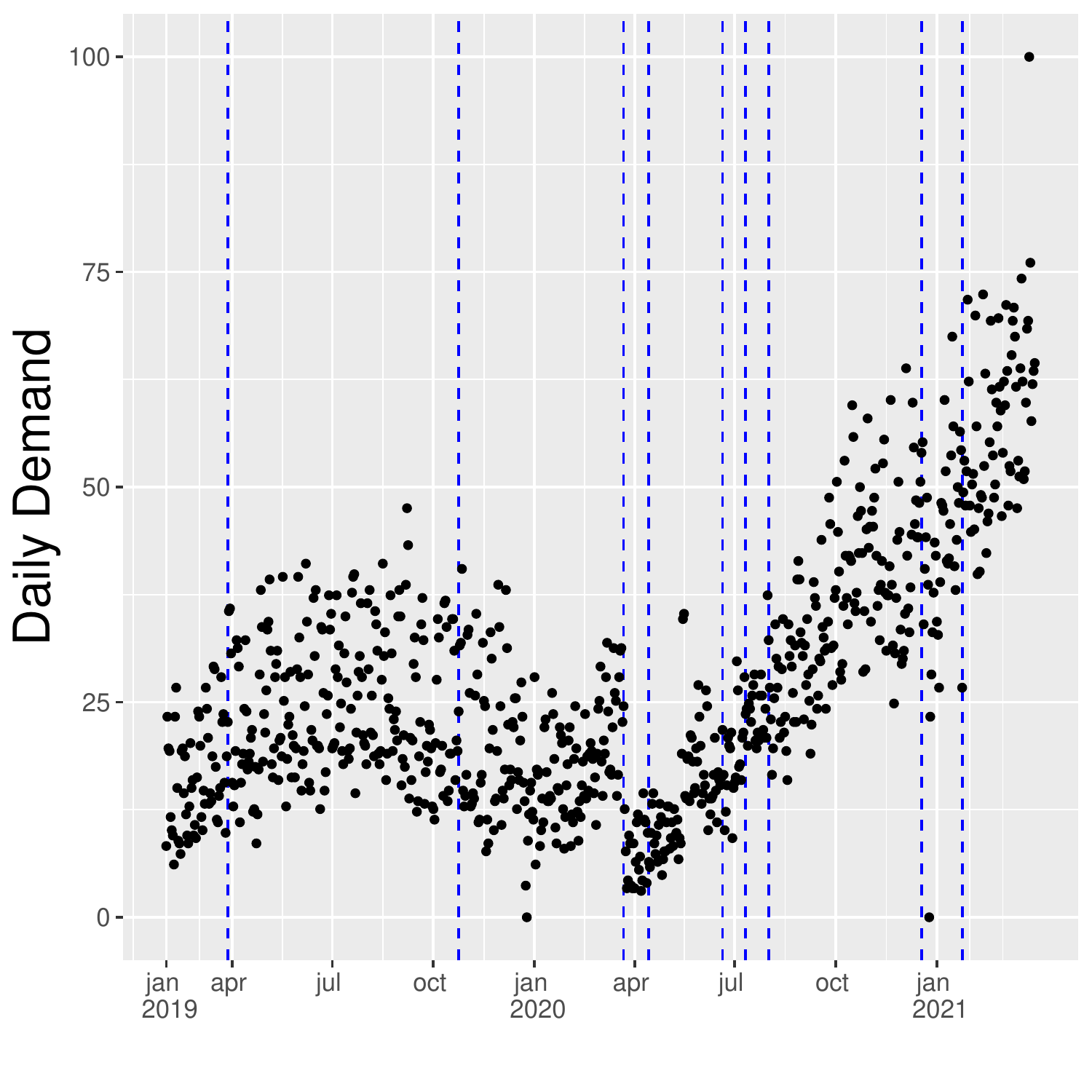}
			\caption{Daily demand \label{fig:Daily_London}}
		\end{subfigure}
		\begin{subfigure}[b]{0.45\textwidth}
			\includegraphics[width=\textwidth,height=7cm]{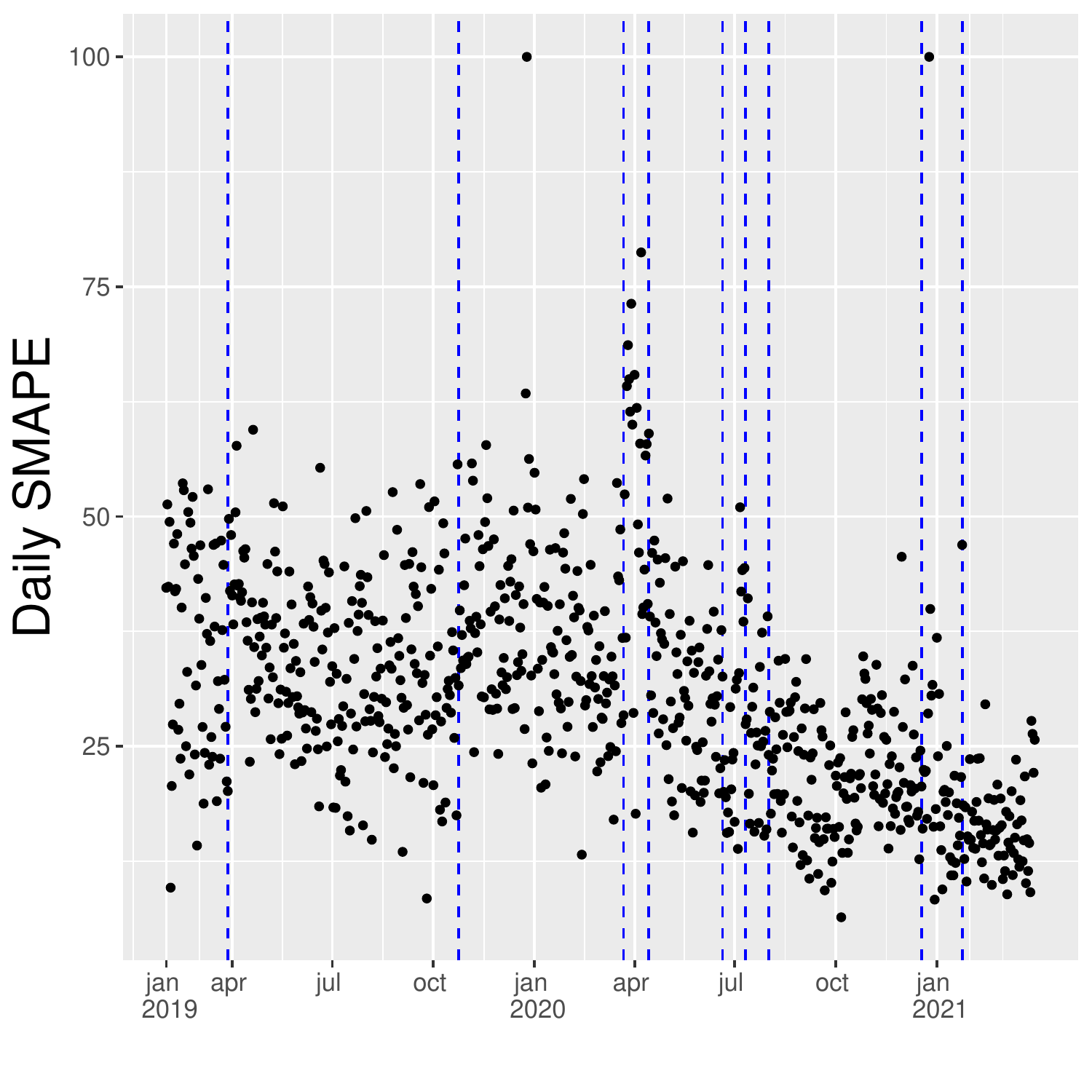}
			\caption{Daily forecast accuracy \label{fig:CentralLondon_Forecastaccuracy}}
		\end{subfigure}
		\begin{subfigure}[b]{0.45\textwidth}
			\includegraphics[width=\textwidth,height=7cm]{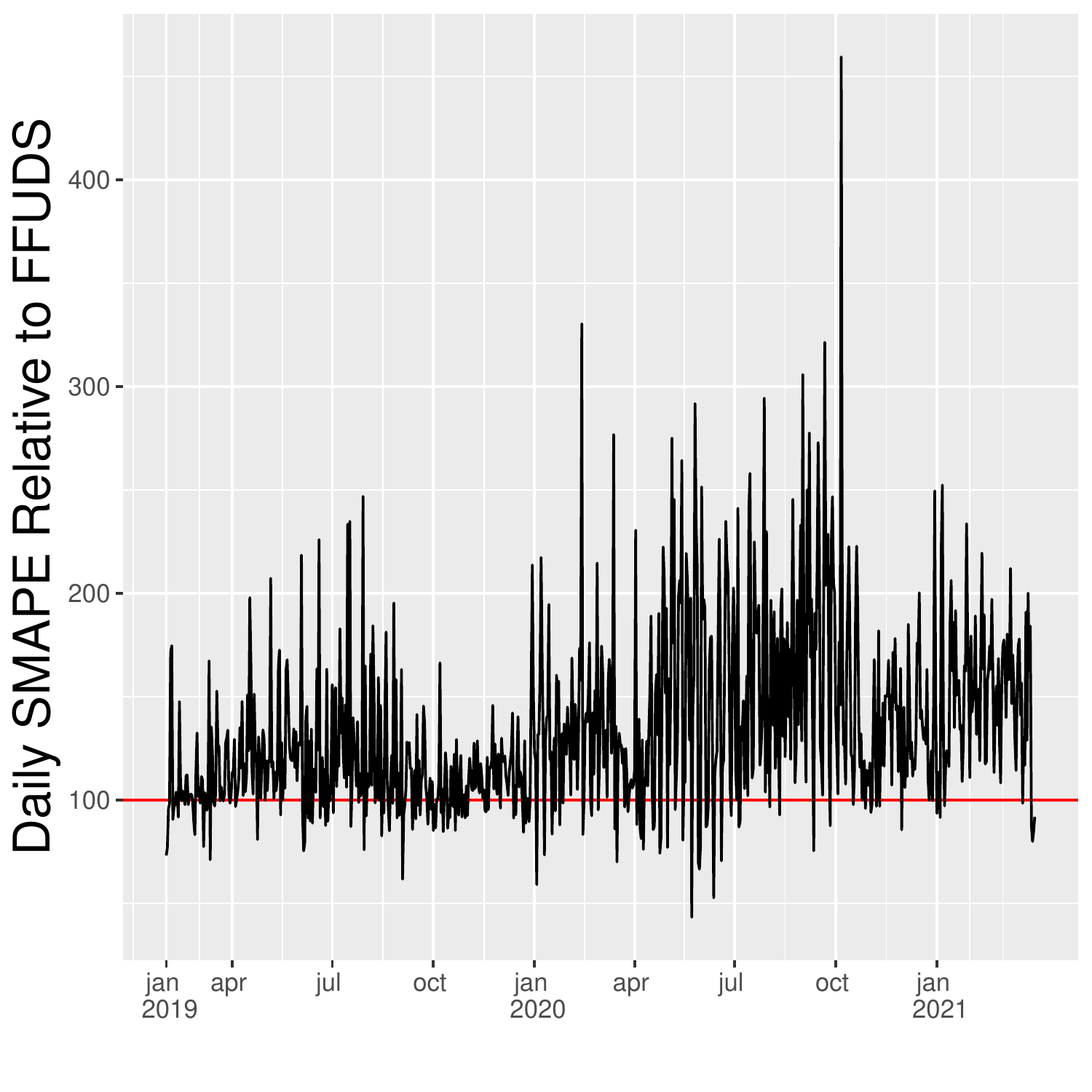}
			\caption{Forecast accuracy Prophet vs. FFUDS \label{fig:CentralLondon_prophet}}
		\end{subfigure}
		\begin{subfigure}[b]{0.45\textwidth}
			\includegraphics[width=\textwidth,height=7cm]{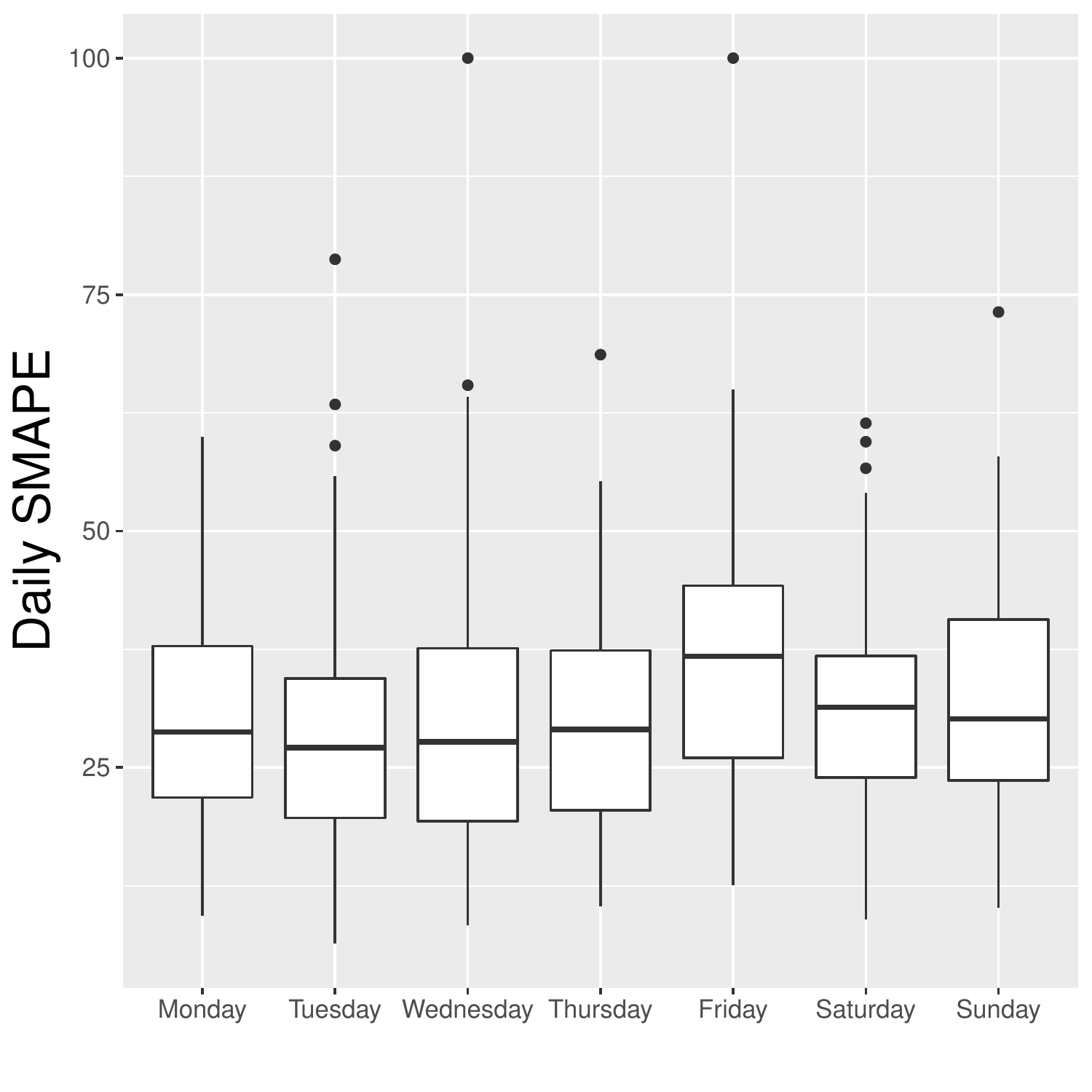}
			\caption{Day of the week forecast accuracy\label{fig:CentralLondon_week}}
		\end{subfigure}
		\caption{Wimbledon  results.  
			\label{fig:wimbledon}}
	\end{figure}
	
	The general results across forecast horizons (over all delivery areas) discussed in the main text reveal that 
	forecast accuracy changes most throughout the business day, with small performance differences between next day and further remote weekdays. 
	For  Wimbledon, we also look at differences in forecast accuracy over the specific  days of the week. Panel (d)  
	shows boxplots of SMAPE  forecast accuracy from Monday until Sunday. Except for Fridays where the median performance is slightly worse, probably due to higher volume and weekend start effects like non-regular sales, forecast performance is stable across days of the week. 

\end{appendices}

\end{document}